\documentclass[twocolumn,trackchanges]{aastex701}

\usepackage{amsmath}
\usepackage{soul}   

\shorttitle{Type Ic Supernovae: SN 2020akf and SN 2021mxx}
\shortauthors{Mulchand Kurre et. al}

\received{December 16, 2025}
\revised{June 10, 2026}
\accepted{June 22, 2026}

\begin{document}
\title{Photometric and Spectroscopic Studies of Type Ic Supernovae SN 2020akf and SN 2021mxx}

\author[0009-0000-4028-1263, gname=Mulchand, sname=Kurre]{Mulchand Kurre}
\altaffiliation{Centre for Mega Projects in Multiwavelength Astronomy, Pt. Ravishankar Shukla University, Raipur}
\affiliation{School of Studies in Physics and Astrophysics, Pt. Ravishankar Shukla University, Raipur 492010, India}
\email[show]{mulkurre123@gmail.com}  

\author[0009-0001-3331-6883, gname='K. R.', sname=Sahu]{K. R. Sahu} 
\affiliation{School of Studies in Physics and Astrophysics, Pt. Ravishankar Shukla University, Raipur 492010, India}
\affiliation{Sant Guru Ghasidas Govt. Post Graduate College, Kurud, Dhamtari, 493663, India}
\email[]{kripa.sahu91@gmail.com}  

\author[gname=Shrutika, sname=Tiwari]{Shrutika Tiwari}
\affiliation{Faculty of Science, The ICFAI University, Raipur 490042, India}
\email{shrutikatiwari7@gmail.com}  

\author[gname=Yogita, sname=Patel]{Yogita Patel}
\affiliation{Department of Astronomy, University of Chile, Santiago 7591245, Chile}
\email[]{yogitapatel2197@gmail.com}  

\author[0009-0007-8332-2812, gname='N. K.', sname=Chakradhari]{N. K. Chakradhari}
\altaffiliation{Centre for Mega Projects in Multiwavelength Astronomy, Pt. Ravishankar Shukla University, Raipur}
\affiliation{School of Studies in Physics and Astrophysics, Pt. Ravishankar Shukla University, Raipur 492010, India}
\email[show]{nkchakradhari@gmail.com}  

\author[0000-0002-6688-0800,gname='D. K.', sname=Sahu]{D. K. Sahu}
\affiliation{Indian Institute of Astrophysics, Koramangala II Block, Bengaluru 560034, India}
\email[]{dks@iiap.res.in}  

\author[0000-0002-0525-0872, gname='Rishabh Singh', sname=Teja]{Rishabh Singh Teja}
\affiliation{Tsung-Dao Lee Institute, Shanghai Jiao Tong University, No.1 Lisuo Road, Pudong New Area, Shanghai, China}
\email[]{rsteja001@gmail.com}  

\author[0000-0003-3533-7183, gname=Anupama, sname='G. C.']{G. C. Anupama}
\affiliation{Indian Institute of Astrophysics, Koramangala II Block, Bengaluru 560034, India}
\email[]{gca@iiap.res.in}  

\author[0000-0001-6706-2749, gname=Mridweeka, sname=Singh]{Mridweeka Singh}
\affiliation{Indian Institute of Astrophysics, Koramangala II Block, Bengaluru 560034, India}
\email[]{yashasvi04@gmail.com}  

\author[0000-0002-7708-3831, gname=Anirban, sname=Dutta]{Anirban Dutta}
\affiliation{Department of Physics and Astronomy, Michigan State University, East Lansing, MI 48824, USA}
\affiliation{Graduate Institute of Astronomy, National Central University, 300 Jhongda Road, 32001 Jhongli, Taiwan}

\email[]{anirbaniamdutta@gmail.com}  

\author[0000-0002-3884-5637, gname=Anjasha, sname=Gangopadhyay]{Anjasha Gangopadhyay}
\affiliation{The Oskar Klein Centre, Department of Astronomy, Stockholm University, Albanova, 106 91 Stockholm, Sweden}
\email[]{anjashagangopadhyay@gmail.com}  

\correspondingauthor{D. K. Sahu}
\email[show]{dks@iiap.res.in}

\begin{abstract}
We present a comprehensive analysis of optical photometry and medium-resolution spectroscopy for  Type Ic supernovae (SNe) SN 2020akf and SN 2021mxx. Our study covers the evolution of SN 2020akf from $-$5 to 109 days and SN 2021mxx from $-$3 to 115 days relative to their $B$-band maximum. 
Their peak quasi-bolometric luminosities are estimated at $\log L_{\text{bol}}^{\text{max}} = 42.37 \pm 0.02$ and $42.21 \pm 0.02\, \text{erg\, {s}}^{-1}$, respectively. The velocities of the  ejecta, derived from the Fe\,\textsc{ii} $\lambda$5169 \AA\ line at maximum light, are approximately $15{,}000\,\text{km\,s}^{-1}$ for SN~2020akf and about $10{,}000\,\text{km\,s}^{-1}$ for SN~2021mxx, which are consistent with those observed in other Type~Ic SNe. 
Using the semianalytical Arnett model, we estimate that SN 2020akf has a kinetic energy of $E_k = 6.33^{+0.68}_{-0.62} \times 10^{51}$ erg and an ejected mass of M$_{\text{ej}} = 4.71^{+0.50}_{-0.46}$\, M$_{\odot}$, while for SN 2021mxx, we get $E_k = 0.54^{+0.08}_{-0.12} \times 10^{51}$ erg and M$_{\text{ej}} = 0.90^{+0.14}_{-0.18}$\,M$_{\odot}$. The mass of $^{56}$Ni synthesized in the explosion is estimated at 0.10 $\pm$ 0.02\,M$_\odot$ for SN 2020akf and 0.05 $\pm$ 0.01\,M$_\odot$ for SN 2021mxx. The metallicities of the host galaxies near the SN regions are $\sim$ 0.81 Z$_\odot$ for SN 2020akf and $\sim$ 0.76 Z$_\odot$  for SN 2021mxx, where Z$_\odot$ indicates solar metallicity. 
Our detailed analysis suggests that SN 2020akf falls into the category of a transitional Type Ic SN, having spectral properties between normal and broad-line Type Ic SNe, while SN 2021mxx is a normal Type Ic SN with an extremely low value of the M$_{\text{ej}}/E_{\mathrm{k}}$ ratio.
  
\end{abstract}


\section{Introduction} 
\label{sec:intro}
Type Ib/c supernovae (SNe) are a subclass of core-collapse SNe (CCSNe), which result from the catastrophic gravitational collapse of the stellar cores in stars with zero-age main-sequence masses 
$\ge$  8\,M$_{\odot}$ \citep{2009smartt}.  Their progenitor stars are believed to have lost their outer envelopes due to strong stellar winds in the case of single massive stars, such as Wolf-Rayet stars \citep{2010yoon, 2011smith, 2015yoon}, or through interactions with a companion in binary systems \citep{2013eldridge}. This stripping process results in the absence of lighter elements, such as hydrogen and helium, in their spectra. Together with the Type IIb subclass, these SNe are collectively called stripped-envelope (SE) CCSNe \citep{1996clocch}. 

The presence or absence of He features, as well as their strength and evolution, are crucial to distinguish between Type Ib and Type Ic SNe. Type Ib SNe display  prominent He\,{\sc i} lines in their spectra, while Type Ic SNe are characterized by the absence of H and He lines
\citep{1986wheeler, 1990harkness, 1990fillippenko, 2001matheson, 2016liu, 2013frey}. 
However, weak He lines have been detected in some Type Ic SNe, such as, 
SN 1994I \citep{Williamson2021},
SN 2007gr \citep{2014chen}, SN 2012ap \citep{mili15}, and SN 2016coi \citep{2017yamanaka, 2018prentice}. 

SE-SNe are characterized by diverse light-curve shapes, luminosities, and spectral features. Their light curves are mainly powered by the radioactive decay chain of $^{56}$Ni,  
with the peak absolute magnitude ranging from $M_V$ = $-$16 to $-$18.5\,mag \citep{Taddia2018}. 
The expansion velocity of the ejecta is found to be ($\sim$ 5,000 to 25,000\,km\,s$^{-1}$). 
A small fraction ($\sim$ 4\%) of Type Ic SNe \citep{2017shivvers}, exhibit broad features corresponding to extremely high ejecta velocities ranging from $\sim$ 15,000 to 35,000\,km\,s$^{-1}$ and significantly higher kinetic energies ($E_k > 10^{52}$\,erg), far exceeding that of typical SESNe. Such highly energetic events are referred to as "hypernovae" \citep{1998iwamoto}, representing a subtype of Type Ic SNe known as broad-lined Type Ic (Ic-BL) SNe.

Type Ic-BL events are sometimes linked to long gamma-ray bursts. The discovery of SN 1998bw associated with GRB 980425 \citep{1998galama}, followed by several SN–GRB pairs \citep{2003hjorth, 2003stanek, 2004malesani, 2006pain, 2012bufano, 2016toy}, established a strong connection between some SNe Ic-BL and GRBs or X-ray flashes. However, late-time radio studies show that not all Ic-BL explosions are associated with GRBs \citep{2002berger,2011corsi}. In contrast, progenitors of normal Type Ic SNe are less extreme, most likely moderately massive Wolf–Rayet stars or interacting binaries. Extensive investigations into the progenitor pathways of Type Ibc SNe further support this picture \citep{Eldridge2013,Kuncarayakti2018,Vartanyan2021,Bersten2014,Eldridge2015}.

Over the past few decades, several distinct subclasses of SE-SNe have been discovered and investigated, including double-peaked Type Ib/c SNe \citep{Das2024}, Ca-rich transients \citep{2010perets, 2012kasliwal, 2017Taubenberger} and interacting SNe, such as Type Ibn \citep{Foley2007, Pastorello2007, Smith2008} and Type Icn \citep{Fraser2021, GalYam2022, Pellegrino22, Perley2022}. 
The growing diversity of SE-SNe poses significant challenges to their study, but also provides an opportunity to better understand their progenitor systems, explosion mechanisms, and environmental influences. Comprehensive, object-specific studies are therefore crucial in advancing our understanding of SE-SNe.

\begin{figure}[ht!]
\centering
\includegraphics[width=0.8\columnwidth]{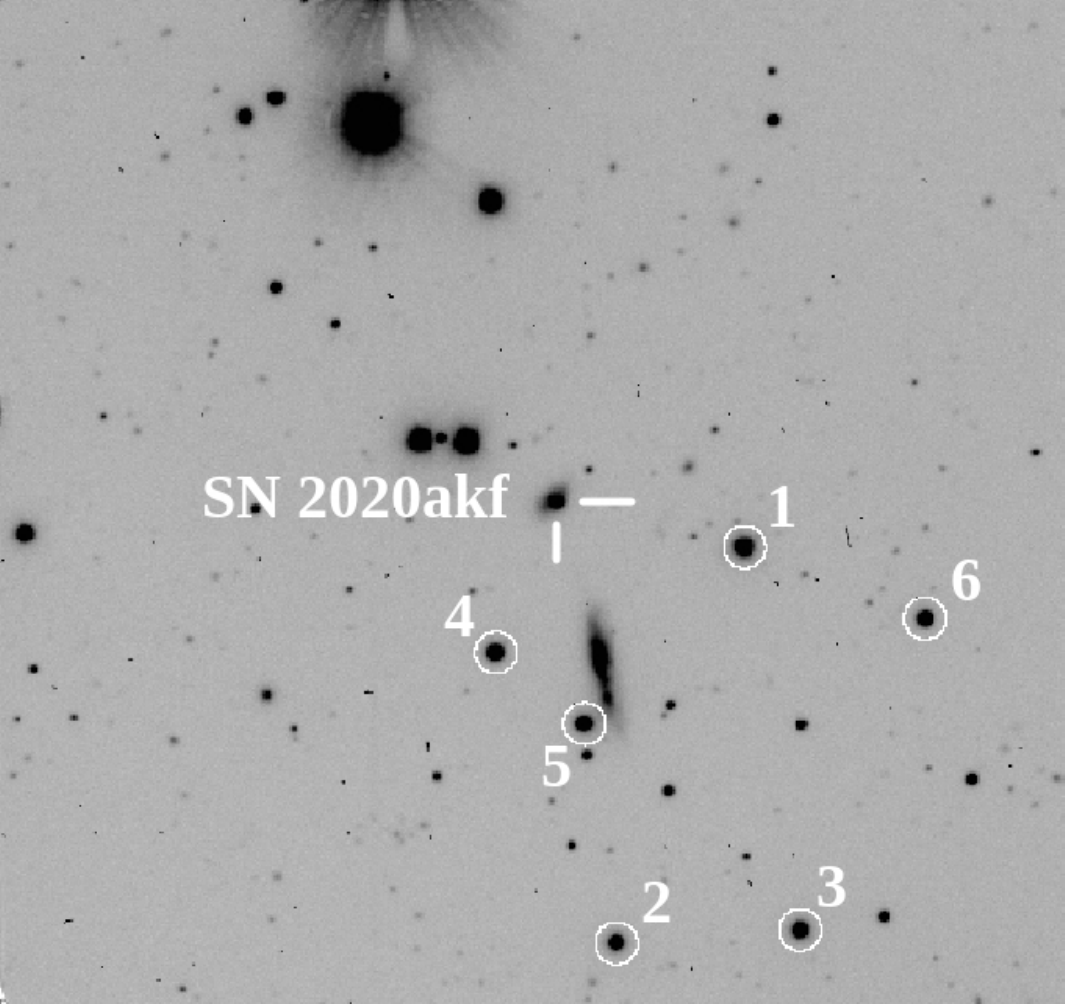}
\includegraphics[width=0.8\columnwidth]{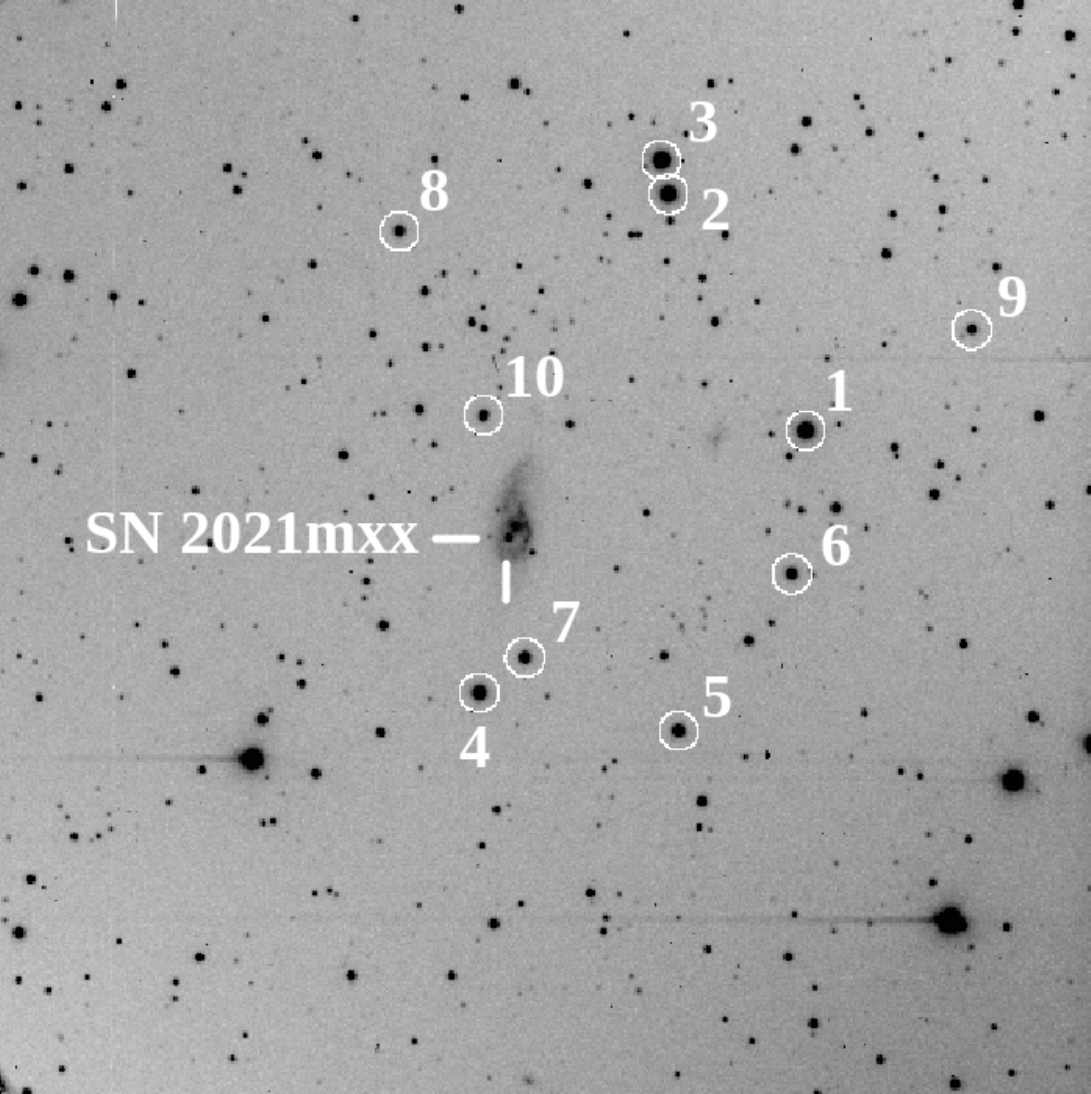}
\caption{Identification charts with North oriented upwards and East to the left, for SNe 2020akf (top) and 2021mxx (bottom). Stars marked were used as secondary standards.} 
\label{fig_sn_field}
\end{figure}

This paper presents a detailed optical photometric and spectroscopic analysis of two Type Ic events, SN 2020akf and SN 2021mxx. SN 2020akf was reported by Asteroid Terrestrial-impact Last Alert System (ATLAS; \citealt{2018Tonry}) on January 20, 2020, with the internal name ATLAS20bmm \citep{2020Tonry}. It was located at RA = 09:28:39.6, DEC = +38:33:47.1, within the host galaxy KUG 0925+387B 
($z$ = 0.012445 $\pm$ 0.000011; source: \href{https://ned.ipac.caltech.edu/}{NED}\footnote{NASA/IPAC Extragalactic Database}, \citealt{alba17}).
SN 2021mxx was discovered  on May 22, 2021 \citep{2021ZTFmxx}, 
by the Zwicky Transient Facility (ZTF; \citealt{ZTFBellm}) at a magnitude of 17.45 in the ZTF-r band and was given the internal designation ZTF21abcgaln. 
The SN was located in the host galaxy UGC 11380 
($z$ = 0.009650 $\pm$ 0.000060; source: \href{https://ned.ipac.caltech.edu/}{NED}, \citealt{bott93}) at RA = 18:56:51.3, DEC = +36:37:20.4. The identification charts for SN 2020akf and SN 2021mxx are displayed in Figure \ref{fig_sn_field}. Some important parameters of SN 2020akf, SN 2021mxx and their host
galaxies (KUG 0925+387B, UGC 11380) are given
in Appendix/Table \ref{tab_sn20akf_sn2021mxx_host}.

Initially, both SN 2020akf \citep{2020akfGSP} and SN 2021mxx \citep{2021mxxGSP} were classified as Type Ia SNe.
However, SN 2021mxx was later reclassified as a Type Ic SN by A. Dahiwale and C. Fremling on behalf of ZTF\footnote{\url{https://www.wis-tns.org/object/2021mxx}}. 
Applying the SNID (Supernova Identification) code to near-maximum light spectra, \citet{MW2023} confirmed that both SNe are indeed Type Ic SNe.

This paper is organized as follows: Section \ref{sec_phot_studies} provides an overview of the photometric observations and data reduction techniques, followed by detailed discussions on the light curve, colour curve, reddening, distance, and absolute magnitude. Section \ref{sec_bol_luminosity} outlines the evolution of the quasi-bolometric light curve.
Section \ref{sec_spec} focuses on spectroscopic observations, data analysis, and the evolution of the spectra and photospheric velocity. Section  \ref{sec_oxygen_mass} presents the estimation of oxygen mass, and Section \ref{sec_metallicity} examines the metallicity of the supernova regions within the host galaxy.  The key explosion parameters are derived in Section \ref{sec_explosion_parameter}. Finally, Section \ref{sec_summary} provides a discussion and summary of this work.

\section{Photometric Studies }
\label{sec_phot_studies}
\subsection{Observations and data reduction}
The optical imaging of SN 2020akf and SN 2021mxx was performed in Bessell’s $UBVRI$ bands using the Himalaya Faint Object Spectrograph Camera (HFOSC), mounted on the 2-meter Himalayan Chandra Telescope (HCT) at the Indian Astronomical Observatory (IAO),  Hanle. The central 2K $\times$ 2K pixels of the 2K $\times$ 4K pixels SITe CCD chip is used for imaging observations. With a plate scale of 0.296 arcsec per pixel, this provides a field of view of 10 $\times$ 10 arcmin$^2$. The CCD has a gain of 1.22 e$^{-}$/ADU and a readout noise of 4.87 e$^{-}$. More detailed information on the telescope and instrument can be found in \citet{Prabhu2000} and the IIA-IAO webpage\footnote{\url{http://www.iiap.res.in/centers/iao}}. 

SN 2020akf was monitored from January 26 to May 19, 2020, while SN 2021mxx was monitored from May 25 to September 20, 2021. In addition to the SNe frames, several calibration frames, including twilight flats and bias frames, were obtained. 
  
The standard star fields PG0231+051, PG0918+029, PG0942-029, PG1047+033, PG1323-086, PG1528+062, and PG1633+099 from \citet{land09}
were observed on photometric nights during the observations of SN 2020akf on January 26, February 3, and April 16, 2020. Similarly, for SN 2021mxx, the standard star fields PG1633+099, PG1657+078, PG2213-006, and PG1528+062 were observed on March 13, 2023.
These standard fields were used to calibrate a set of secondary standards within the supernova fields of SN 2020akf and SN 2021mxx, as marked in Figure \ref{fig_sn_field}. 

Standard data processing steps, including bias subtraction and flat-field correction, were carried out using the Image Reduction and Analysis Facility ({\sc iraf}\footnote{\url{https://iraf-community.github.io/}}) package.  

Aperture photometry was performed on the Landolt standard stars. The photometric zero points were calculated by taking the average atmospheric extinction for  IAO-Hanle site and the average colour terms for the HFOSC. The sequence of secondary standards, calibrated using the calculated zero points and average colour terms, is listed in  Appendix/Table \ref{tab_sec_std}.

SN 2020akf and SN 2021mxx both occurred near the nuclei of their respective host galaxies, in regions with high and variable background levels. To reduce the host galaxy's contribution to the supernova flux, a template-subtraction method was used. Deep $UBVRI$ template images of the KUG 0925+387B and UGC 11380  fields were obtained using the same instrumental setup once the SNe faded sufficiently below the detection threshold. These template images were subtracted from the individual supernova frames following a standard procedure. Aperture photometry was then performed on the supernova left in the template-subtracted frames, which was differentially calibrated relative to secondary standard stars. The final magnitudes of SN 2020akf and SN 2021mxx are provided in Appendix/Tables \ref{tab_mag_sn2020akf} and \ref{tab_mag_sn2021mxx}, respectively.

For this study, we also utilized photometric data in the $g$ and $r$ bands from AleRCE/ZTF \citep{ZTFBellm}, and the $c$-cyan and $o$-orange bands from ATLAS forced photometry \citep{2018Tonry,2020ATLSmith} database. 
ZTF photometry is performed using the ZTF pipeline \citep{ZTFDATA2019, ZTFDATA2023}, which employs difference-imaging analysis and calibrated to the ZTF system using Pan-STARRS1 reference stars. The ATLAS $c$- and $o$-band photometric measurements are derived from difference images with host-galaxy subtraction applied and are calibrated to the ATLAS photometric system, accounting for filter–detector response \citep{2018Tonry}

\subsection{Light curve and colour curve evolution}
\label{sec_light_colour_evol}
\begin{figure}
\centering
\includegraphics[width=0.9\columnwidth, trim={0 0.4cm 0 0},clip]{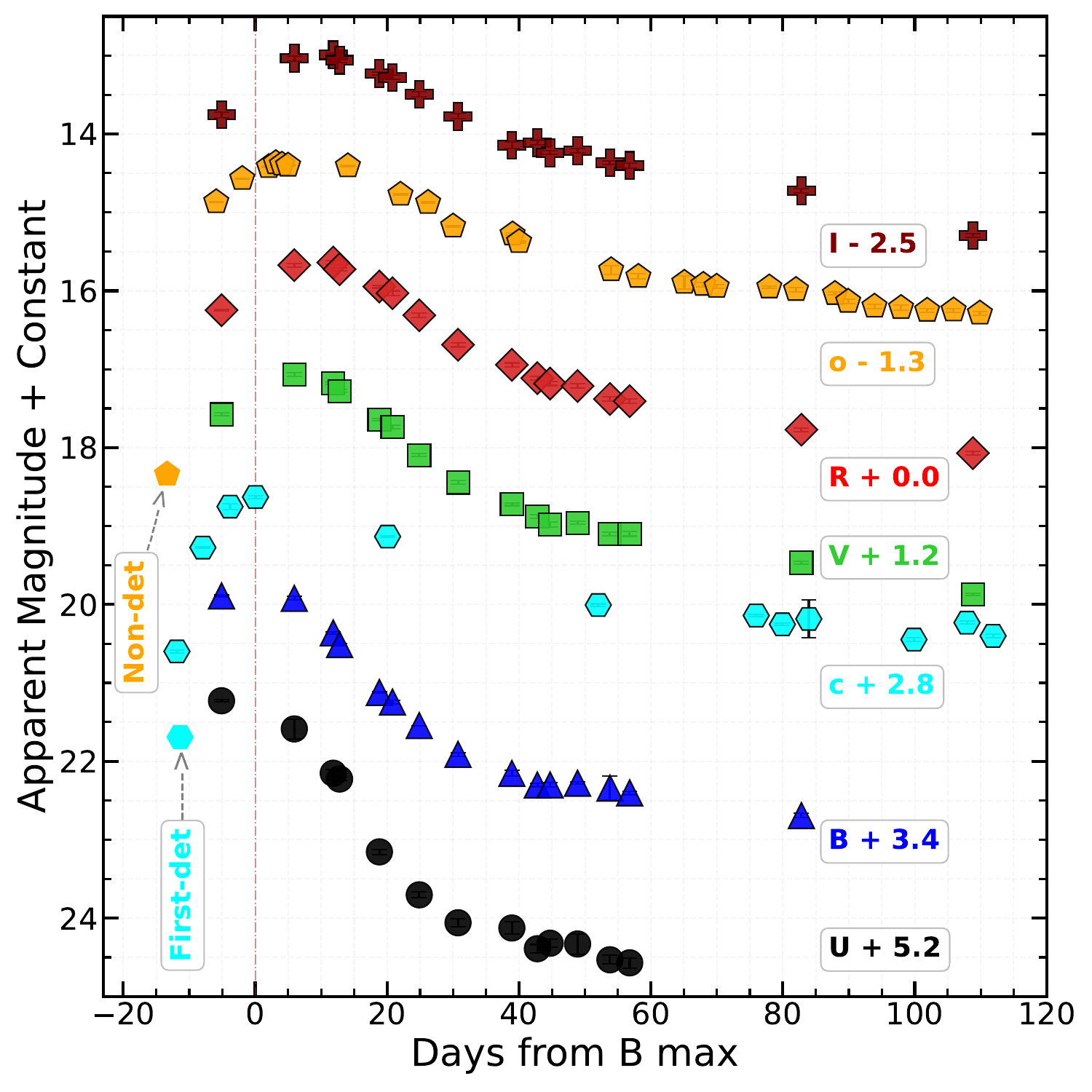}
\includegraphics[width=0.9\columnwidth]{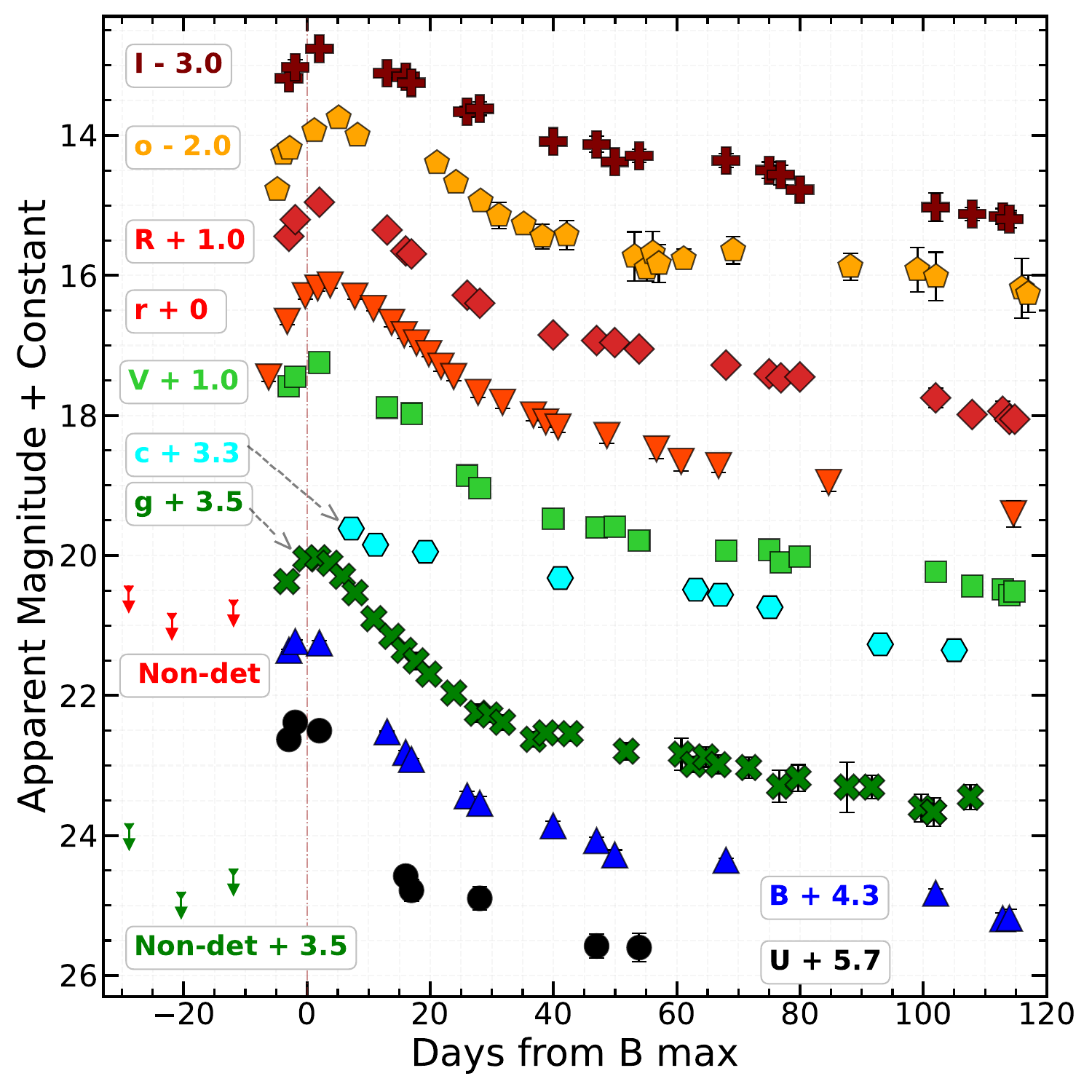}
\caption{The multi-band light curves of SN~2020akf (top) and SN~2021mxx (bottom).
Each light curve is vertically shifted by the amount indicated in the legend for clarity. Phases are measured in days from the $B$-band maximum, marked by a vertical line in each panel. The pre-discovery non-detection limits for SN 2021mxx in the ZTF $g$ and $r$ bands, binned at $\sim$10 days, are also shown.}
\label{fig_lc_2020akf_2021mxx}
\end{figure}

The optical light curves of SN 2020akf are shown in the top panel of Figure \ref{fig_lc_2020akf_2021mxx}.  
The pre-discovery limiting magnitude in the ATLAS $o$-band and the discovery magnitude in the $c$-band are indicated. Similarly, the $UBVRI$ band light curves of SN 2021mxx, along with the data points in the $c$ and $o$-bands from ATLAS and the $g$ and $r$-bands from ZTF, are displayed in the bottom panel of Figure \ref{fig_lc_2020akf_2021mxx}.
The pre-discovery non-detection limits in the $g$ and $r$ bands are also shown.

The date of maximum brightness in the $UBVRI$ bands for each supernova was estimated by iteratively fitting a cubic spline to the observed data points around peak brightness. The estimated values are given in Appendix/Table \ref{tab_lc_parameter}. For SN 2020akf, the $B$-band reached its maximum brightness on JD 245\,8880.4 $\pm$ 0.5 with an apparent magnitude of 16.34 $\pm$ 0.02\,mag. The $U$-band peaked $\sim$ 2 days before, while the $V$-band peaked around 6 days, and the $R$ and $I$-bands peaked $\sim$ 9 days after the $B$-band maximum. For SN 2021mxx, the $B$-band reached its peak on JD 245\,9363.3 $\pm$ 0.5 with an apparent magnitude of 16.90 $\pm$ 0.03\,mag. The $U$-band peaked $\sim$ 0.5 days earlier. The $V$-band peak around $\sim$2 days, and the $R$ and $I$-bands peak occurred  $\sim$ 3.5 days after the $B$-band maximum. 

SN 2020akf was discovered on January 20, 2020, at 12:04:19 UTC (JD 245\,8869.0), with an apparent magnitude of 18.89\,mag. Its last non-detection was on January 18, 2020, at 12:43:12 UTC (JD 245\,8867.0), with a limiting magnitude of 19.64\,mag \citep{2020Tonry}. 
We estimated the explosion epoch to be JD$_{\text{exp}} = 245\,8864.46^{+0.28}_{-0.30}$ for SN 2020akf and JD$_{\text{exp}} = 245\,9352.86^{+0.74}_{-0.57}$ days for SN 2021mxx (refer to Section \ref{Arnett_model_akf_mxx}). The observed time of maximum brightness suggests that SN 2020akf reached peak brightness with a rise time of $t_r \sim 16 \pm 1$ days in the $B$-band, $t_r \sim 22 \pm 1$ days in the $V$-band, and $t_r \sim 24.5 \pm 1$ days in the $R$-band. For SN 2021mxx, the rise time was $\sim 10.44 \pm 1$, $12.2 \pm 1$, and $14 \pm 1$ days in the $B$, $V$ and  $R$ bands, respectively. SN 2021mxx reached peak brightness faster than SN 2020akf in all bands. 

The post-peak light curve evolution of both SNe shows a rapid decline in the $U$ and $B$ bands, while the $V$, $R$, and $I$ band light curves exhibit a slower decline. In the $B$-band, the decline in magnitude 15 days after peak brightness, represented by the $\Delta m_{15}(B)$ parameter, is estimated as 0.98 $\pm$ 0.02 for SN 2020akf and 1.52 $\pm$ 0.03 for SN 2021mxx. The declines in the other bands are listed in Appendix/Table \ref{tab_lc_parameter}.

\begin{figure}
\centering
\includegraphics[width=0.95\columnwidth]{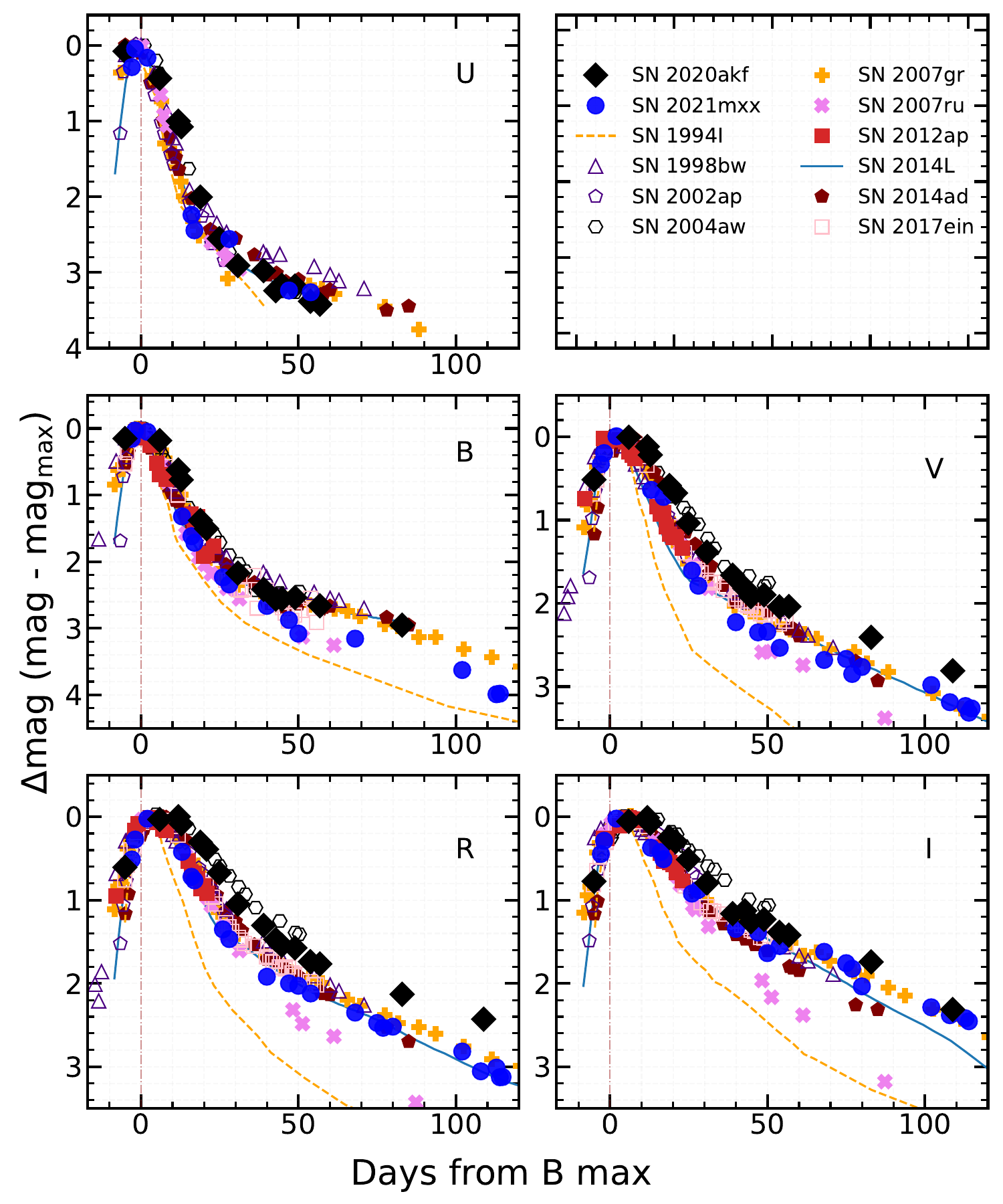}
\caption{Comparison of the $UBVRI$ light curves of SN 2020akf and SN 2021mxx with those of other well-studied Type Ic SNe. The light curves are shifted to align with their peak brightness and the epoch of the $B$-band maximum.} 
\label{fig_lc_comparison}
\end{figure}

The $UBVRI$ light curves of SN 2020akf and SN 2021mxx are compared in Figure \ref{fig_lc_comparison} with those of several well-studied SNe Ic, including 
SN 1994I \citep{1994IRich},
SN 2007gr \citep{2014chen},
SN 2014L \citep{2014LZhang},  
narrow-line Ic event SN 2017ein \citep{2017einDan}, 
inter-mediate Ic event SN 2004aw \citep{taub06} and 
Ic-BL events SN 2002ap \citep{fole03}, SN 2007ru \citep{sahu09},
SN 2012ap \citep{mili15},
SN 2014ad \citep{sahu14ad}, 
and GRB980425/SN 1998bw \citep{1998galama}. 

The early-phase light curves of SN 2020akf, SN 2021mxx, and other SNe appear similar in the $U$ band, while in the $BVRI$ bands, the light curves of SN 2020akf are broader than those of SN 2021mxx. 
The light curves of SN 2020akf match well with the broad light curves of  SN 2004aw, with a similar rise time. In contrast, the light curves and the rise to maximum of SN 2021mxx match well with those of SN 2007gr, SN 2012ap, and SN 2014L (see also Figure \ref{fig_bol_lc}).

\begin{figure}
\centering
\includegraphics[width=0.95\columnwidth]{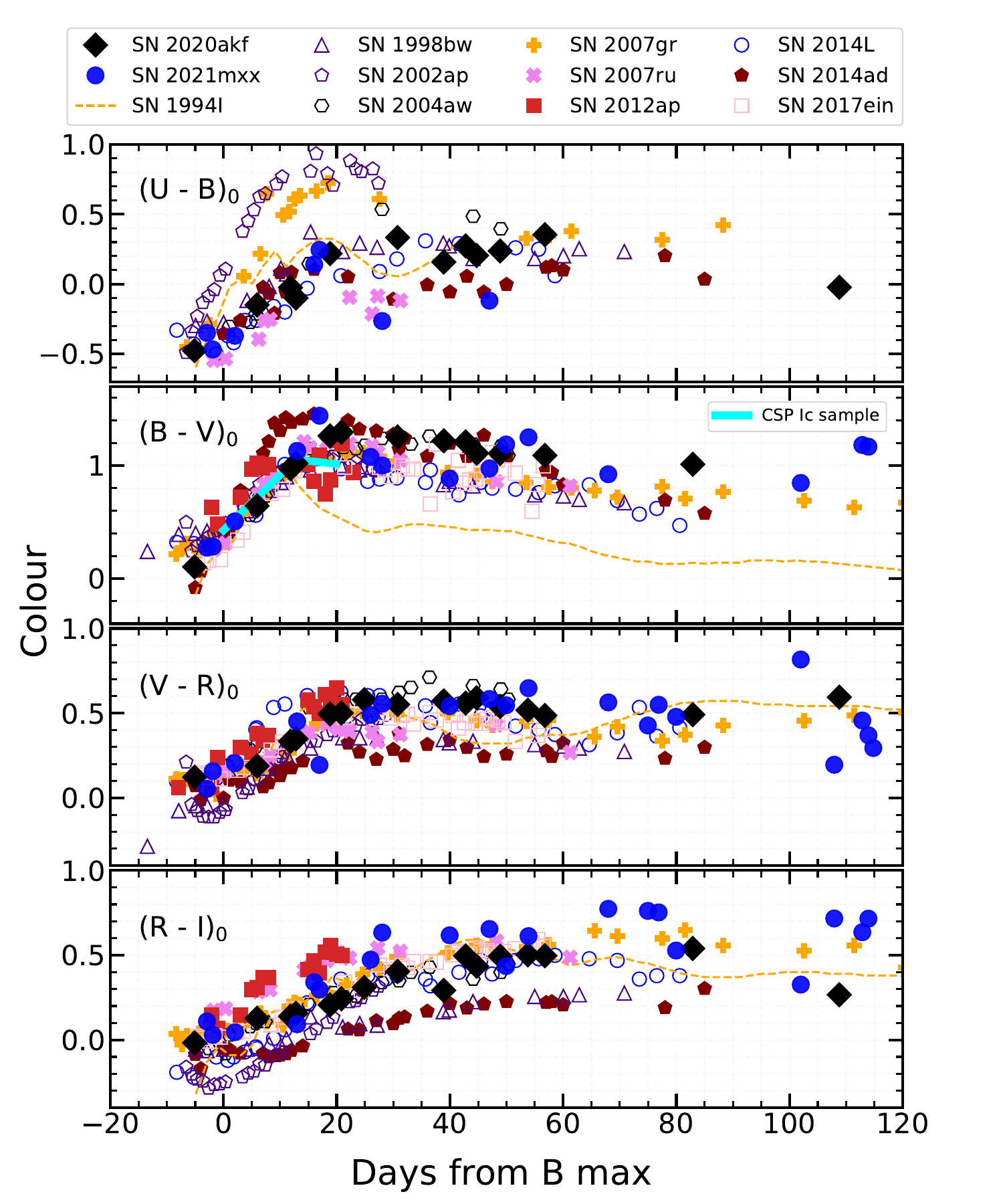}
\caption{The reddening corrected $(U-B)$, $(B-V)$, $(V-R)$ and $(R-I)$ colour curves of SN 2020akf and SN 2021mxx are compared with those of other well-studied Type Ic SNe.
The ($B - V $) colour evolution of minimally reddened sample of Type Ic SNe from the Carnegie Supernova Project (CSP; \citealt{csp2018}) is also shown.}
\label{fig_colour_curve}
\end{figure}

The reddening-corrected colour curves of SN 2020akf and SN 2021mxx are displayed in Figure \ref{fig_colour_curve}, alongside those of other Type Ic SNe. 
The corrections for colour excess were applied using $E(B - V)$ = 0.013\,mag for SN 2020akf and $E(B - V)$ = 0.20\,mag for SN 2021mxx (refer to Section \ref{sec_reddening}). The $E(B - V)$ values for other SNe were obtained from their respective references.

The early $(U-B)$ colour evolution of both SN 2020akf and SN 2021mxx is similar and aligns well with most of the Type Ic SNe used for comparison, except for SN 2002ap and SN 2007gr. However, in later stages, SN 2021mxx becomes bluer. 
The $(B-V)$, $(V-R)$, and $(R-I)$ colours of both SN 2020akf and SN 2021mxx also exhibit similarities with those of most other Type Ic SNe.

\subsection{Reddening, distance and absolute magnitudes}
\label{sec_reddening} 
To estimate the Galactic reddening in the direction of SN 2020akf  and SN 2021mxx, we used the infrared dust map \citep{schl11}. The derived values of $E(B - V)_{\text{Gal}}$ for SN 2020akf and SN 2021mxx are  0.013 $\pm$ 0.001  and  0.08 $\pm$ 0.002\,mag, respectively.  Medium-resolution spectra of SN 2020akf and SN 2021mxx, taken around the epoch of maximum light, do not show the Na\,{\sc i} D absorption line from interstellar matter within the Milky Way or the host galaxies. 

\citet{csp2018} have analyzed the colour evolution of a sample of SESNe from the Carnegie Supernova Project (CSP). They showed that the colour evolution of the minimally reddened sub-sample reveals a high degree of homogeneity during 0 to +20 days with respect to the $B$-band maximum. The ($B-V$) colour curves of SN 2020akf and SN 2021mxx, corrected for Galactic reddening, were compared with the colour template of \citet{csp2018} in Figure \ref{fig_colour_curve} to estimate host reddening. The colour curve of SN 2020akf matches well with the template; hence, Galactic reddening is taken as the total reddening for this event. For SN 2021mxx, a vertical shift of 0.12 mag was required to match the template, giving $E(B - V)_{\text{total}}$ = 0.20 mag. For comparison, we have also plotted the color evolution of some well-studied Type Ic SNe in Figure \ref{fig_colour_curve}.

To estimate the distances to the host galaxies KUG 0925+387B (for SN 2020akf) and UGC 11380 (for SN 2021mxx), we used the radial velocities of the hosts, corrected for local group infall onto the Virgo cluster. The corrected velocities are $v$ = 3,903 $\pm$ 14\,km\,s$^{-1}$ for KUG 0925+387B and $v$ = 3,225 $\pm$ 25\,km\,s$^{-1}$ for UGC 11380 \citep[source: \href{https://ned.ipac.caltech.edu/}{NED},][]{moul00}. Adopting a value of $H_0$ = 72 $\pm$ 5\,km\,s$^{-1}$\,Mpc$^{-1}$ \citep{free01}, we estimated the distance to SN 2020akf to be 54.21 $\pm$ 4\,Mpc, corresponding to a distance modulus of 33.67 $\pm$ 0.15\,mag. For SN 2021mxx, the estimated distance is 44.79 $\pm$ 3\, Mpc, with a distance modulus of 33.26 $\pm$ 0.15\,mag. 

The peak absolute magnitudes of SN 2020akf and SN 2021mxx in various bands were estimated using the calculated reddening values and distance moduli. The values are listed in Appendix/Table \ref{tab_lc_parameter}. The $V$-band peak absolute magnitudes are $-$17.85 $\pm$ 0.15\,mag for SN 2020akf and $-$17.60 $\pm$ 0.15\,mag for SN 2021mxx. SN 2020akf and SN 2021mxx are brighter than SN 2002ap ($M_{V}$ = $-$17.37 $\pm$ 0.05\,mag; \citealt{fole03}; \citealt{pand03}),
SN 2007gr ($M_{V}$ = $-$17.22 $\pm$ 0.18\,mag; \citealt{hunt09}) and SN 2017ein ($M_{V}$ = $-$17.47 $\pm$ 0.35\,mag; \citealt{2017einDan}). Their absolute magnitude is comparable to that of  SN 1994I ($M_{V}$ = $-$17.62 $\pm$ 0.3\,mag; \citealt{1994IRich}; \citealt{saue06}),  SN 2014L ($M_{V}$ = $-$17.73 $\pm$ 0.28\,mag; \citealt{2014LZhang}) but lower than  that of  SN 2004aw ($M_{V}$ = $-$18.02 $\pm$ 0.3\,mag; \citealt{taub06}, SN 2012ap ($M_{V}$ = $-$18.67 $\pm$ 0.08\,mag; \citealt{mili15}) and SN 2014ad ($M_{V}$ = $-$18.86 $\pm$ 0.23\,mag; \citealt{sahu14ad}). 

\subsection{Bolometric Luminosity}
\label{sec_bol_luminosity}

The quasi-bolometric light curves of SN 2020akf and SN 2021mxx were estimated from the $UBVRoI$ and $UBgVrRI$ observations described in Section \ref{sec_phot_studies}. The magnitudes were corrected for reddening, with $E(B-V)$ values of 0.013 and 0.20\,mag, respectively, following the \citet{card89} extinction law. Using the distances derived in Section \ref{sec_reddening}, these corrected magnitudes were used to construct the bolometric light curves with the SuperBol code \citep{Nicholl18}. 

\begin{figure}[ht!]
\centering
\includegraphics[width=0.9\columnwidth]{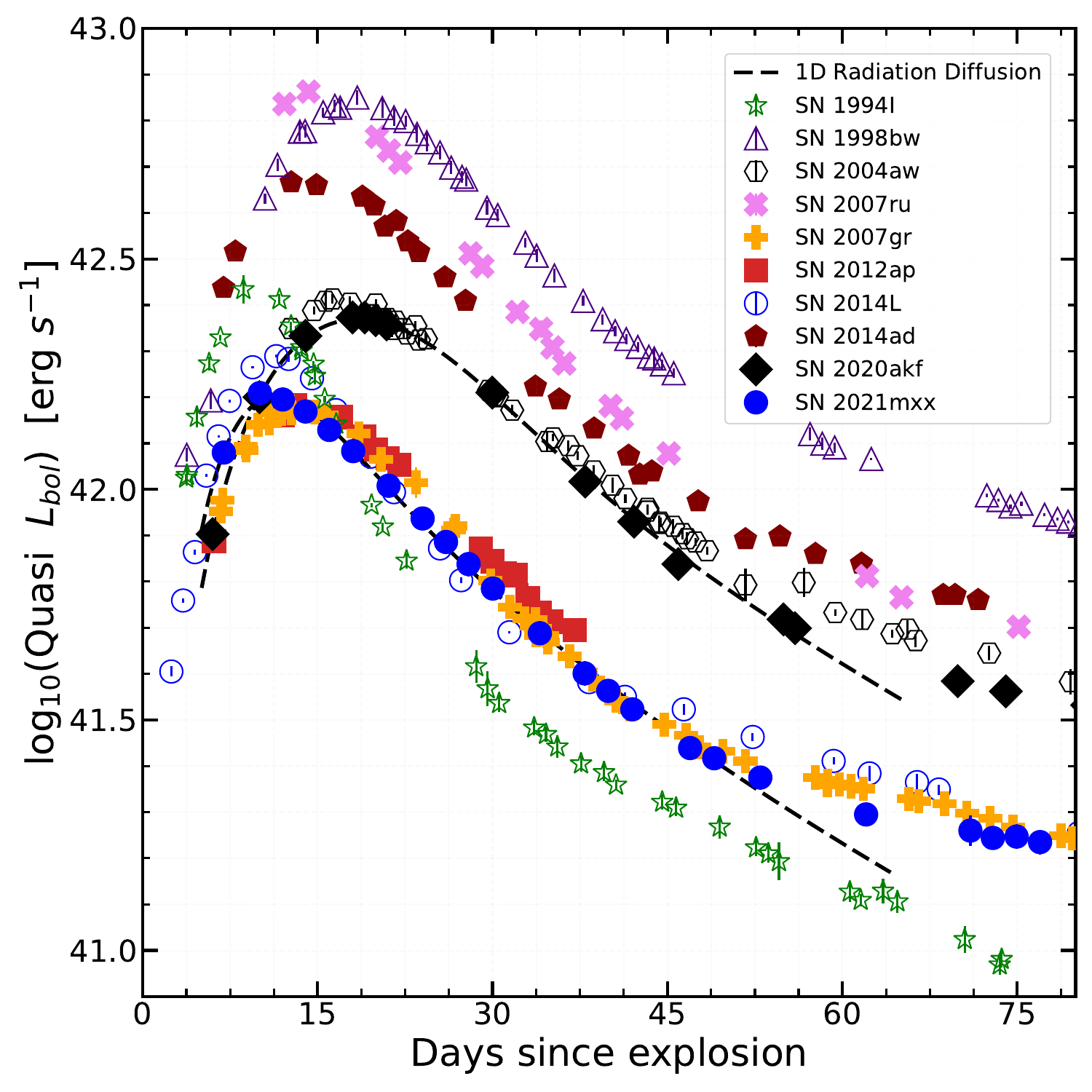}
\caption{The quasi-bolometric light curves of SN 2020akf ($UBVRoI$) and SN 2021mxx ($UBgVrRI$) are compared with the quasi-bolometric ($UBVRI$) light curves of other Type Ic SNe, all derived using the SuperBol code \citep{Nicholl18}. The dashed lines represent the Arnett-Valenti model fit (Section \ref{Arnett_model_akf_mxx}) to the quasi-bolometric light curves of SN 2020akf and SN 2021mxx.}  
\label{fig_bol_lc}
\end{figure}

\begin{figure}
\centering
\includegraphics[width=0.9\columnwidth]{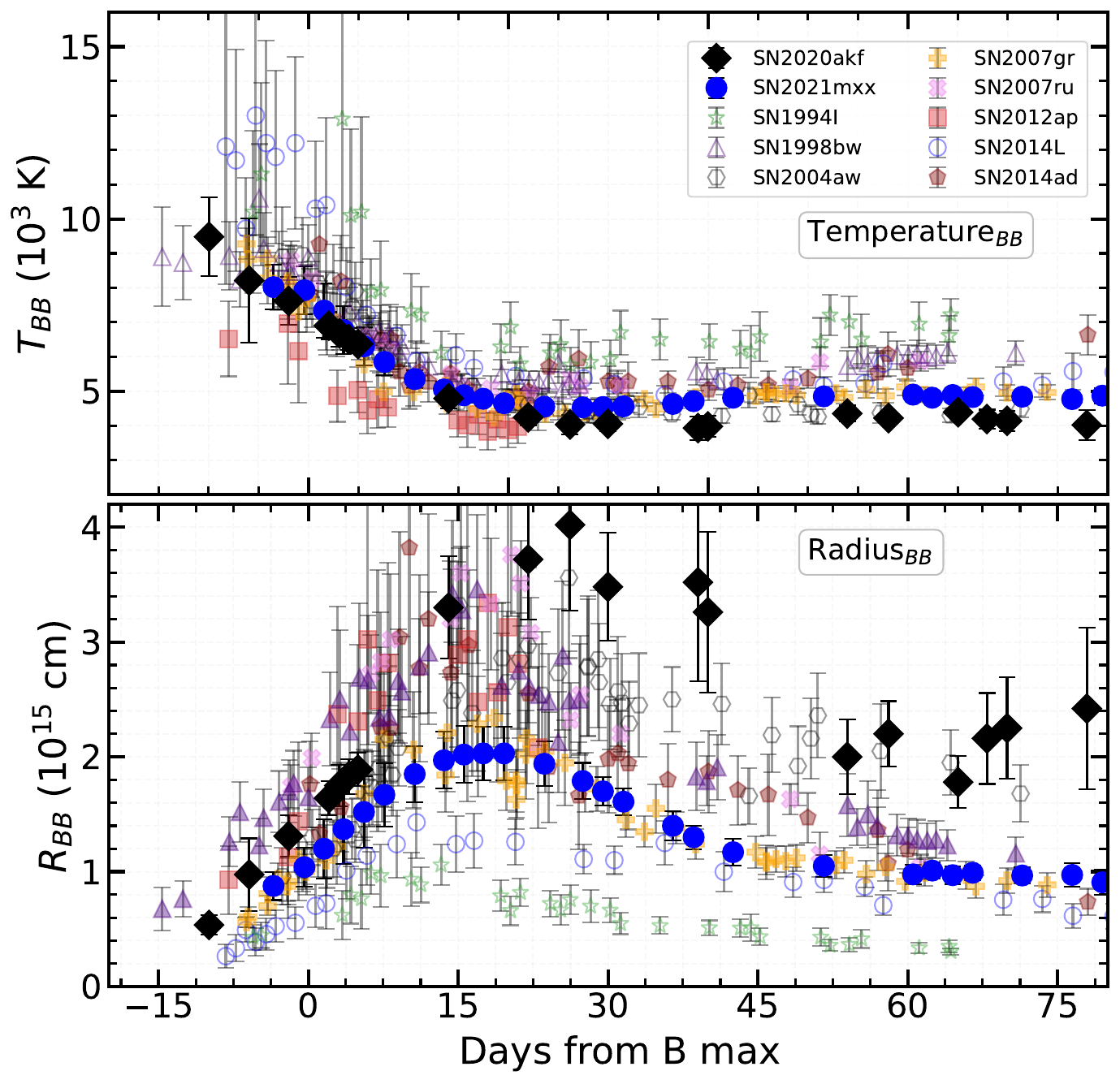}
\caption{Photospheric temperature and radius evolution of SN 2020akf and SN 2021mxx, along with other Type Ic SNe derived using the SuperBol Code \citep{Nicholl18}.}   
\label{fig_temp_radius_evol}
\end{figure}

A well-sampled light curve is first selected as the reference band. The remaining bands are then interpolated and, when necessary, extrapolated to match the temporal coverage of this reference. Interpolation is performed iteratively using polynomial fits (typically third-order), with visual inspection at each step to ensure a satisfactory representation of the observed data. Extrapolation is applied only when required to align the boundary epochs. In such cases, a constant colour offset relative to the reference band is assumed, determined from the nearest available data points. To minimize extrapolation, we limited the data to epochs within 70--75 days of the $B$-band maximum. The resulting light curves were then used to construct the quasi-bolometric light curves.

The quasi-bolometric light curves of SN 2020akf and SN 2021mxx are shown in Figure \ref{fig_bol_lc}. 
The peak quasi-bolometric luminosities were estimated to be (2.36 $\pm$ 0.11) $\times$ 10$^{42}$\, erg\,s$^{-1}$ for SN 2020akf and (1.62 $\pm$ 0.08) $\times$ 10$^{42}$\,erg\,s$^{-1}$ for SN 2021mxx. 

Based on the study by \citet{prentice2016}, we included the near-infrared (NIR) and ultraviolet (UV) contributions to the $UBVRI$ luminosity for Type Ib/c SNe, adopting typical values of $\sim$ 15\% for the NIR and $\sim$ 10\% for the UV contributions at peak brightness.
The total peak bolometric luminosities were estimated to be 
$L_{\text{UVOIR}}$ = (2.95 $\pm$ 0.13) $\times 10^{42}$\,erg\,s$^{-1}$ for SN 2020akf and $L_{\text{UVOIR}}$ = (2.02 $\pm$ 0.09) $\times 10^{42}$\,erg\,s$^{-1}$ for SN 2021mxx. 

The quasi-bolometric light curves of SN 2020akf and SN 2021mxx are compared with those of other well-studied Type Ic SNe in Figure \ref{fig_bol_lc}. 
The quasi-bolometric light curve of SN 2020akf resembles that of SN 2004aw, whereas the light curve of SN 2021mxx is similar to those of SN 2007gr and SN 2012ap. 
Both SN 2020akf and SN 2021mxx exhibit significantly lower quasi-bolometric luminosities than Type Ic-BL SNe, such as SN 1998bw, SN 2007ru, and SN 2014ad. The early post-maximum decline, $\Delta M_{15}$, of the quasi-bolometric magnitude light curves was approximately  0.57\,mag for SN 2020akf and  0.77\,mag for SN 2021mxx. During the late phase, decline rates were found to be  1.40\,mag\,(100\,d)$^{-1}$ for SN 2020akf and 1.61\,mag\,(100\,d)$^{-1}$ for SN 2021mxx. 
The decline of the quasi-bolometric light curve of both SNe in the late phase is significantly steeper than the decay rate of $^{56}$Co to $^{56}$Fe, indicating substantial gamma-ray leakage through the ejecta.

Figure \ref{fig_temp_radius_evol} displays the blackbody temperature and radius evolution for both SNe, as inferred from SuperBol. 
While the temperature evolution is remarkably similar for both events, SN 2020akf exhibits a larger photospheric radius at peak brightness and in the later phases. As a result, SN 2020akf attains a higher luminosity compared to SN 2021mxx. 

\section{Spectroscopic studies}
\label{sec_spec}
\subsection{Spectroscopic observations and data reduction}
\label{sec_spec_obs}

Optical spectra for SN 2020akf and SN 2021mxx were obtained using HFOSC mounted on HCT. We employed Gr7 (wavelength range 3800--7000 \AA) and Gr8 (5200--9200 \AA) grisms, with a resolution of approximately 7 \AA. Detailed information about spectroscopic observations is provided in Appendix/Table \ref{tab_spec_2020akf_2021mxx}. Our spectroscopic coverage spans from $\sim -$5 to $+$57 days for SN 2020akf and $\sim -$3 to $+$113 days for SN 2021mxx, with respect to $B$-maximum.  

The spectroscopic data were processed in a standard manner using various tasks available in {\sc iraf}.  
A one-dimensional spectrum was extracted using the optimal extraction method.
Dispersion solution and wavelength calibration were achieved using FeAr and FeNe arc lamp spectra. The wavelength calibration accuracy was validated by comparing it against night-sky emission lines, with minor adjustments applied as needed. We preferably used spectrophotometric standard stars observed on the same night to correct for the instrumental response and perform flux calibration. In the absence of a spectrophotometric standard star observed on the same night, observations from adjacent nights were utilized. The spectra from the blue and red regions (Gr7 and Gr8) were combined and scaled to a weighted mean, resulting in the final spectrum on a relative flux scale. The combined spectra were finally scaled with respect to the photometric flux to bring them to an absolute flux scale. 

We applied reddening corrections in the spectra, with $E(B - V)_{\text{total}} = 0.013$\,mag for SN 2020akf and $E(B - V)_{\text{total}} = 0.20$\,mag for SN 2021mxx.
The spectra of SN 2020akf and SN 2021mxx were corrected for the redshift of their host galaxies, using values of $z$ = 0.012445 and $z$ = 0.000965, respectively, as obtained from \href{https://ned.ipac.caltech.edu/}{NED}. 

\subsection{Spectral evolution in pre-maximum phase}
\label{sec_spec_pre_max}

\begin{figure}[ht!]
\centering
\includegraphics[width=0.9\columnwidth, trim={0 1.6cm 0 0},clip]{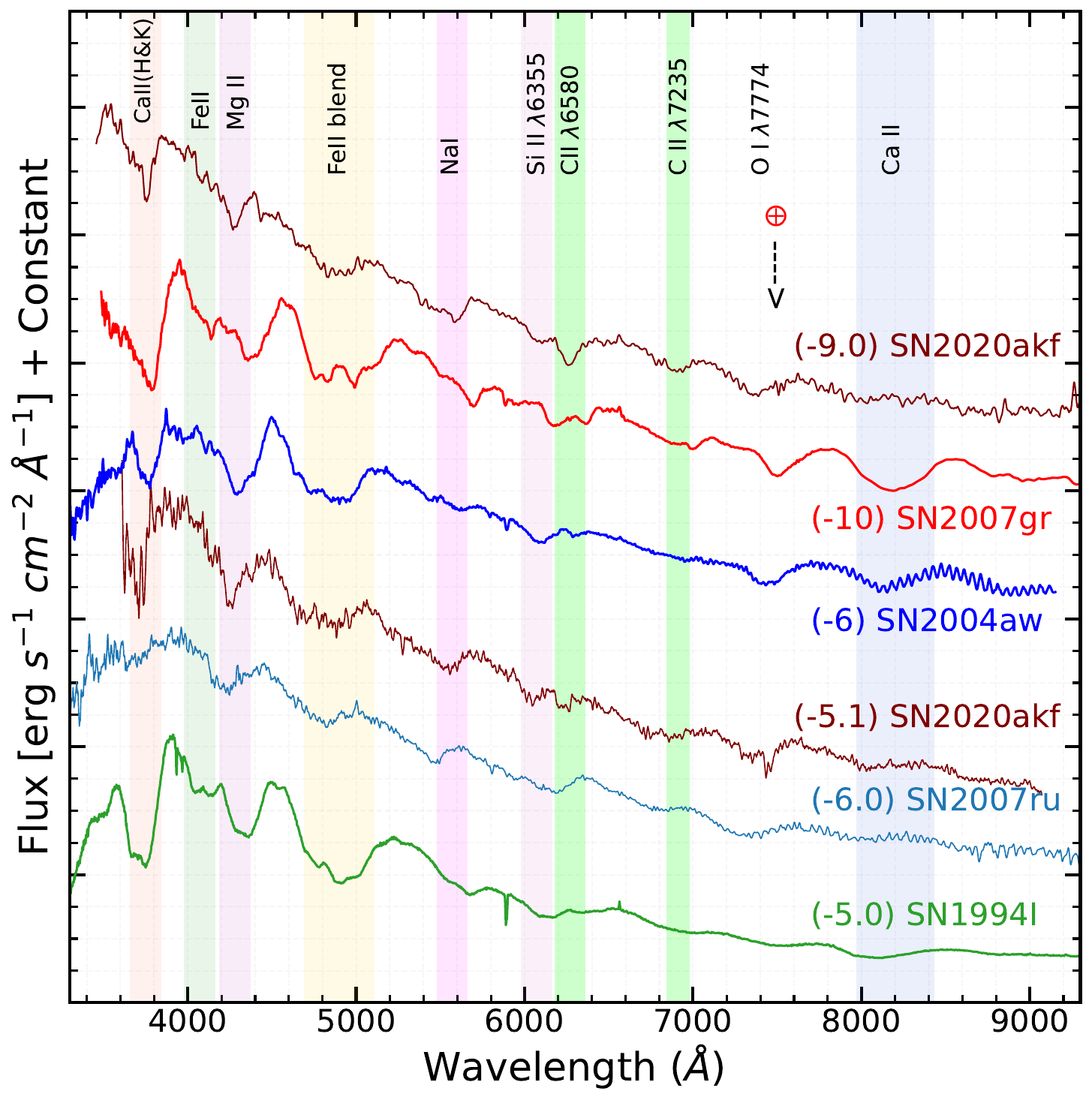}
\includegraphics[width=0.9\columnwidth]{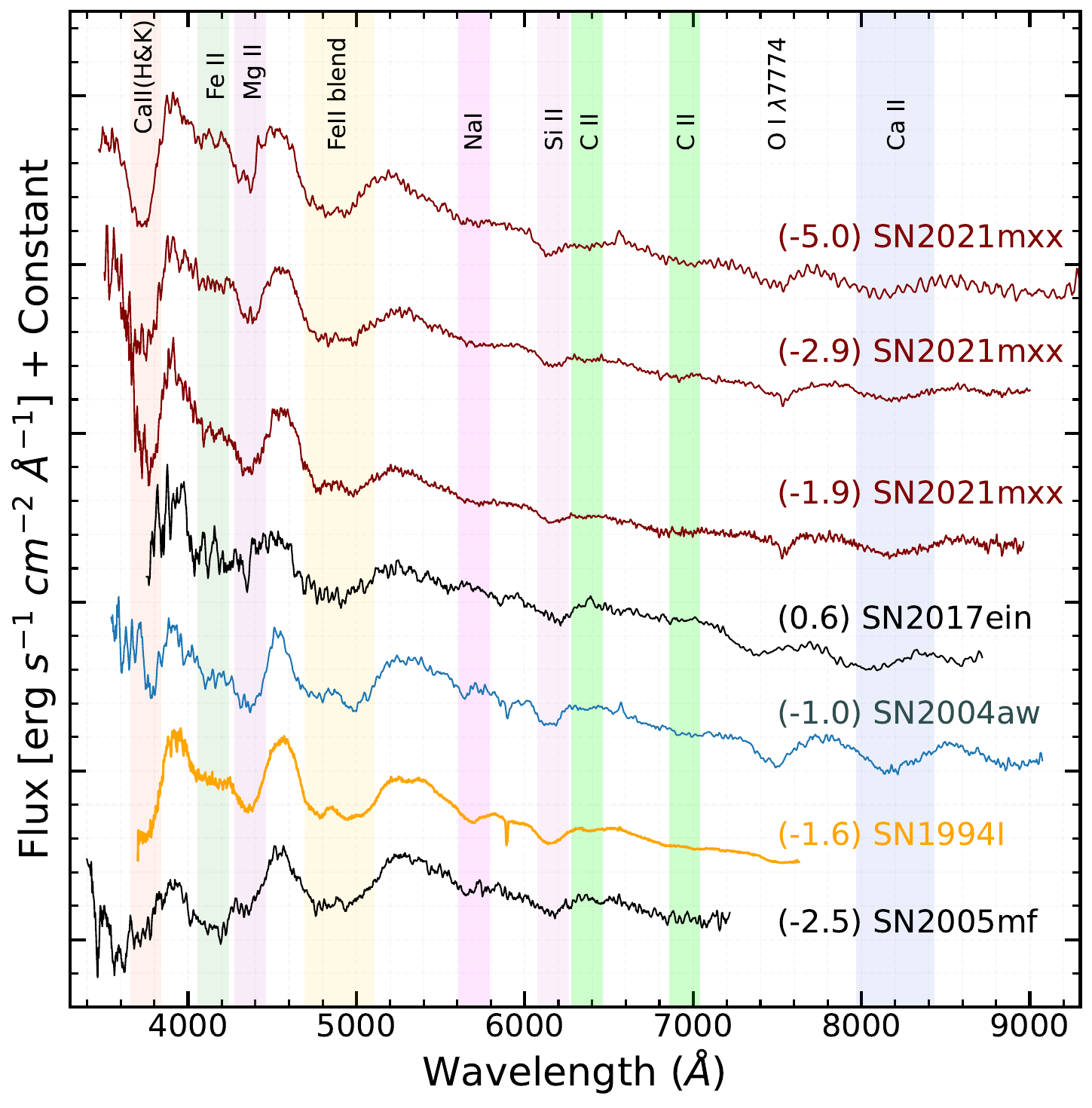}
\caption{Pre-maximum spectral evolution of SN 2020akf (top) and SN 2021mxx (bottom) are compared with those of other well-studied Type Ic SNe. The prominent features in the spectra are marked. To enhance clarity, the spectra have been vertically shifted. Telluric lines are indicated by a circled plus sign.}
\label{fig_spec_pre_sn20akf}
\end{figure}

Spectral evolution of SN 2020akf (top) and SN 2021mxx (bottom) during the pre-maximum phase is shown in Figure \ref{fig_spec_pre_sn20akf}, along with spectra of other Type Ic SNe, including SN 1994I, SN 2004aw,  SN 2005mf, SN 2007gr, SN 2007ru and SN 2017ein, at comparable epochs, available publicly on WISeREP\footnote{\url{https://www.wiserep.org/}} \citep{yaro12}. 
The spectrum of SN 2020akf in the pre-maximum phase was obtained at $-$5 days, while pre-maximum spectra of SN 2021mxx were obtained at $-$3 and $-$2 days relative to the $B$-band maximum.
To investigate the early characteristics, we made use of pre-maximum spectra of SN 2020akf at $-$9 days \citep{2020akfGSP} and SN 2021mxx at $-$5 days \citep{2021mxxGSP} from WISeREP/TNS, and included in Figure \ref{fig_spec_pre_sn20akf}.

Both SNe 2020akf and  2021mxx exhibit a blue continuum with prominent absorption features near $\sim$ 3800 and $\sim$ 4400\,\r{A}, attributed to Ca\,{\sc ii} (H\&K) and Mg\,{\sc ii}, respectively. Spectra of both SNe show a blend of Fe\,{\sc ii} lines at approximately 4900\,\r{A}. The presence of the Si\,{\sc ii} $\lambda$6355 line around $\sim6150$\,\r{A} and the Ca\,{\sc ii} line near $\sim8200$\,\r{A} is clearly identifiable in the pre-maximum spectra of SN 2021mxx, whereas the Ca\,{\sc ii} line in SN 2020akf is still developing. 
The spectra of both SN 2020akf and SN 2021mxx show C\,{\sc ii} lines near $\sim6300$\,\r{A}; this feature is notably strong in SN 2020akf at $-$9 days but weakens significantly in later phases. A similar feature is observed in the spectra of SN 2007gr \citep{2014chen} and SN 2004aw. Overall, the spectral features of SN 2020akf and SN 2021mxx are consistent with those of other Type Ic SNe. 

\subsection{Spectral evolution in post-maximum phase} 
\label{sec_spec_post_max}

\begin{figure}[ht!]
\centering
\includegraphics[width=0.9\columnwidth, trim={0 1.4cm 0 0},clip]{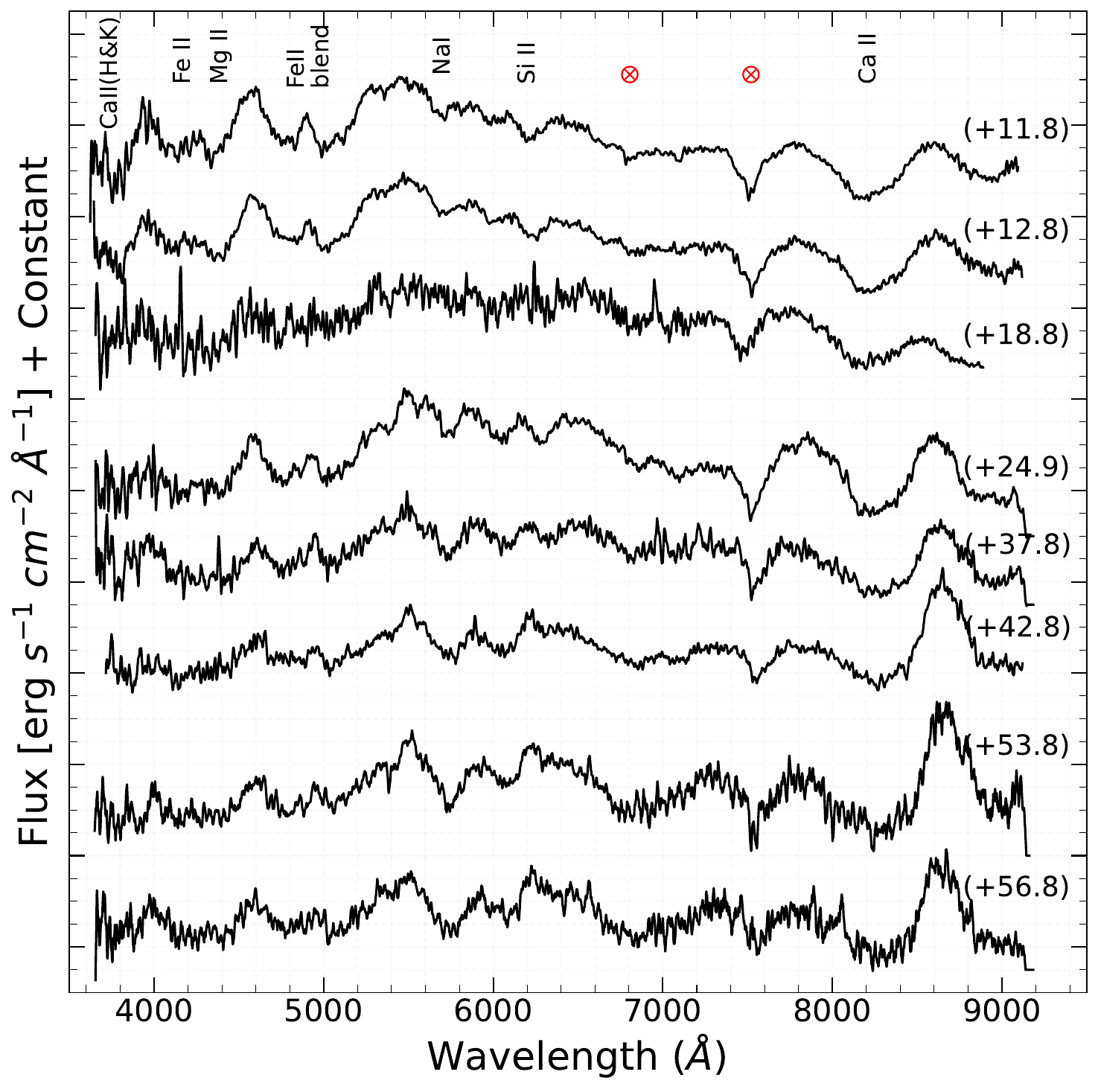}
\includegraphics[width=0.9\columnwidth]{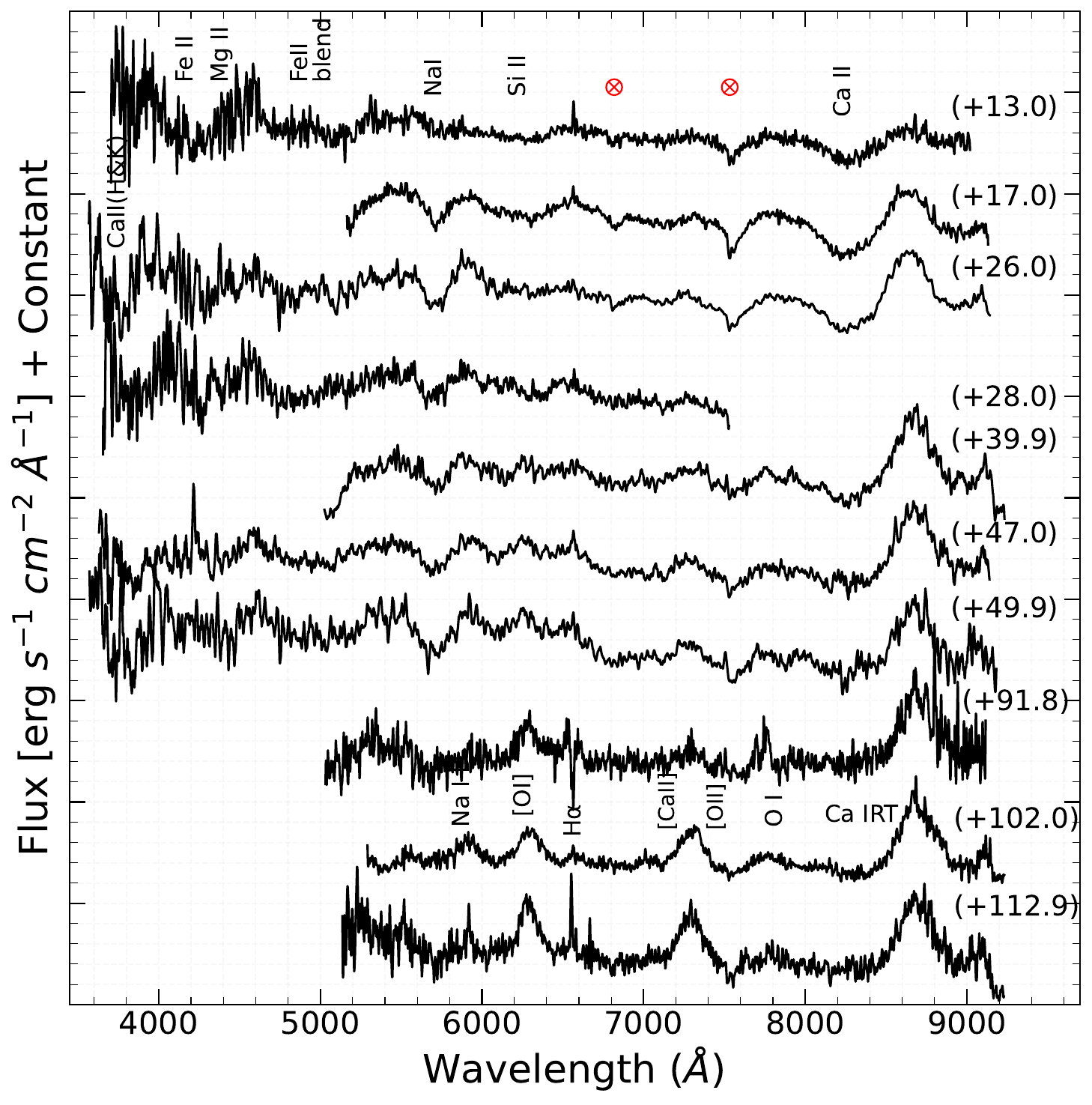}
\caption{Spectral evolution of SN 2020akf (top panel) from $\sim$ +12 to +57 days and for SN 2021mxx (bottom panel) from +13 to +113 days. To enhance clarity, the spectra have been vertically shifted. Telluric lines are indicated by a circled plus sign.}
\label{fig_20akf_21mxx_spec_evo}
\end{figure}

The post-maximum spectral evolution of SN 2020akf (top panel) and SN 2021mxx (bottom panel) is presented in Figure \ref{fig_20akf_21mxx_spec_evo}.
Up to $\sim$57 days, the evolution of SN 2020akf reveals the distinctive `W' feature of the Mg\,{\sc ii} ($\sim4400$\,\r{A}) line, along with Fe\,{\sc ii} ($\sim4745$\,\r{A}) and Fe\,{\sc ii} ($\sim4960$\,\r{A}) lines. The evolution of the Na\,{\sc i} line is also apparent. The Mg\,{\sc ii} and Si\,{\sc ii} lines gradually fade, and another `W' feature of the Fe\,{\sc ii} lines ($\sim$5900 \AA) begins to emerge around 25--40 days after maximum light. The Ca\,{\sc ii} NIR triplet's transition from an absorption to an emission feature occurs $\sim$ 25 days after maximum light. 

In the spectral evolution of SN 2021mxx (Figure \ref{fig_20akf_21mxx_spec_evo}, bottom panel), the spectrum at 13 days shows the Fe\,{\sc ii} ($\sim4745$\,\r{A}) and Fe\,{\sc ii} ($\sim4960$\,\r{A}) lines, along with the Mg\,{\sc ii} ($\sim4400$\,\r{A}) line, all of which are clearly developed. However, the Na\,{\sc i} lines continue to evolve and become prominent after $\sim$17 days from maximum light. The Mg\,{\sc ii} and Si\,{\sc ii} $\lambda$6355 lines gradually weaken and begin to merge into the characteristic `W' feature of Fe\,{\sc ii}. The Ca\,{\sc ii} NIR absorption feature also evolves significantly, eventually becoming emission-dominated.

\begin{figure}[ht!]
\centering
\includegraphics[width=0.9\columnwidth, trim={0 1.4cm 0 0},clip]{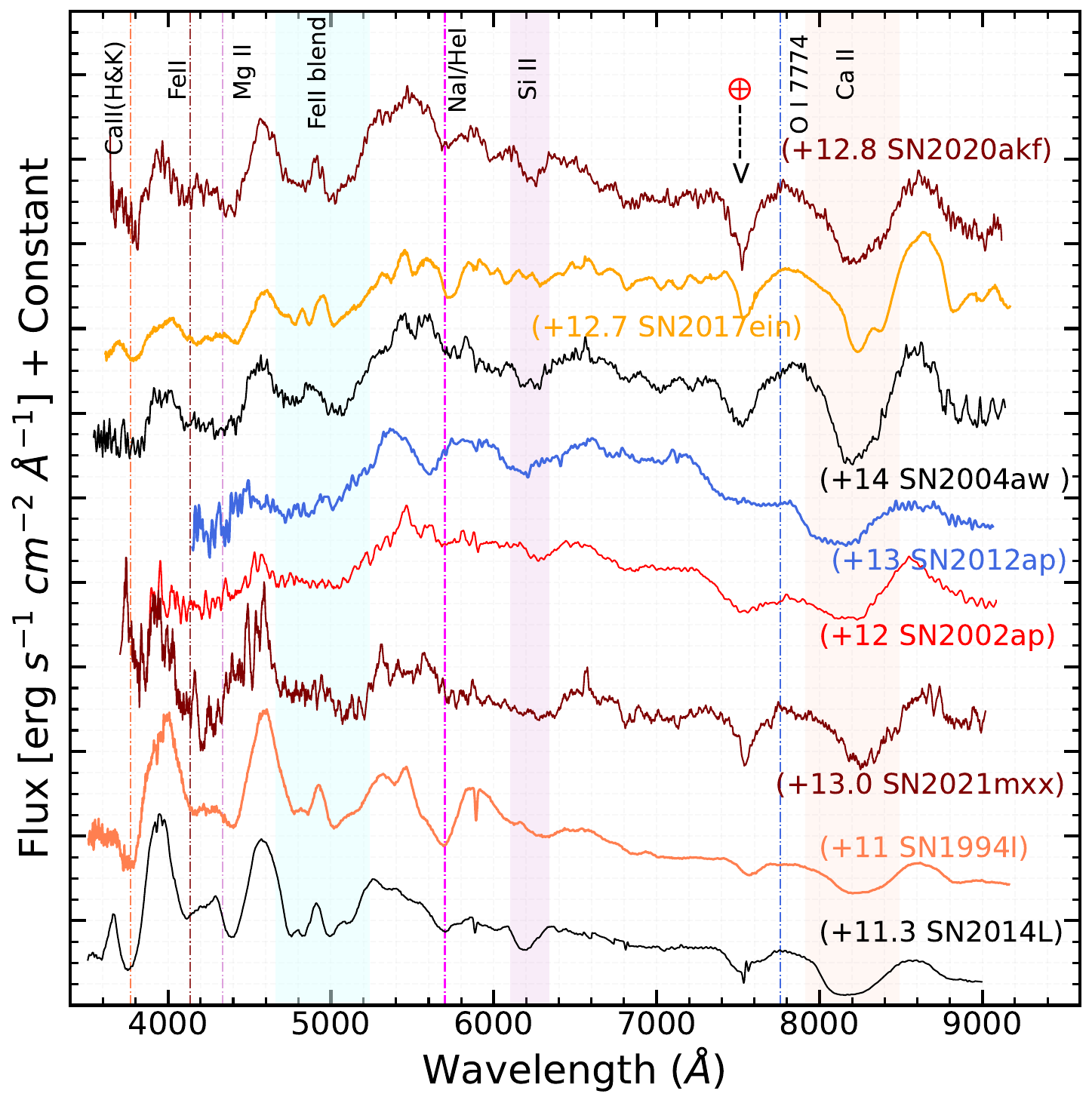}
\includegraphics[width=0.9\columnwidth]{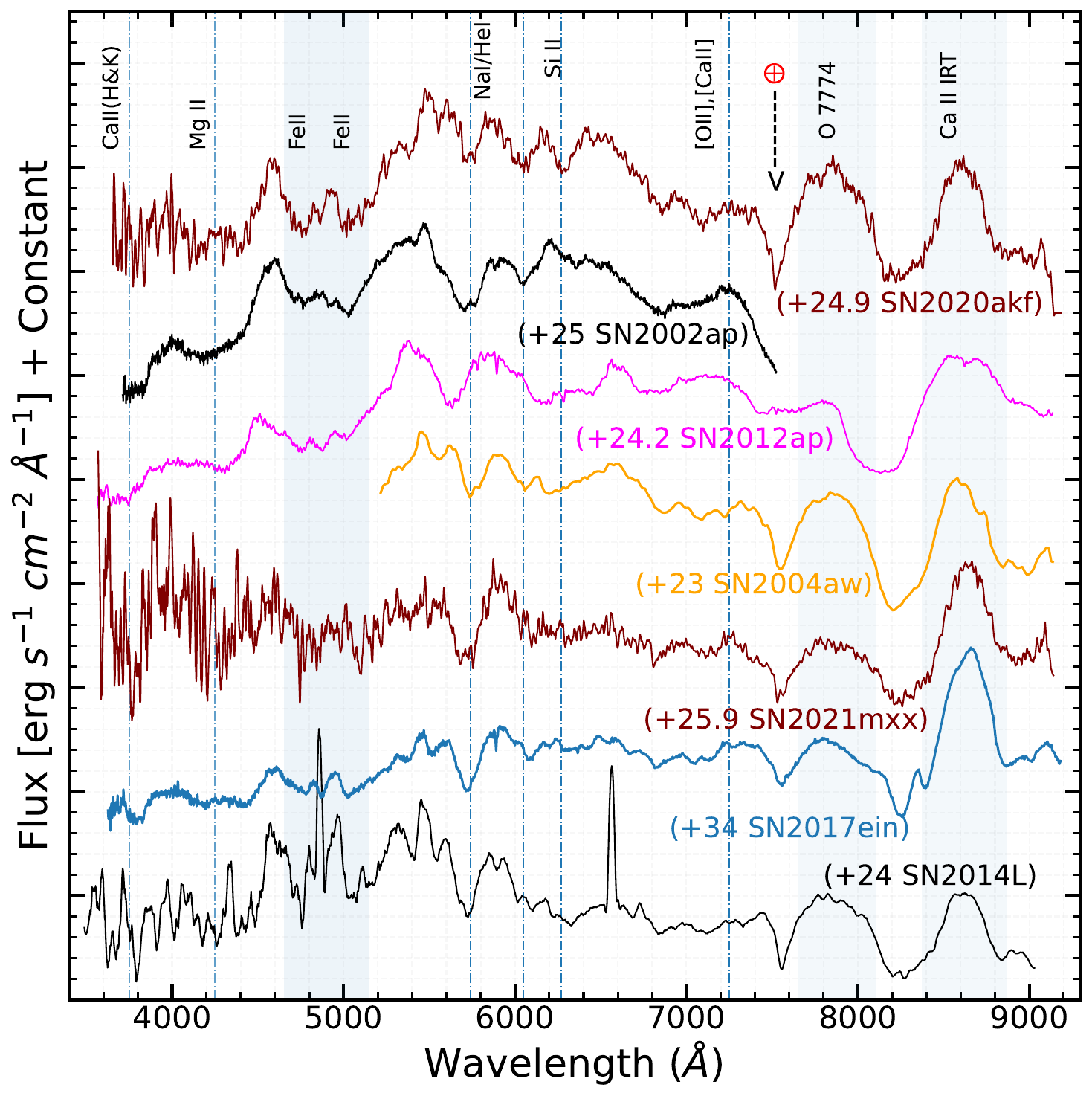}
\caption{Comparison of spectra of SN 2020akf and SN 2021mxx at $\sim$13 days (top panel) and $\sim$25 days (bottom panel) with those of other Type Ic SNe at similar epochs.  The prominent features in the spectra are marked. Telluric lines are indicated by a circled plus sign.}
\label{fig_20akf_21mxx_spec_com}
\end{figure}

Figures \ref{fig_20akf_21mxx_spec_com} show the spectra of SN 2020akf and SN 2021mxx obtained at approximately $\sim$13  (top panel) and 25 days  (bottom panel), alongside other Type Ic SNe at similar epochs for comparison. In SN 2020akf, the `W' feature of Fe\,{\sc ii} ($\sim4745$\,\r{A}) and Fe\,{\sc ii} ($\sim4960$\,\r{A}), as well as the Na\,{\sc i} lines, resemble those seen in other SNe, including SN 1994I, SN 2014L and SN 2017ein. The central portion of the SN 2020akf spectrum appears somewhat bumpy, a feature shared with SNe such as SN 2002ap, SN 2004aw, SN 2012ap, and SN 2017ein.

In SN 2021mxx, the blended Fe\,{\sc ii} lines are comparable to those in SN 2002ap, SN 2004aw, and SN 2012ap. However, the overall shape of SN 2021mxx's spectrum is flat, resembling those of SNe 1994I and 2014L.

\subsection{Spectral evolution in nebular phase} 
\label{sec_spec_nebular_phase}
\begin{figure}[ht!]
\centering
\includegraphics[width=0.9\columnwidth]{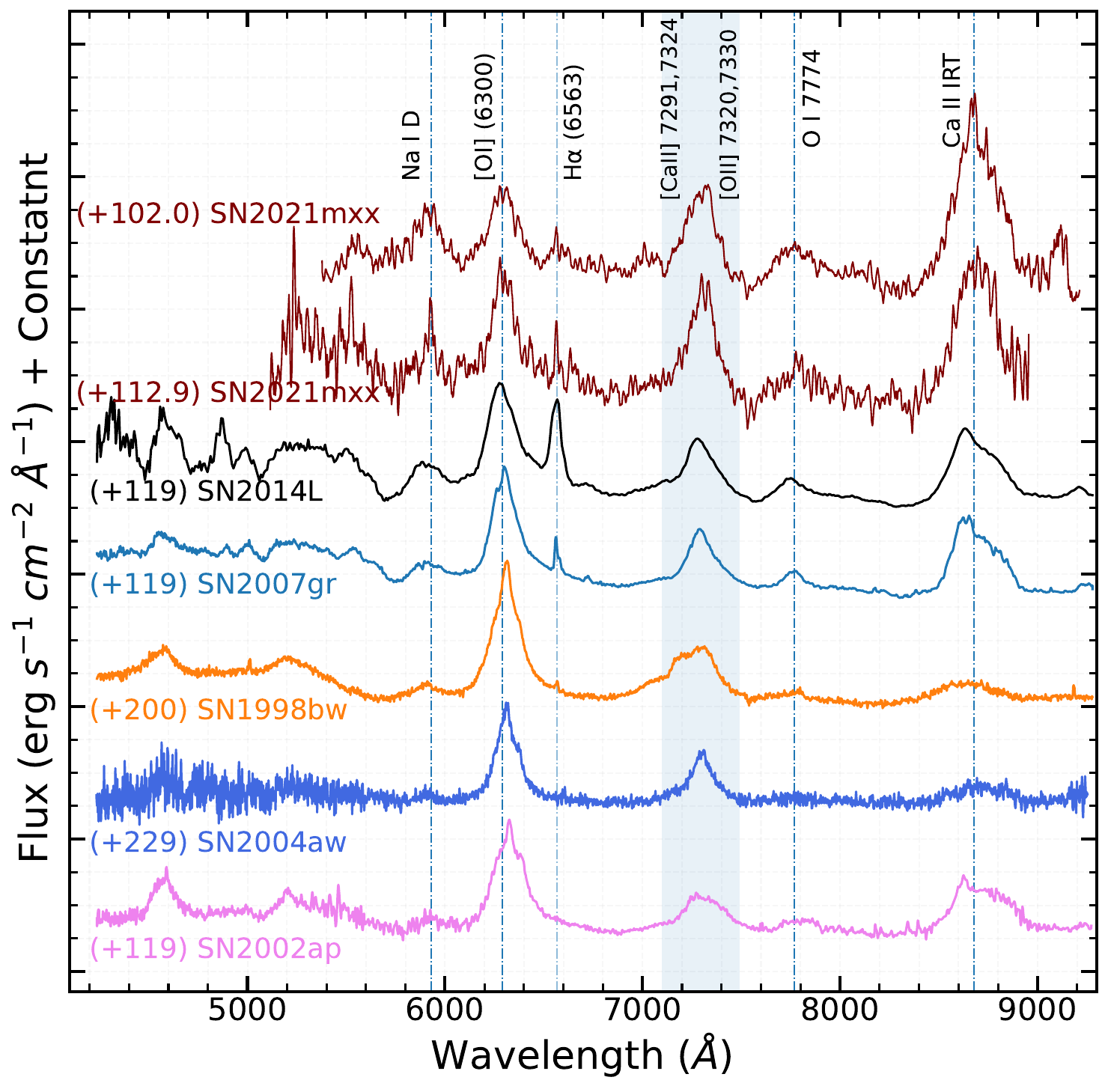}
\caption{Nebular phase spectra of SN 2021mxx at +102 and +113 days are compared with those of other Type Ic SNe at similar epochs. The prominent features in the spectra are marked. }
\label{fig_sp_comp_102d}
\end{figure}

Figure \ref{fig_sp_comp_102d} presents the nebular-phase spectra of SN 2021mxx, obtained around 102 and 113 days, and compared with nebular spectra of other Type Ic SNe. These spectra are dominated by [O\,{\sc i}] $\lambda$$\lambda$6300, 6363 and [Ca\,{\sc ii}] $\lambda$$\lambda$7291, 7324 emission lines. The [Ca\,{\sc ii}] $\lambda$$\lambda$7291, 7324 lines are blended with [Fe\,{\sc ii}] $\lambda$$\lambda$$\lambda$$\lambda$7155, 7175, 7388, and 7452 as well as with [O\,{\sc ii}] $\lambda$$\lambda$7320, 7330 forbidden emission lines \citep{stritzinger2009}. The spectra also display a prominent Ca\,{\sc ii} NIR triplet emission line. Narrow H$_\alpha$ emission line from the host galaxy is clearly identifiable.

The nebular-phase spectra of SN 2021mxx are compared with those of Type Ic SNe, SN 2007gr, SN 2014L, SN 2004aw,  broad-lined  SN 2002ap, and SN 1998bw  around similar epochs. Prominent emission lines in SN 2021mxx, such as [O\,{\sc i}] $\lambda$$\lambda$6300, 6363, [Ca\,{\sc ii}] $\lambda$$\lambda$7291, 7324, and the Ca\,{\sc ii} NIR triplet, appear similar to those in other SNe. The  [Ca\,{\sc ii}] $\lambda$$\lambda$7291, 7324 line in SN 2021mxx and other typical Ic SNe is sharper, whereas it appears broad with a flat top in broad-lined Ic SN 1998bw and SN 2002ap. This broadening is attributed to blending of [Ca\,{\sc ii}] $\lambda$$\lambda$7291, 7324  line with [Fe\,{\sc ii}], [O\,{\sc ii}], and [Co\,{\sc ii}] lines \citep[][see Fig.~01]{Maeda2006}.

We used the \texttt{splot} task in {\sc iraf} to derive line-profile characteristics, e.g., line widths and line fluxes. For each emission feature, a local continuum was defined and subtracted, and the profile was fitted with a Gaussian function. The FWHM was derived from the fitted width and converted to velocity using the Doppler relation. Uncertainties were estimated by repeating the fits with reasonable variations in continuum placement and fitting regions. The line flux was obtained by integrating the best-fitting Gaussian profile.

The full width at half maximum (FWHM) velocities for [O\,{\sc i}] $\lambda$$\lambda$6300, 6363 in the spectra of SN 2021mxx at 102 and 113 days are approximately 9,150 $\pm$ 550 and 6,150 $\pm$ 535 \,km\,s$^{-1}$, respectively. Similarly, the FWHM velocities for [Ca\,{\sc ii}] $\lambda$$\lambda$7291, 7324 are around 
7,615 $\pm$ 500 and 7,100 $\pm$ 590\,km\,s$^{-1}$, respectively. The FWHM velocities of these lines are higher than the core velocity values of 1,000 - 5,000\,km\,s$^{-1}$ suggested by \cite{fransson1989} depending on the envelope mass and the explosion energy. The higher core velocity may result from a more energetic explosion, significant mass loss before the explosion, and mixing of the ejecta. This mixing can result in broader emission line profiles. Additionally, high core velocities could be associated with hydrodynamic instabilities, such as Rayleigh-Taylor instabilities, which would further influence the density distribution \citep{fransson1989, filli1985}.

The line ratio of [O\,{\sc i}] $\lambda$$\lambda$6300, 6363 to [Ca\,{\sc ii}] $\lambda$$\lambda$7291, 7324 is a good indicator for zero age main-sequence mass of progenitor star \citep{Nomoto2006}, and known to be insensitive to both density and temperature but increases with progenitor mass \citep{fransson1989, 2004Elmhamdi}.
For SN 2021mxx, the [O\,{\sc i}]/[Ca\,{\sc ii}] line ratio is approximately 0.96, similar to SN 2012ap (0.9) and slightly more than  SN 2009bb (0.8). However, it is notably lower than other Type Ic SNe, such as SN 2007gr (1.4), SN 2014L (1.5), 
as well as Ic-BL SNe like  SN 1998bw (1.7), SN 2002ap (2), SN 2007ru (1.6), and SN 2014ad (1.54). 

\citet{kuncarayakti2015} investigated the [O\,{\sc i}]/[Ca\,{\sc ii}] flux ratio in several CCSNe and linked it to progenitor properties.  For SN 2021mxx, the measured flux ratio of 0.96 places it within the region associated with binary progenitors of Type Ib/c SNe. This suggests that SN 2021mxx likely originated from a lower-mass progenitor in a binary system.

\begin{figure}[ht!]
	\centering
	\includegraphics[width=1.0\columnwidth]{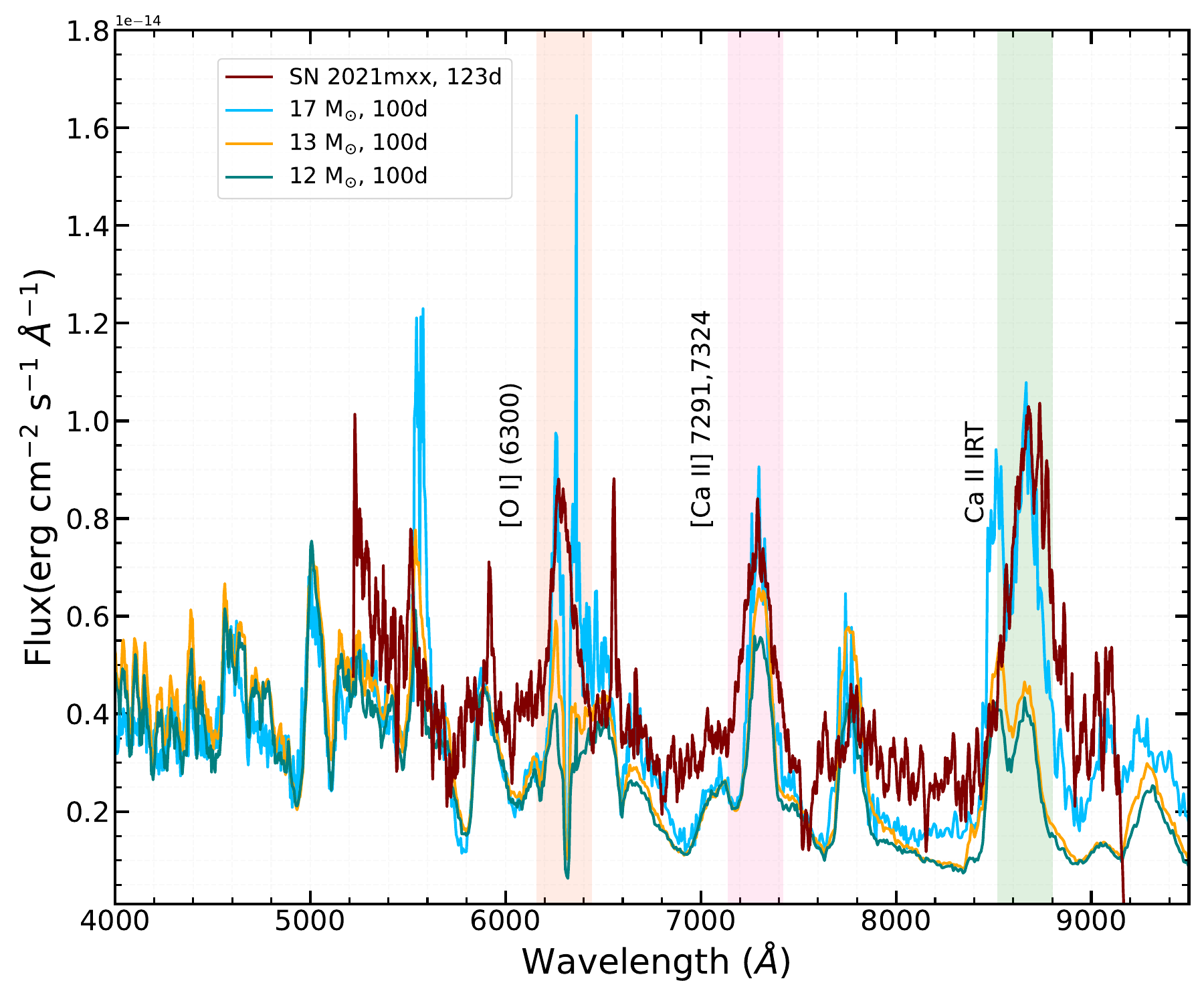}
	\caption{The nebular phase spectrum of SN 2021mxx, observed 123 days after explosion, is compared with the models of  \cite{2015Jerkstrand} for different progenitor masses. }   
	\label{fig_jerjstrand_model_fit_21mxx}
\end{figure}

\cite{2015Jerkstrand} have provided models for the nebular spectra of Type IIb SNe,  based on single-star evolutionary channels with progenitor mass of 12, 13, and 17\,M$_{\odot}$.
 These model spectra depend on parameters such as the initial $^{56}$Ni mass, distance, and time. The models assume a $^{56}$Ni mass of 0.075\,M$_{\odot}$ and a distance of 7.8 Mpc.
In Figure \ref{fig_jerjstrand_model_fit_21mxx} we compare the nebular-phase spectrum of SN~2021mxx with the models from \cite{2015Jerkstrand} at 100 days post-explosion. 
The observed spectrum of SN 2021mxx at 123 days post-explosion has been corrected for reddening and redshift, then scaled using equation\,(2) from \cite{Bostroem2019}, with a $^{56}$Ni mass of 0.07\,M$_{\odot}$ (estimated from complete bolometric luminosity), a distance of 44.79 Mpc, and the corresponding phase adjusted to match the model spectrum. A comparison of the nebular spectrum of SN 2021mxx with various models indicates that its [O\,{\sc i}] luminosity falls within the range predicted for progenitor masses of 13--17\,M$_{\odot}$.
Similar analyses have been performed for Type Ibc SNe by \cite{Fremling2016, Anjasha2020, MS2021}.

However, binary interaction prior to explosion can substantially modify the core structure and lower the final core progenitor mass of SESNe \cite{Bersten2014, Fremling2016}. Consequently, comparisons with single-star nebular models may systematically underestimate the zero-age main-sequence progenitor mass in systems that have experienced binary stripping. The mass estimates presented here should thus be regarded as lower limits.

\subsection{Spectral Modelling with \texttt{TARDIS}}
\label{sec_tardis_fit}

We use the one-dimensional radiative transfer code \texttt{TARDIS}\footnote{\url{https://github.com/tardis-sn/tardis}} \citep{Kerzendorf2014, Vogl2019} to model the spectra and explore the physical conditions of the ejecta, including density structure, chemical composition, ionization state, and temperature distribution for SN 2020akf and SN 2021mxx. Our modeling approach follows a configuration similar to \citet{2014adKwok} and \citet{MW2023}, utilizing publicly available setup  files\footnote{\url{https://github.com/tardis-sn/tardis-setups/tree/master/2022/williamson_19ewu}}. We select a power-law density profile following homologous expansion with:

\begin{equation}
\rho(v,t_\text{exp}) = \rho_0 \left( \frac{v}{v_0} \right)^{-\alpha} \left( \frac{t_0}{t_\text{exp}} \right)^{3}
\end{equation}

where $\rho_0$ is the inner density at the reference time $t_0$, $v_0$ is the reference velocity that governs the density slope, $v$ is the photospheric velocity, $t_\text{exp}$ is the time since explosion, and $\alpha$ is the power law index. We used $\alpha = 6$, 30 iterations, and 300,000 packets for our analysis.

\begin{figure*}[ht!]
\centering
\includegraphics[width=0.45\textwidth]{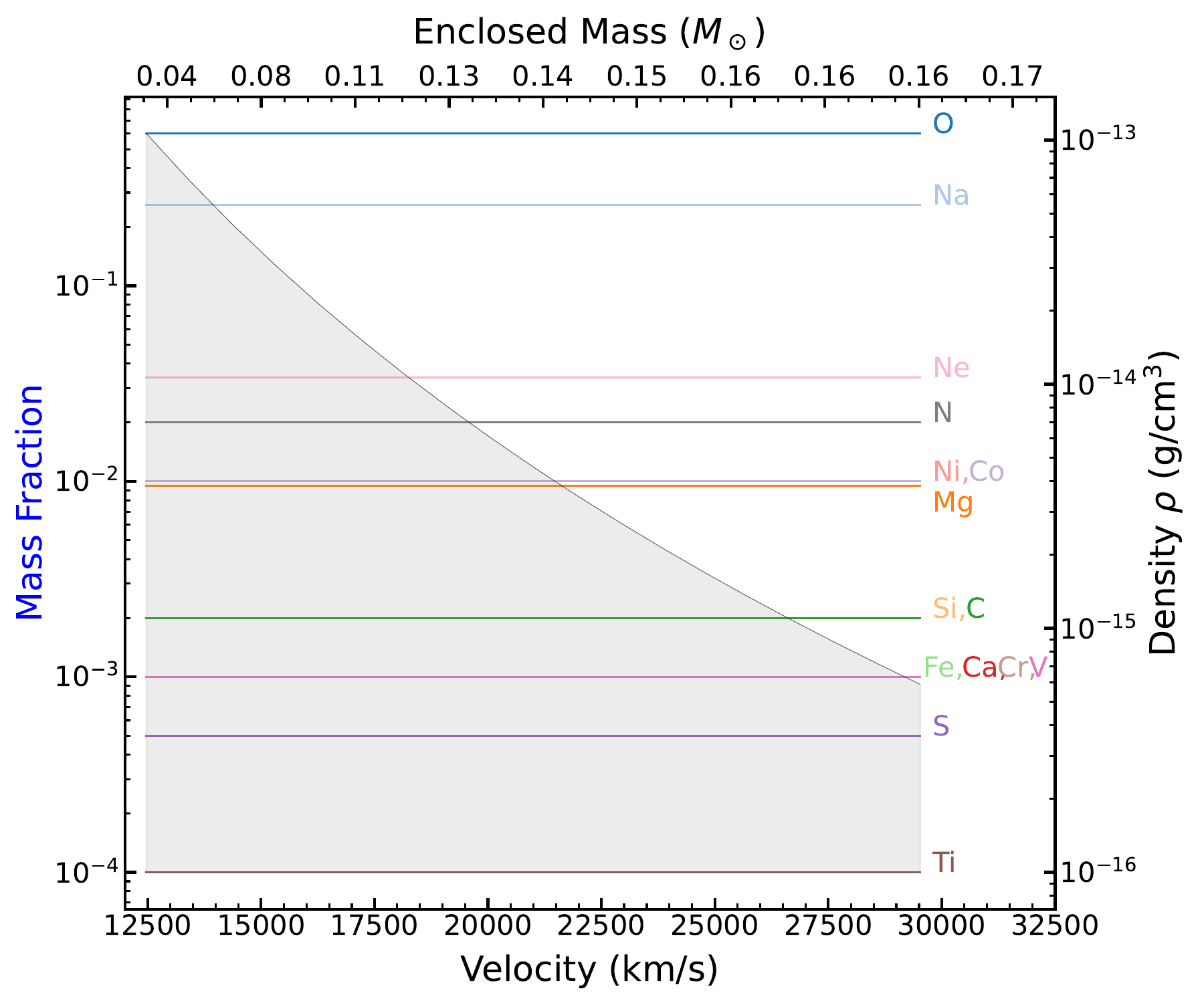}
\includegraphics[width=0.45\textwidth]{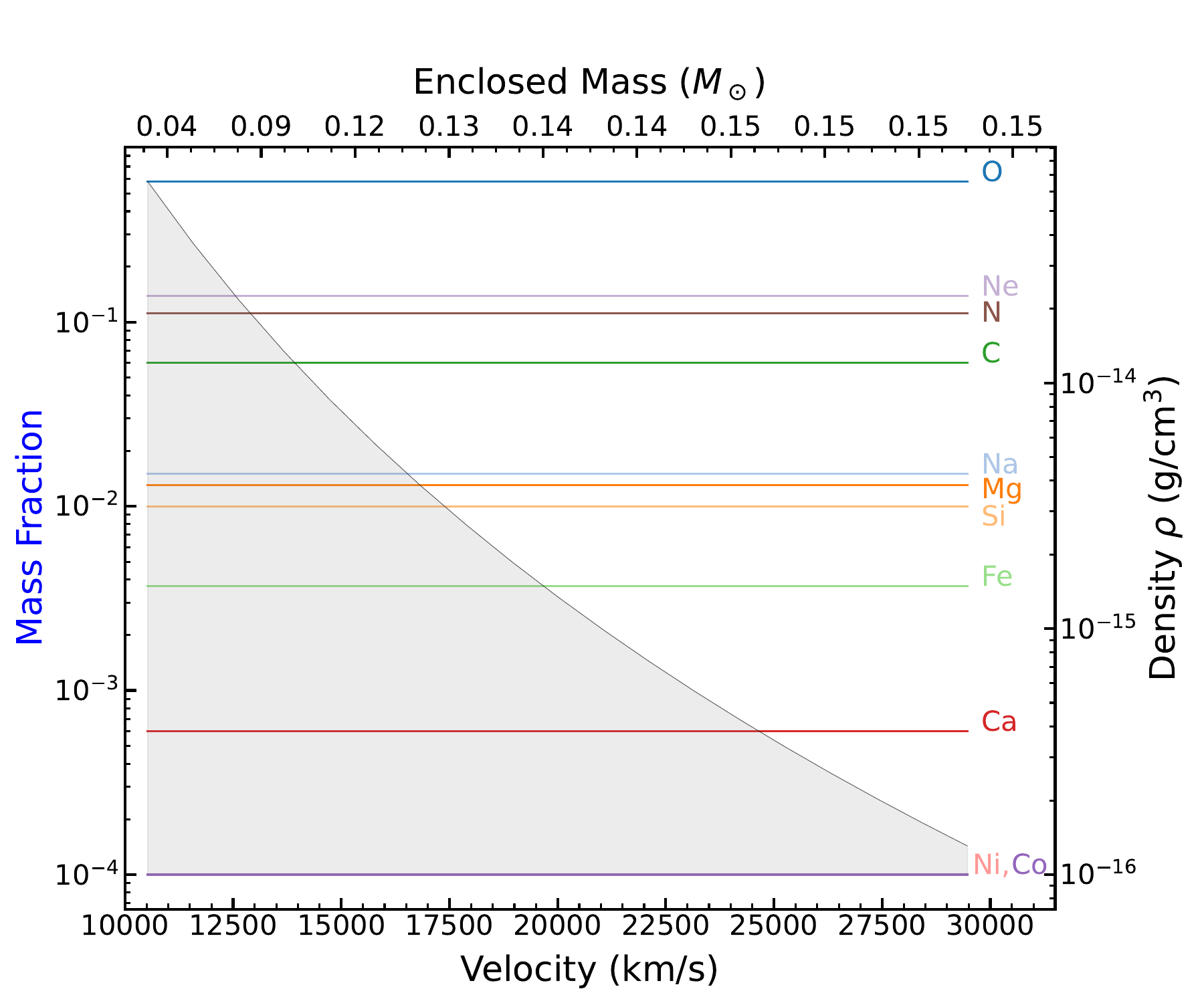}
\caption{The \texttt{TARDIS} density profile (gray shaded) and the constant mass fraction of elements for SN~2020akf (left panel) at $-$5 days and SN 2021mxx (right panel) at $-$3 days are shown as a function of velocity. The top axis indicates the corresponding enclosed mass in both panels.}
\label{fig_mass_fraction_20akf_21mxx}
\end{figure*}

We assume a uniform fractional elemental abundance and  
use very simple elemental configurations of Ca, Fe, Ni, Mg, Si, Na, O, and C for SN 2020akf and SN 2021mxx. 
Our adopted density and elemental abundance profiles, along with the enclosed mass, that best fit the observed spectra are shown in Figure \ref{fig_mass_fraction_20akf_21mxx}.

\begin{figure*}[ht!]
    \centering
    \includegraphics[width=\textwidth]{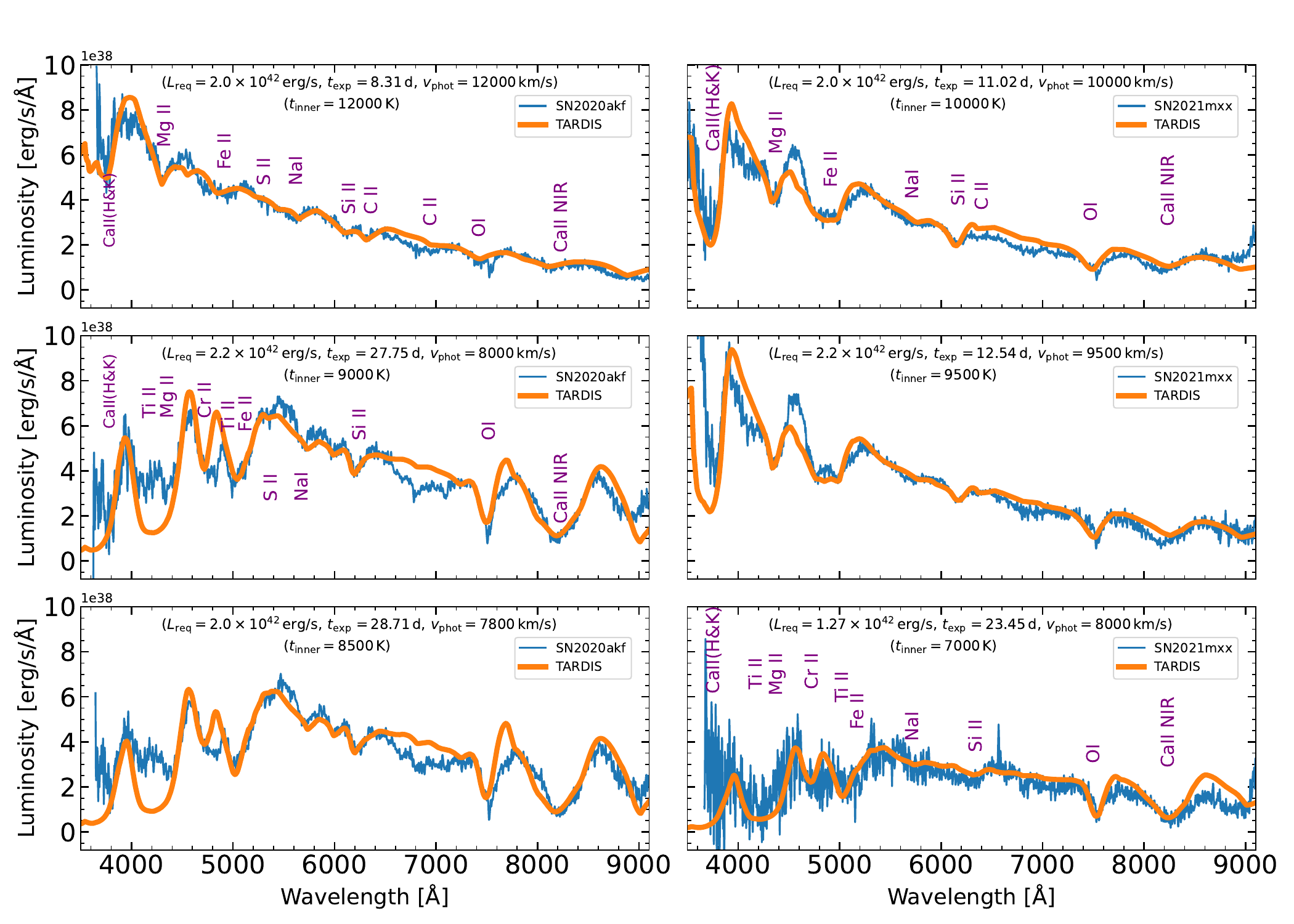}
   \caption{Observed spectral series of SN 2020akf and SN 2021mxx (stealblue) are compared with the corresponding best-fit \texttt{TARDIS} synthetic spectra (orange). For each spectrum, the fitted parameters, luminosity ($L_{\mathrm{req}}$), explosion time ($t_{\mathrm{exp}}$), photospheric velocity ($v_{\mathrm{phot}}$), and inner boundary temperature ($T_{\mathrm{inner}}$) are indicated. }
  \label{fig_tardis_akf_mxx}
\end{figure*}
 
The model fitting process was performed using the chi-by-eye method, similar to that described in  \cite{Barna2021, 2014adKwok, MS2024}. 
The best-fit parameters are $\rho_0$ = 8.5 $\times$ 10$^{-16}$\,g\,cm$^{-3}$, $t_0$ = 16\,days and $v_0$ = 20,000\,km\,s$^{-1}$ for SN 2020akf, and $\rho_0$ = 1.0 $\times$ 10$^{-14}$\,g\,cm$^{-3}$, $t_0$ = 10\,days and $v_0$ = 15,000\,km\,s$^{-1}$ for SN 2021mxx. These best-fit parameters were fixed and used consistently to model the spectra at three epochs for each supernova, successfully reproducing the key spectral features of SN 2020akf at $-$5, +12, and +13 days (left column) and SN 2021mxx at $-$3, $-$2, and +13 days (right column) as shown in Figure \ref{fig_tardis_akf_mxx}. 

To estimate the enclosed mass within the velocity range  $v_{\mathrm{min}}$ and $v_{\mathrm{max}}$ for SN 2020akf at $-$5 days, the \texttt{TARDIS} model density profile was integrated between 12,000 and 30,000\,km\,s$^{-1}$. A similar exercise was done for SN 2021mxx at $-$3 days within the velocity range 10,000--30,000\,km\,s$^{-1}$.
The total enclosed mass was computed within 30 discrete velocity shells. The resulting velocity-dependent enclosed mass distributions for both SNe are shown in Figure \ref{fig_mass_fraction_20akf_21mxx}. SN 2020akf has a higher density than SN 2021mxx, resulting in a greater ejecta mass. 

We examined the convergence of our \texttt{TARDIS} model by tracking the evolution of the radiation temperature ($T_{\text{rad}}$) and dilution factor ($W$) across 30 iterations, for SN 2020akf and SN 2021mxx. The deviation in iteration is minimal after the third iteration, and both parameters stabilize by reaching the velocity $\sim$29,000\,km\,s$^{-1}$. Beyond this point, the values vary negligibly with velocity throughout the iteration, indicating that the model has reached a stable solution. This trend was observed in both SNe. The convergence plots for the radiation temperature and dilution factor of SN 2020akf and SN 2021mxx are shown in the Appendix/Figure \ref{fig_converge_TW}. 

\subsection{Photospheric expansion velocity}
\label{sec_velocity_evolution}

In Type Ib/c SNe, Fe\,{\sc ii} lines near 5000\,\r{A} are often employed to estimate the photospheric expansion velocity of the ejecta. 
However, in the cases where the Fe\,{\sc ii} lines are blended, the Si\,{\sc ii} $\lambda$6355 line serves as a reliable alternative indicator for photospheric velocity.

\begin{figure}[ht!]
\centering
\includegraphics[width=0.9\columnwidth, trim={0 1.2cm 0 0},clip]{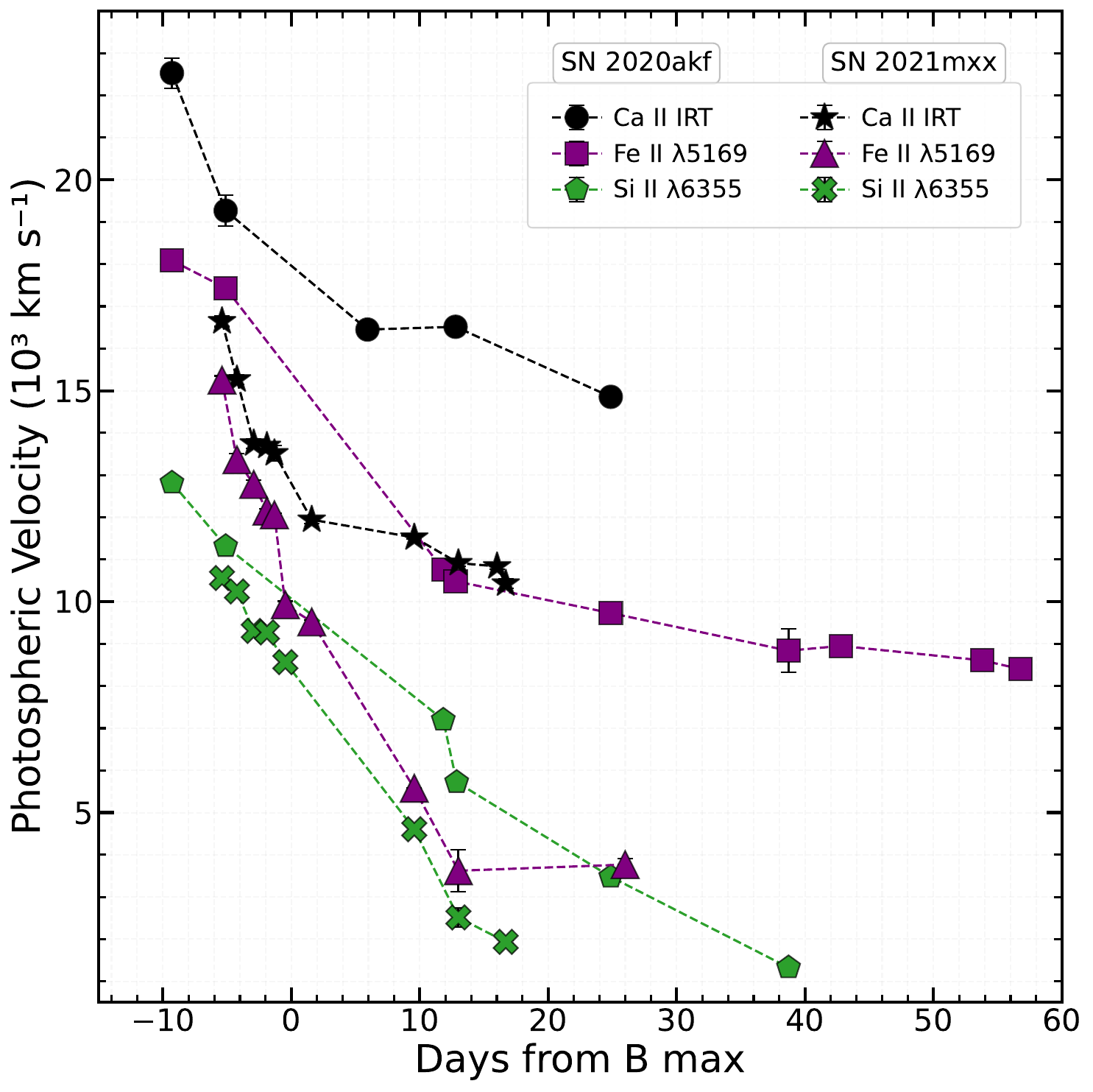}
\includegraphics[width=0.88\columnwidth]{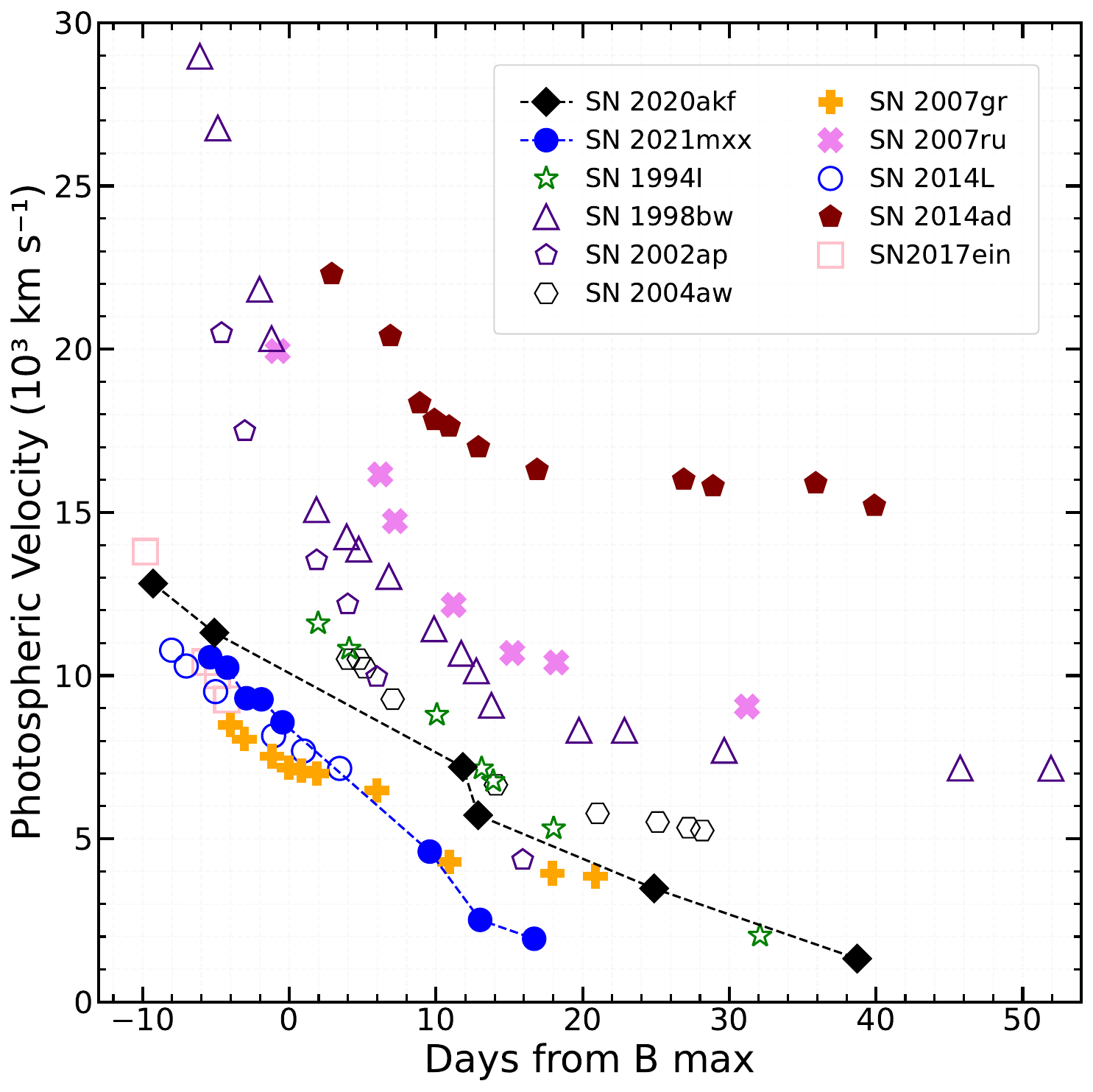}
\caption{Top panel: Evolution of the photospheric expansion velocities of SN 2020akf and SN 2021mxx, obtained from the absorption troughs of Si\,{\sc ii}, Ca\,{\sc ii}, and Fe\,{\sc ii}. Bottom panel: Comparison of the photospheric velocity evolution of SN 2020akf and SN 2021mxx derived from Si\,{\sc ii} with other Type Ic SNe. 
The velocities of the comparison SNe are taken from their respective literature \citep{taub06,sahu09,hunt09,sahu14ad,2014LZhang,2017einDan}}.
\label{fig_20akf_21mxx_vel_velcom}
\end{figure}

For both SN 2020akf and SN 2021mxx, we estimated the expansion velocity of the ejecta using the Fe\,{\sc ii} $\lambda$5169,  Si\,{\sc ii} $\lambda$6355 and Ca\,{\sc ii} NIR triplet lines by fitting Gaussian profile to their absorption trough in the redshift corrected spectra and are presented in Figure \ref{fig_20akf_21mxx_vel_velcom} (top panel).  The velocity estimated using Ca\,{\sc ii} NIR triplet and Fe\,{\sc ii} is found to be higher than Si\,{\sc ii} line velocity, as seen in other Type Ic SNe. 
For SN 2020akf, Ca\,{\sc ii} velocity declines from 
22,500 $\pm$ 760\,km\,s$^{-1}$ at $-$9 days to 
17,000 $\pm$ 525\,km\,s$^{-1}$ at +10 days. Similarly, for SN 2021mxx, the Ca\,{\sc ii} velocity declines from 
16,500 $\pm$ 525\,km\,s$^{-1}$ at $-$5 days to 
11,000 $\pm$ 590\,km\,s$^{-1}$ at +10 days. The higher velocity of the Ca\,{\sc ii} line indicates that calcium was distributed up to the outermost layers of the ejecta. The Fe\,{\sc ii} line velocity declines from 
18,000 $\pm$ 570\,km\,s$^{-1}$ at $-$9 days to 
10,000 $\pm$ 550\,km\,s$^{-1}$ at +20 days for SN 202akf.  In SN 2021mxx, the Fe\,{\sc ii} velocity is lower, declining from 
15,000 $\pm$ 550\,km\,s$^{-1}$ at $-$5 days to 
4,000 $\pm$ 550\,km\,s$^{-1}$ at +20 days.

In Figure \ref{fig_20akf_21mxx_vel_velcom} (bottom panel), the expansion velocities of SN 2020akf and SN 2021mxx measured using the Si\,{\sc ii} line are compared with other well-studied Type Ic SNe.
The Si\,{\sc ii} expansion velocities of both SN 2020akf and SN 2021mxx are within the typical range observed for normal Type Ic SNe and much lower than SNe Ic-BL. 
At the peak brightness, the velocities are around 
10,000 $\pm$ 530\,km\,s$^{-1}$ and 
8,500 $\pm$ 590\,km\,s$^{-1}$, for SN 2020akf and SN 20221mxx, respectively.

The expansion velocity of SN 2021mxx is comparable to those of SN 2007gr, SN 2014L, and SN 2017ein up to 10 days relative to $B$-maximum. SN 2020akf exhibits a higher velocity than SN 2021mxx, and its late-phase evolution closely matches that of SN 1994I.

\section{Oxygen mass}
\label{sec_oxygen_mass}

The nebular spectra of SN 2021mxx, obtained at approximately 102 and 113 days after $B$-maximum, display prominent [O\,{\sc i}] 6300 \AA \  emission lines. It is supposed to be the primary cooling agent in the late phases of Type Ib/c SNe \citep{1986Uomo, 1987Frans}.
The oxygen mass in the ejecta can be estimated from the flux of the [O\,{\sc i}] emission line using the expression provided by \citet{1986Uomo}:
\begin{equation}
    \text{M(O)} = 10^8 \times D^2 \times F([\text{O}\,\textsc{i}]) \times \exp{\left(\frac{2.28}{T_4}\right)} \ \text{M}_\odot
\end{equation}
where,  $F$([O\,{\sc i}]) represents the flux of the [O\,{\sc i}] $\lambda$$\lambda$6300, 6364 lines in units of erg\,s$^{-1}$\,cm$^{-2}$, $D$ is the distance to the supernova in Mpc, and $T_4$ denotes the temperature of the oxygen-emitting region in units of 10$^4$\,K. Ideally, the temperature of the oxygen-emitting region can be determined from the ratio of [O\,{\sc i}] $\lambda$5577 to [O\,{\sc i}] $\lambda$$\lambda$6300, 6364 lines. 

In the nebular spectrum of SN 2021mxx, the [O\,{\sc i}] $\lambda$5577 line is absent, constraining the line flux ratio [O\,{\sc i}] $\lambda$5577 / [O\,{\sc i}] $\lambda$$\lambda$6300, 6364 to $\leq$ 0.1. According to \citet{1989oster}, this ratio depends on the electron density and temperature of the emitting regions. The emitting region should either be at a relatively low temperature $(T_4 \leq 0.4)$ in the high-density limit or at a low electron density $(n_e \leq 5 \times 10^6 \text{ cm}^{-3})$ with temperature $T_4 = 1$ \citep{2007Maeda}. 
In the high-density regime, using the absolute flux of [O\,{\sc i}] $\lambda$$\lambda$6300, 6364 lines in the 113-days spectrum (1.53 $\times$ 10$^{-14}$\,erg\,s$^{-1}$\,cm$^{-2}$) and adopting a distance of 44.79 Mpc, the oxygen mass in SN 2021mxx is estimated to be 0.92\,M$_\odot$. 

\citet{2010Mazzalli} showed that oxygen mass estimated using flux of [O\,{\sc i}] $\lambda$$\lambda$6300, 6364 line can be regarded as a lower limit of the total oxygen mass ejected during the explosion. Hence, 0.92\,M$_\odot$ can be taken as a lower limit of the oxygen mass ejected in the explosion of SN 2021mxx.

The oxygen mass in the ejecta provides crucial insights into the progenitor's initial mass. 
\citet{kuncarayakti2015} compared the derived oxygen masses of SNe with nucleosynthesis yields predicted for massive stars of various initial masses based on models from \citet{nomoto1997}, \citet{rauscher2002}, and \citet{limongi2003}. For SN 2021mxx, the estimated oxygen mass of 0.92\,M$_\odot$ suggests a progenitor mass in the range of 15--18\,M$_\odot$ across all three models.

\section{Metallicity of the host galaxy}
\label{sec_metallicity}
To better understand the diverse properties of the supernova region in the host galaxy, we obtained a spectrum of UGC 11380 (the host galaxy of SN 2021mxx) on September 9, 2021. For SN 2020akf, we utilized the spectrum of its host galaxy, KUG 0925+387B, obtained from the SDSS Science Archive Server (\href{https://dr18.sdss.org/optical/spectrum/view?plateid=1214&mjd=52731&fiberid=156&run2d=26&zwarning=0&matches=any}{SAS}\footnote{\url{https://dr18.sdss.org/optical/spectrum/view?plateid=1214&mjd=52731&fiberid=156&run2d=26&zwarning=0&matches=any}}), which was observed on December 13, 2003.

In the host galaxy spectrum of SN 2021mxx, we identified several prominent narrow emission lines, including [O\,{\sc iii}] $\lambda4959$ and $\lambda5007$, H$_\alpha$, H$_\beta$, [N\,{\sc ii}] $\lambda6583$, [S\,{\sc ii}] $\lambda6716$ and $\lambda6730$. We used the fluxes from these narrow lines to estimate the metallicity of the host galaxy, employing well-established empirical relations from the literature, such as the N2 and O3N2 indices \citep{pett04}. The N2 index is defined as N2 
$\equiv \log([\text{N\,{\sc ii}}] \lambda6583/\text{H}_\alpha)$, and the O3N2 index as O3N2 $\equiv \log(([\text{O\,{\sc iii}}] \lambda5007/\text{H}_\beta)/([\text{N\,{\sc ii}}] \lambda6583/\text{H}_\alpha))$. 

For the galaxy KUG 0925+387B, the N2 index yielded an oxygen abundance of 12 + $\log$(O/H) = 8.68 in the host galaxy environment of SN 2020akf, while the O3N2 index gave a value of 8.59. Averaging these results, we find an oxygen abundance of 12 + $\log$(O/H) = 8.64 $\pm$ 0.06, corresponding to a metallicity of 0.81\,Z$_\odot$. Similarly, the oxygen abundance, 12 + $\log$(O/H), at the location of SN 2021mxx was estimated to be 8.60 using the N2 index and 8.59 using the O3N2 index. These values agree well with the oxygen abundance derived from the N2 index, based on host galaxy emission lines in the nebular spectrum of SN 2021mxx at $\sim$113 days (refer to Section \ref{sec_spec_nebular_phase}), which is around 8.64. The average oxygen abundance, 12 + $\log$(O/H) = 8.61 $\pm$ 0.03, corresponds to a metallicity of 0.76\,Z$_{\odot}$. This analysis suggests that the metallicity in the host galaxy environments of SN 2020akf and SN 2021mxx is close to solar metallicity ($12 + \log(\mathrm{O/H})_{\odot} = 8.73$, \citealt{Asplund2021}). 
\cite{2025ganss} have analyzed a large sample of Type Ib/Ic and Type IIP SNe to study their environments. They found mean oxygen abundances of approximately 12 + $\log$(O/H) = 8.50 for Type Ic SNe using both N2 and O3N2 calibrations.  Thus, the host environments of SN 2020akf and SN 2021mxx fall within the observed range for Type Ic SNe.

\section{Light curve modelling and explosion parameters}
\label{sec_explosion_parameter}

\subsection{Arnett–Valenti model}
\label{Arnett_model_akf_mxx}

We estimated the synthesized $^{56}$Ni mass using a modified radiation diffusion model \citep{arne82, vale08, Chatzopoulos2012}. This model applies to SNe with light curves powered by radioactive decay and assumes homologous, spherical expansion of the ejecta during the photospheric phase. The model also assumes that $^{56}$Ni is centrally located without mixing into the outer layers, a constant opacity, and ejecta dominated by radiation pressure. We used the modified luminosity formula from \citet{vale08}, which includes the energy contribution from $^{56}$Co to $^{56}$Fe decay and incorporates the Gamma-ray leakage from the ejecta. The $^{56}$Ni mass and effective diffusion time ($\tau_d$), which governs the width of the bolometric light curve, were treated as free parameters. 

The produced luminosity is expressed as:
\begin{eqnarray}
L(t) = {\text{M}}_{\text{Ni}} e^{-x^2} \left[ (\epsilon_{\text{Ni}} - \epsilon_{\text{Co}}) \int_0^x 2z e^{z^2 -2zy} \, dz \right. \notag \\ + \left. \epsilon_{\text{Co}} \int_0^x 2z e^{z^2 -2zy+2zs} \, dz \right] (1 - e^{-(\frac{t_\gamma}{t})^2})
\end{eqnarray}
where \( x \equiv t_{\text{exp}} / t_{\text{d}} \), $t_{\text{exp}}$ is the time since explosion (in days) and \( t_{\text{d}} \) is the diffusion timescale (days). \( y \equiv t_{\text{d}} / (2 t_{\text{Ni}}) \) with \( t_{\text{Ni}} = 8.8 \) days, \( s \equiv [t_{\text{d}} (t_{\text{Co}} - t_{\text{Ni}})] / (2 t_{\text{Co}} t_{\text{Ni}}) \) with \( t_{\text{Co}} = 111.3 \) days. M$_{\text{Ni}}$ is the Ni mass and \( t_{\gamma} \) is the Gamma ray leakage timescale (in days). 
$\epsilon_{\text{Ni}}$ = 3.9 $\times$ 10$^{10}$\,erg\,s$^{-1}$\,g$^{-1}$ and $\epsilon_{\text{Co}}$ = 6.8 $\times$ 10$^9$\,erg\,s$^{-1}$\,g$^{-1}$ represent the energy generation rates from the radioactive decay of $^{56}$Ni and $^{56}$Co, respectively. 

For uniform density, the kinetic energy of the ejecta ($E_k$) and the effective diffusion time scale ($\tau_d$) were derived following \citet{arne82, arne96}, given by

\begin{equation}
\label{equ:energy}
E_{k}\approx \frac{3}{5} \frac{{\text{M}_{\rm ej}v_{\rm ph}^{2}}}{2}
\end{equation} 

\begin{equation}
\label{equ:tau} 
\tau_{d} = \left(\frac{\kappa}{\beta c}\right)^{1/2} \left(\frac{{6 \text{M}_{\rm ej}^{3}}}{{5 E_{k}}}\right)^{1/4}
\end{equation} 
where, $\kappa$ is the optical opacity, taken as 0.07\,cm$^{2}$\,g$^{-1}$ \citep{chug00}, $\beta \approx$ 13.8 is a constant of integration \citep{arne82}, and $c$ is the speed of light.

\begin{figure}[htbp]
\centering
\includegraphics[width=0.9\columnwidth, trim={0 1.2cm 0 0},clip]{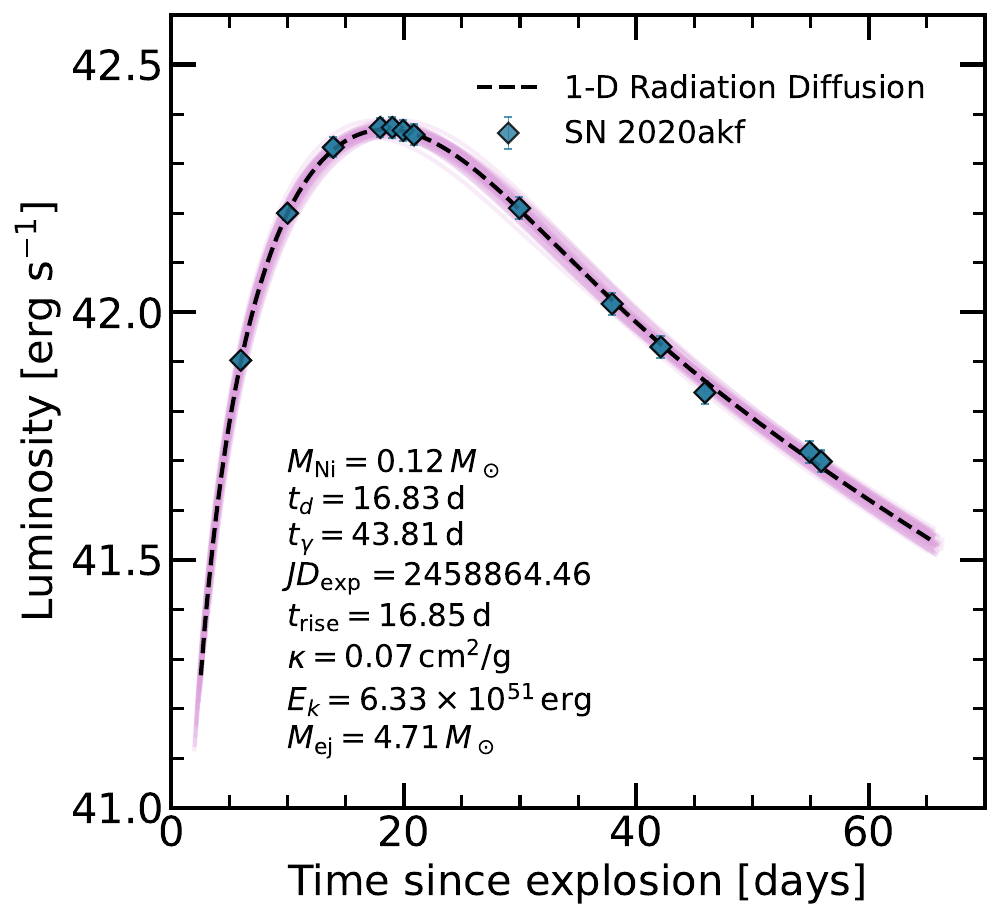}
\includegraphics[width=0.9\columnwidth]{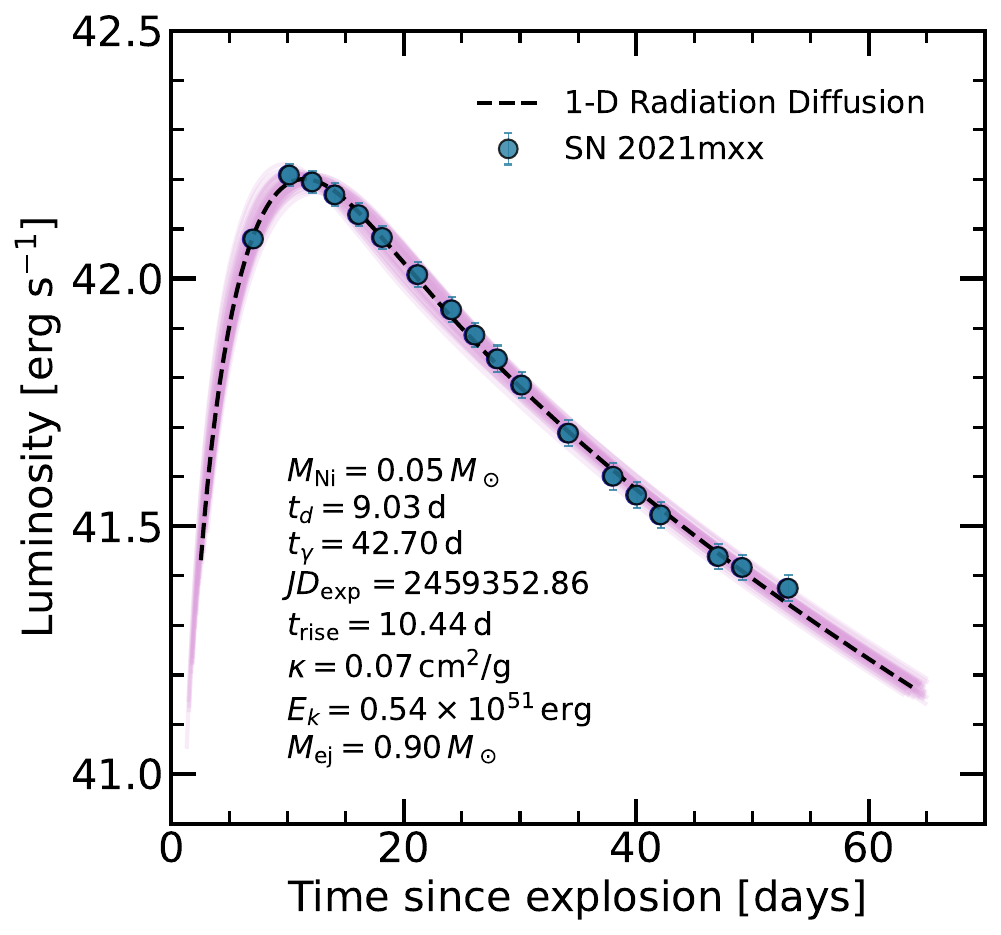}
\caption{The quasi-bolometric light curves of SN 2020akf (top panel) and SN 2021mxx (bottom panel) are fitted using the \textit{Arnett–Valenti} model. The estimated parameters are presented below the fit.}
\label{fig_20akf_21mxx_arnett_model}
\end{figure}

To fit the quasi-bolometric light curves of SN 2020akf and SN 2021mxx, we employed a Markov Chain Monte Carlo (MCMC) approach. We utilized the \texttt{emcee} package in Python to optimize the model parameters and sample the posterior distribution, following the \citet{Foreman-Mackey2013}. The model fitting process incorporated four key parameters: $t_{\text{exp}}$ (the explosion epoch), M$_{\text{Ni}}$ (the mass of $^{56}$Ni synthesized), $t_{\text{d}}$ (the diffusion timescale) and $t_{\gamma}$ (the Gamma-ray leakage timescale). For a detailed description of the model fit procedure, we refer to \citet{Anirban2022}. 

From the fit, we derived the following parameters for SN 2020akf, M$_{\text{Ni}}$ = 0.12$^{+0.01}_{-0.01}$\,M$_{\odot}$, $t_{\text{d}}$ = 16.83$^{+0.97}_{-0.82}$\,days, $t_{\gamma}$ = 43.81$^{+2.04}_{-2.17}$\,days and JD$_{\text{exp}}$ = 245\,8864.46$^{+0.28}_{-0.30}$. Similarly, for SN 2021mxx, the estimated parameters are M$_{\text{Ni}}$ = 0.05$^{+0.001}_{-0.001}$\,M$_{\odot}$, $t_{\text{d}}$ = 9.03$^{+0.69}_{-0.89}$\,days, $t_{\gamma}$ = 42.70$^{+1.31}_{-1.22}$\,days, and JD$_{\text{exp}}$ = 245\,9352.86$^{+0.74}_{-0.57}$. 
The model fits for SN 2020akf (top panel) and SN 2021mxx (bottom panel) are shown in Figure \ref{fig_20akf_21mxx_arnett_model} and the corresponding corner plots are shown in the Appendix/Figure \ref{fig_20akf_21mxx_corner_arnett}.

Following \citet{Taddia2018,cano2013,lyma16}, we used the photospheric Fe\,{\sc ii} line velocities to estimate the ejecta mass (M$_{\text{ej}}$) and kinetic energy ($E_k$) of the explosion. Taking $v_{\text{ph}}$ = 15,000 $\pm$ 700\,km\,s$^{-1}$ for SN 2020akf and 10,000 $\pm$ 600\,km\,s$^{-1}$ for SN 2021mxx at peak brightness, and applying 
equations \ref{equ:energy} and \ref{equ:tau}, we obtained M$_{\text{ej}}$ = 4.71$^{+0.50}_{-0.46}$\,M$_\odot$ and $E_k$ = 6.33$^{+0.68}_{-0.62}$ $\times$ 10$^{51}$\,erg for SN 2020akf, and M$_{\text{ej}}$ = 0.90$^{+0.14}_{-0.18}$\,M$_\odot$ and $E_k$ = 0.54$^{+0.08}_{-0.12}$ $\times$ 10$^{51}$\,erg for SN 2021mxx.

\subsection{MOSFiT model}

We performed multi-colour light curve modeling for SN 2020akf and SN 2021mxx using the Modular Open-Source Fitter for Transients (\texttt{MOSFiT}, \citealt{mosfit2018}) with its \texttt{default} model, which incorporates the radioactive decay of nickel and cobalt \citep{nady94}. We employed the dynamic nested sampling method, dynesty \citep{2020speagle}, available within \texttt{MOSFiT}, and assessed the fit quality using the likelihood score ($\log$\,(Z); \citealt{watanabe2010}). 
The model contains 10 free parameters: the host-galaxy hydrogen column density ($n_{\rm H,host}$), the total ejecta mass (M$_{\rm ej}$), the $^{56}$Ni mass fraction ($f_{\rm Ni} \equiv \rm M_{Ni}/M_{ej}$), the ejecta velocity ($v_{\rm ej}$), the $\gamma$-ray opacity ($\kappa_\gamma$), 
the explosion epoch relative to the first observation ($t_{\rm exp}$), the minimum photospheric temperature ($T_{\min}$), a white-noise variance term ($\sigma$), and the temporal and wavelength correlation scales ($\Delta t$ and $\Delta\lambda$) that describe the Gaussian-process covariance used to account for additional correlated scatter in the light-curve data.
To fit the light curve, well-constrained priors were adopted, including flat, log-flat, and Gaussian. We fixed a few important parameters, such as redshift, $E(B-V)_{MW}$, and luminosity-distance, for the light curve fitting of each SN. For the better convergence\footnote{\url{https://mosfit.readthedocs.io/en/latest/assessing.html}}, \texttt{MOSFiT} was run with \texttt{-R 1.05}, ensuring convergence once the Potential Scale Reduction Factor (\texttt{PSRF}) reached 1.05. The Gelman-Rubin statistic was used to assess chain mixing, with \texttt{PSRF} values near 1 indicating well-mixed chains. This procedure improved posterior sampling by enforcing proper mixing. The sampling utilized 2000 walkers until convergence was achieved. A detailed description of the model-fitting procedure is provided in \citet{mosfit2018} and recent studies, such as \citet{SS2023, Davis2023, Gkini2024, Moore2025}.

\begin{figure}[ht!]
\centering
\includegraphics[width=0.90\columnwidth, trim={0 0 0 0},clip]{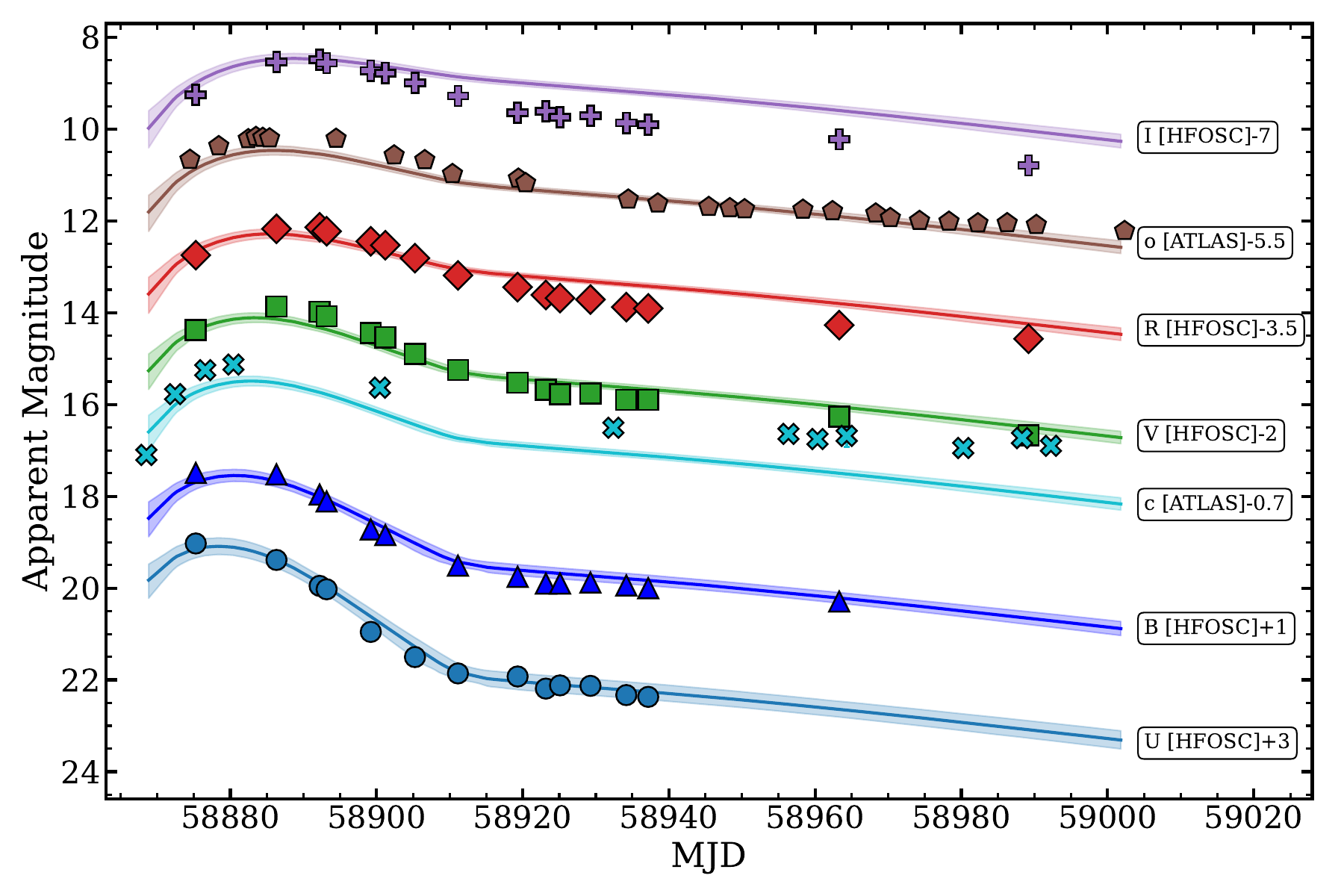}
\includegraphics[width=0.91\columnwidth, trim={0.5cm 0 0 0},clip]{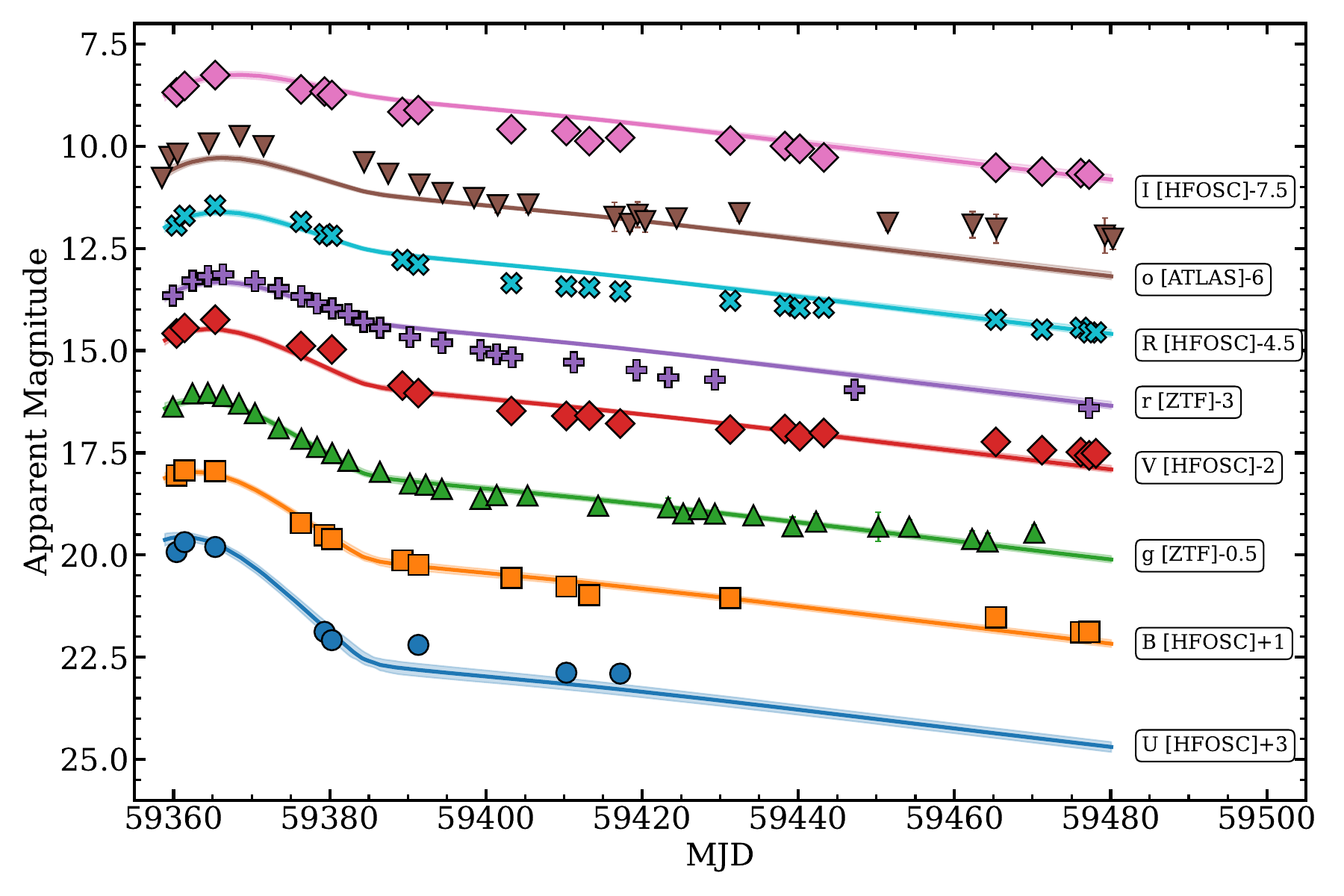}
\caption{The multi-band $UBcVRoI$ light curves of SN~2020akf (top panel) and $UBgVrRoI$ light curves of SN~2021mxx (bottom panel) are fitted with the default \texttt{MOSFiT} model. The plotted lines show the median \texttt{MOSFiT} light curve fits, and the shaded regions indicate the 5--95\% confidence intervals for each band during the first 120 days. }
\label{fig_20akf_21mxx_lc_mosfit}
\end{figure}

The multi-colour light curves of SN 2020akf (top panel) and SN 2021mxx (bottom panel) with fitted models are displayed in Figures \ref{fig_20akf_21mxx_lc_mosfit}. The associated corner plots are shown in the Appendix/Figures \ref{fig_cakf_Mosfit} and \ref{fig_cmxx_Mosfit}, respectively. 
The optimized model reproduced the observed light curves of both SNe reasonably well in the early phase $\sim$ 30 days after maximum, except $c$ and $o$ band in SN 2020akf, and $o$ band in SN 2021mxx. However, in the later phases, some deviation is seen, especially in the redder bands, such as $o$ and $I$ bands.  From the \texttt{MOSFiT} model, for SN 2020akf, the synthesized $^{56}$Ni mass (M$_{\mathrm{Ni}} =  f_{\rm Ni} \times \rm{M_{ej}}$), and the ejecta masses (M$_{\text{ej}}$) are estimated to be $0.20^{+0.07}_{-0.05}$\,M$_{\odot}$ and $3.16^{+0.55}_{-0.47}$\,M$_{\odot}$, respectively. For SN 2021mxx, the corresponding quantities are $0.09^{+0.02}_{-0.01}$\,M$_{\odot}$ and $1.53^{+0.18}_{-0.17}$\,M$_{\odot}$. The other estimated parameters and priors for both SNe are summarized in Appendix/Table \ref{tab:mosfit_akf_parameters}. 
There is a slight difference in the M$_{\rm Ni}$ and M$_{\mathrm{ej}}$ values estimated using 
\texttt{MOSFiT} than those from the \cite{arne82} model, which may be due to variations in $t_{\mathrm{exp}}$ and $v_{\mathrm{ej}}$ in the two models. However, they are consistent within $3\sigma$ uncertainties. 

\section{Discussion and summary}
\label{sec_summary}

We present a comprehensive optical photometric and spectroscopic analysis of two Type Ic SNe, 2020akf and 2021mxx. 

\subsection{SN 2020akf}

It exhibits broader light curves across the $UBVRI$ bands, indicating a slower decline in brightness. The absolute $V$-band magnitude of SN 2020akf ($M_{V}$ = $-$17.85 $\pm$ 0.15\,mag) and its quasi-bolometric luminosity ($\log L_\text{bol}^\text{max} = 42.37 \pm 0.02$\,erg\,s$^{-1}$) are on the higher side compared to many well-sampled Type Ic SNe but lower than those of Type Ic-BL SNe. This places SN 2020akf in the category of transitional Type Ic SN, between the normal Type Ic and Type Ic-BL SNe, similar to SN 2004aw \citep{taub06}.
The spectral evolution of SN 2020akf exhibits prominent Si\,{\sc ii} and C\,{\sc ii} lines, which we could reproduce using \texttt{TARDIS}. Similar spectral features have been observed in the early spectra of SN 2007gr \citep{2014chen} and SN 2004aw \citep{taub06}.

Previous investigations by \citet{MW2023} noted spectroscopic similarities between SN 2020akf and two stripped-envelope superluminous SNe (SLSNe Ic), iPTF13ajg \citep{Vreeswijk2014} and SSS120810-23 \citep{nicholl2014}, and also discussed the presence of O\,{\sc ii} features in the early spectra of SN 2020akf, which are generally found in SLSNe Ic. Using \texttt{TARDIS}, we examined the presence of O\,{\sc ii} line and found that the observed features are more consistent with Mg\,{\sc ii} lines.
Comparing the photometric and spectroscopic properties of SN 2020akf with those of well-studied SLSNe Ic (e.g., \citealt{liu2017, mazzali2016, yan2015}) suggests that it does not exhibit the features of superluminous events, while it shares the majority of characteristics of Type Ic SNe (e.g., \citealt{prentice2016}).
  
To understand the progenitor properties, we estimated the explosion parameters using the Ni decay model of \cite{arne82,arne96} and the \texttt{MOSFiT} tool. With the updated Arnet-Valenti model, 
the amount of radioactive $^{56}$Ni synthesized is M$_{\text{Ni}}$ = 0.12$^{+0.01}_{-0.01}$\,M$_{\odot}$, 
the ejected mass during the explosion is M$_{\text{ej}}$ = 4.71$^{+0.50}_{-0.46}$\,M$_\odot$, and kinetic energy is $E_k = 6.33^{+0.68}_{-0.62} \times 10^{51}$ erg.  The diffusion time scale is estimated to be $t_{\text{d}}$ = 16.83$^{+0.97}_{-0.82}$\,days. The higher diffusion time for SN 2020akf indicates a denser, higher-opacity environment consistent with its larger ejected mass. 
Using the \textit{default} model in \texttt{MOSFiT},  we estimated M$_\text{Ni}$ = 0.20$^{+0.07}_{-0.05}$\,M$_\odot$ and  M$_\text{ej}$ = 3.16$^{+0.55}_{-0.47}$\,M$_\odot$. From both modelling approaches, we find that the inferred explosion parameters are consistent within the $3\sigma$ uncertainties. The metallicity estimates based on the N2 and O3N2 indices suggest that SN 2020akf exploded in a region with nearly solar metallicity. Due to the non-availability of late-phase data of SN 2020akf, it is difficult to estimate the mass of its progenitor star. However, taking into account the estimated mass of the ejecta $\sim$ 4 M$_\odot$ and assuming that the mass of the neutron star formed after explosion is $\sim$ 2 M$_\odot$, the mass of the core at the time of explosion would be $\sim$ 6 M$_\odot$. 

\subsection{SN 2021mxx}

SN 2021mxx exhibits narrower light curves with a rapid rise and a faster decline, similar to those of Type Ic SNe 1994I, 2014L, and 2007gr. It attained peak $V$-band absolute magnitude ($M_{V} = -17.60 \pm 0.15$\,mag) and quasi-bolometric luminosity ($\log L_\text{bol}^\text{max}$ = 42.21 $\pm$ 0.02\,erg\,s$^{-1}$). The Arnet-Valenti model fit gives M$_{\text{Ni}}$ = 0.05$^{+0.00}_{-0.00}$\,M$_{\odot}$, M$_{\text{ej}} = 0.90^{+0.14}_{-0.18}$\,M$_\odot$, and $E_k = 0.54^{+0.08}_{-0.12} \times 10^{51}$\,erg. The estimated diffusion time $t_{\text{d}}$ = 9.03$^{+0.69}_{-0.89}$\,days suggests that SN 2021mxx exhibits a rapid rise. The \textit{default} model in \texttt{MOSFiT} gives M$_\text{Ni} = 0.09^{+0.02}_{-0.02}$\,M$_\odot$, and M$_\text{ej} = 1.53^{+0.18}_{-0.17}$\,M$_\odot$, indicating that the inferred explosion parameters are consistent within the uncertainties $3\sigma$ in both the model.

 The prominent features identified in the spectra of SN 2021mxx are identical to those of Type Ic events SN 1994I, SN 2004aw,  SN 2017ein, and were successfully reproduced using \texttt{TARDIS}. The estimated ejecta velocity at peak brightness and its evolution are comparable to those of SN 2007gr and SN 2014L. 
 
 Analysis of the nebular-phase spectrum of SN 2021mxx, observed 123 days post-explosion, reveals [O\,{\sc i}]/[Ca\,{\sc ii}] line ratio of approximately 0.96, indicating a lower mass progenitor for this supernova. Using the absolute flux of the [O\,{\sc i}] $\lambda6300$ line, the estimated mass of oxygen in the ejecta is $\sim$ 0.92\,M$_\odot$, which indicates a ZAMS mass of the progenitor star to be in the range 15--18 M$_\odot$. We compared the nebular-phase spectrum of SN 2021mxx  with the nebular-phase model spectra from \citet{2015Jerkstrand}. The best-fitting model indicates that the progenitor mass is in the range of 13--17 M$_\odot$. Hence, the mass range of the progenitor star could be taken as 13--18 M$_\odot$. The metallicity of the host galaxy in the supernova region is found to be near-solar.

\begin{figure}[ht!]
\centering
\includegraphics[width=0.9\columnwidth]{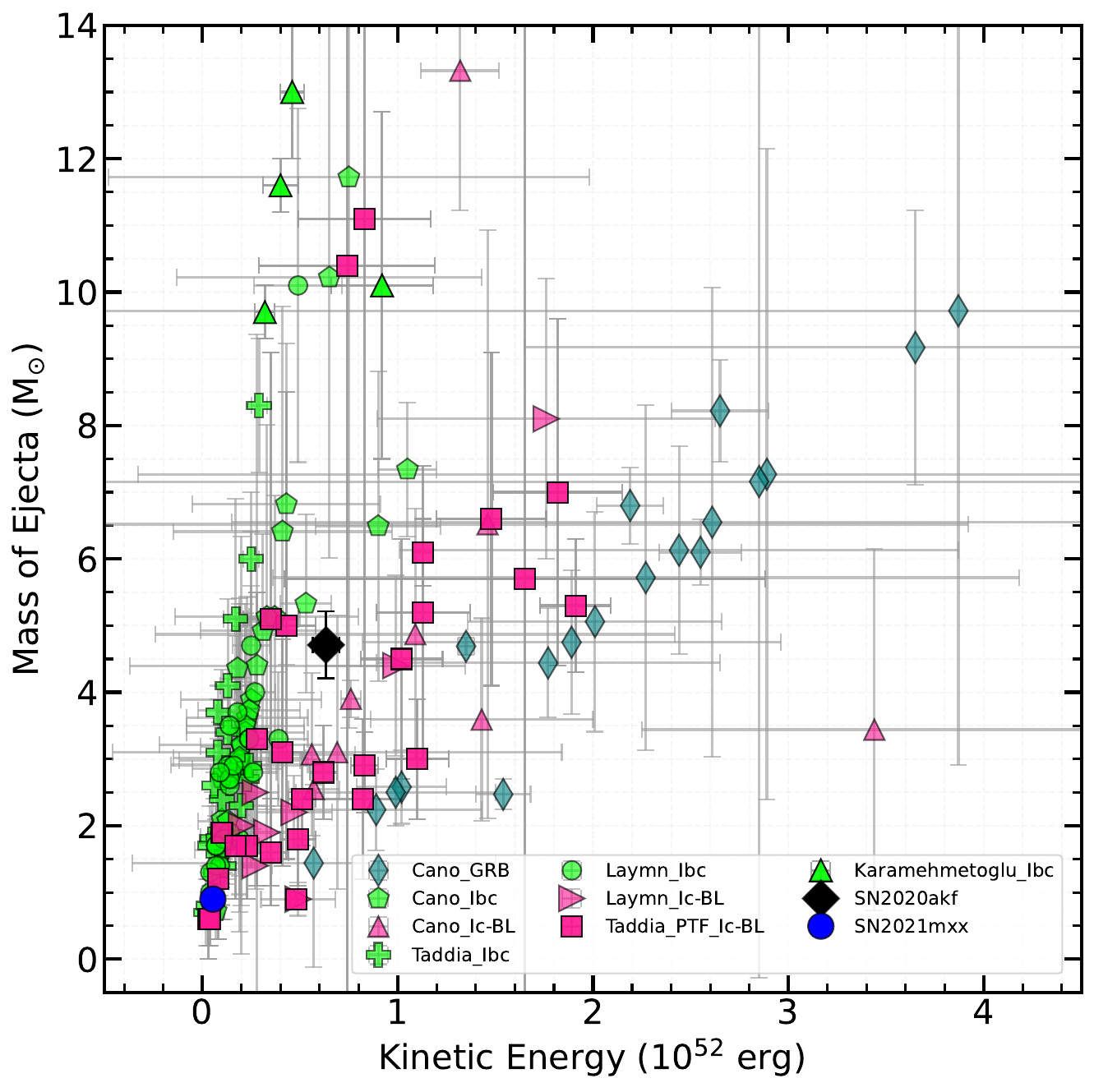}
\caption{The ejecta mass (M$_{\text{ej}}$) and kinetic energy ($E_k$) for SN 2020akf and SN 2021mxx are plotted along with those of other well-studied Type Ic, Type Ic-BL, and GRB/XRF SNe. The Sample data are taken from \cite{cano2013}, \cite{lyma16}, \cite{Taddia2018}, \cite{Taddia2019}, and \cite{Karame2023}. }
\label{fig_Mej_Kinetic_E}
\end{figure}

In Figure \ref{fig_Mej_Kinetic_E}, the distribution of M$_\text{ej}$--$E_{k}$ for well-studied Type Ibc, Type Ic-BL, and GRB/XRF SNe,  taken from \cite{cano2013}, \cite{lyma16}, \cite{Taddia2018}, \cite{Taddia2019}, and \cite{Karame2023},  has been plotted along with SN 2020akf and SN 2021mxx. 
\citet{cano2013} found that Type Ibc SNe typically exhibit a median ejecta mass of M$_{\rm ej}$ $\sim$ 3.4--3.9\,M$_{\odot}$, a kinetic energy of 
$E_{k}$ $\sim$ 0.2 $\times$ 10$^{52}$\,erg, and a characteristic $E_{k}$/M$_{\rm ej}$ ratio of $\sim$ 0.06 in unit of (10$^{52}$\,erg/M$_{\odot}$). 
For SN~2020akf, we derive $E_{k}$/M$_{\rm ej}$ $\sim$ 0.134, 
which is comparable to that of SN~2004aw (0.134). However, with $E_{k}$/M$_{\rm ej}$ $\sim$ 0.06,  SN~2021mxx shows value more consistent with the well-studied, typical Type~Ic SNe, 
such as SN~2005mf (0.063), SN~2007gr (0.064), and SN~2011bm (0.063).
The explosion parameters of both SNe are broadly consistent with those of Type Ib/c SNe and fall well within the region occupied by this class. However, SN 2021mxx lies distinctly in the parameter space, characterized by very low M$_\text{ej}$ and $E_{k}$.

\begin{acknowledgments}
We thank the referee for carefully reading the manuscript and providing constructive suggestions which helped in improving the manuscript. We express our gratitude to the Director and Dean of IIA, Bengaluru, for their generous hospitality and the facilities provided. We appreciate  assistance of IIA staff at CREST and IAO during our observations, as well as the observers of the 2-m HCT who kindly allocated part of their observing time for supernova studies. D.K.S. acknowledges financial support from the CRG/ 2022/007688 project grant. 
M.S. acknowledges financial support provided under the National Post Doctoral Fellowship (N-PDF; File Number: PDF/2023/002244) by the Science \& Engineering Research Board (SERB), Anusandhan National Research Foundation (ANRF), Government of India.
This work has utilized data from the NASA Astrophysics Data System (ADS), the NASA/IPAC Infrared Science Archive (IRSA), and the NASA/IPAC Extragalactic Database (NED), which are operated by the Jet Propulsion Laboratory, California Institute of Technology, under contract with the National Aeronautics and Space Administration.  We have utilized the Weizmann Interactive Supernova Data Repository (WISeREP) and data from ZTF, ATLAS, and SDSS. 
\end{acknowledgments}

\section*{Data Availability}

The data presented in this article will be shared on request to the corresponding author. All spectra will be made publicly available through  WISeREP. 

\facility{2-meter Himalayan Chandra Telescope (HCT) at the Indian Astronomical Observatory (IAO),  Hanle, India. }

\software{IRAF \citep{IRAF}, TARDIS \citep{Kerzendorf2014, Vogl2019}, MOSFiT \citep{mosfit2018}, SuperBol \citep{Nicholl18}, Matplotlib \citep{matplotlib}, Numpy \citep{numpy}, Astropy \citep{astropy,Astropy2018,Astropy2022}, Scipy \citep{scipy}, scikit-learn \citep{scikit}, corner \citep{corner}, emcee \citep{emcee}. 
}

\bibliography{references}{}
\bibliographystyle{aasjournalv7}

\appendix

\begin{deluxetable*}{llclc}
\tabletypesize{\footnotesize}
\tablewidth{0pt}
\tablecaption{Important parameters of SN 2020akf, SN 2021mxx and their host galaxies. \label{tab_sn20akf_sn2021mxx_host} }
\tablehead{
\colhead{Parameters} & \colhead{Values} & \colhead{Ref} & \colhead{Values} & \colhead{Ref} 
}
\startdata
               & {\bf SN 2020akf}  &   & {\bf SN 2021mxx}  & \\
Discovery Date & 2020-01-20 12:04:19 & 1 & 2021-05-22 09:22:03 & 6 \\
               & (JD 245\,8869.0)    &   & (JD 245\,9356.9) & \\
RA (J2000)  & 09:28:39.6 & 1 & 18:56:51.3 & 6\\
DEC (J2000) & 38:33:47.1 & 1 & 36:37:20.4 & 6\\
Explosion Date & JD 245\,8864.46$^{+0.28}_{-0.30}$ & 2 & JD 245\,9352.86$^{+0.74}_{-0.57}$ & 2 \\
Date of $B$-maximum & JD 245\,8880.4 $\pm$ 0.5 & 2 & JD 245\,9363.3 $\pm$ 0.5  & 2 \\
$\Delta$m$_{15}$(B) & 0.98 $\pm$ 0.02   & 2 & 1.52 $\pm$ 0.03 & 2 \\
Galactic reddening & 0.013 $\pm$ 0.001 mag  & 3 & 0.08 $\pm$ 0.002 mag  & 3 \\
Host reddening & Negligible   & 2 & 0.12 $\pm$ 0.03 mag & 2 \\
$E(B - V)_{\text{total}}$ & 0.013 $\pm$ 0.001 mag   & 2 & 0.20 $\pm$ 0.03 mag & 2 \\
$m_B^{\text{max}}$ & 16.34 $\pm$ 0.02 mag   & 2 & 16.90 $\pm$ 0.03 mag & 2 \\
$M_V^{\text{max}}$ & $-$17.85 $\pm$ 0.15 mag   & 2 & $-$17.60 $\pm$ 0.15 mag  & 2 \\
$\log L^{\text{max}}_{\text{quasi-bol}}$ & 42.37 $\pm$ 0.02 erg\,s$^{-1}$   & 2 & 42.21 $\pm$ 0.02 erg\,s$^{-1}$ & 2 \\
$^{56}$Ni mass & 0.10 $\pm$ 0.02 M$_\odot$   & 2 & 0.05 $\pm$ 0.01 M$_\odot$  & 2 \\
Ejected mass (M$_{\text{ej}}$) & 4.71$^{+0.50}_{-0.46}$\,M$_\odot$   & 2 &  0.90$^{+0.14}_{-0.18}$\,M$_\odot$  & 2 \\
$v_{\text{max}}$(Fe\,{\sc ii} 5169) & 15,000\,km\,s$^{-1}$  & 2 & 10,000 \,km\,s$^{-1}$  & 2 \\
Kinetic energy ($E_k$) & 6.33$^{+0.68}_{-0.62} \times 10^{51}$ erg  & 2 & 0.54$^{+0.08}_{-0.12} \times 10^{51}$ erg  & 2 \\
\\[-2.5ex]
            & {\bf KUG 0925+387B} & & {\bf UGC 11380}  & \\
Type & Spiral-Sc   & 4 &  Spiral-Sab & 4 \\
RA (J2000) & 09:28:39.8  & 5 & 18:56:51.1  & 5 \\
DEC (J2000) & 38:33:49.0   & 5 & 36:37:26.3  & 5 \\
$z_{\text{helio}}$ & 0.012445 $\pm$ 0.000011  & 5 & 0.009650 $\pm$ 0.000060 & 5 \\
$v_{\text{Virgo}}$ & 3,903 $\pm$ 14\,km\,s$^{-1}$   & 5 & 3,225 $\pm$ 25\,km\,s$^{-1}$ & 5 \\
Distance & 54.21 $\pm$ 4 Mpc  & 5$^a$ & 44.79 ± 3 Mpc & 5$^a$ \\
Distance modulus & 33.67 $\pm$ 0.15 mag    & 5$^a$ & 33.26 $\pm$ 0.15 mag & 5$^a$ \\
\enddata
\tablecomments{1. \citet{2020Tonry} 2. This work 3. \citet{schl11} \\
4. \href{http://atlas.obs-hp.fr/hyperleda/}{Hyperleda} 5. \href{https://ned.ipac.caltech.edu/}{NED} 5$^a$. Using $H_0$ = 72\,km\,s$^{-1}$\,Mpc$^{-1}$ 6. \citet{2021ZTFmxx} \\
}
\end{deluxetable*}

\begin{deluxetable*}{lccccccc}
\tabletypesize{\scriptsize}
\tablewidth{0pt}
\tablecaption{Magnitudes of secondary standard stars in the field of SN 2020akf and SN 2021mxx. The stars are marked in Figure \ref{fig_sn_field}. \label{tab_sec_std} }
\tablehead{
\colhead{ID} & \colhead{RA} & \colhead{DEC} & \colhead{$U$} & \colhead{$B$}  & \colhead{$V$} &  \colhead{$R$} & \colhead{$I$}
}
\startdata
\multicolumn{8}{c}{\textbf{SN 2020akf}} \\
        1 & 09 28 30.7 & +38 33 22.7 & 14.43 $\pm$ 0.04  & 14.56 $\pm$ 0.07 & 14.09 $\pm$ 0.06 & 13.77 $\pm$ 0.08 & 13.46 $\pm$ 0.09 \\
        2 & 09 28 36.6 & +38 29 41.8 & 5.35 $\pm$ 0.05  & 15.40 $\pm$ 0.07 & 14.82 $\pm$ 0.07 & 14.44 $\pm$ 0.09 & 14.08 $\pm$ 0.12 \\
        3 & 09 28 27.9 & +38 29 50.4 & 16.48 $\pm$ 0.06  & 15.85 $\pm$ 0.09 & 14.94 $\pm$ 0.08 & 14.42 $\pm$ 0.11 & 13.94 $\pm$ 0.14 \\
        4 & 09 28 42.5 & +38 32 23.3 & 16.65 $\pm$ 0.04  & 15.92 $\pm$ 0.08 & 15.01 $\pm$ 0.07 & 14.51 $\pm$ 0.10 & 14.07 $\pm$ 0.12 \\
        5 & 09 28 38.2 & +38 31 44.0 & 16.64 $\pm$ 0.05  & 15.98 $\pm$ 0.08 & 15.10 $\pm$ 0.08 & 14.58 $\pm$ 0.08 & 14.10 $\pm$ 0.10 \\
        6 & 09 28 22.1 & +38 32 45.1 & 16.42 $\pm$ 0.05  & 16.08 $\pm$ 0.08 & 15.24 $\pm$ 0.07 & 14.75 $\pm$ 0.11 & 14.28 $\pm$ 0.13 \\
        \multicolumn{8}{c}{\textbf{SN 2021mxx}} \\
        1  & 18 56 38.1 & +36 38 17.3 & 15.56 $\pm$ 0.02 & 15.16 $\pm$ 0.02 & 14.37 $\pm$ 0.02 & 13.90 $\pm$ 0.02 & 13.46 $\pm$ 0.02 \\
        2  & 18 56 44.2 & +36 40 23.7 & 15.19 $\pm$ 0.02 & 15.01 $\pm$ 0.02 & 14.36 $\pm$ 0.01 & 13.96 $\pm$ 0.02 & 13.61 $\pm$ 0.01 \\
        3  & 18 56 44.6 & +36 40 41.7 & 15.72 $\pm$ 0.02 & 15.54 $\pm$ 0.02 & 14.86 $\pm$ 0.01 & 14.46 $\pm$ 0.03 & 14.07 $\pm$ 0.03 \\
        4  & 18 56 52.6 & +36 35 57.2 & 16.08 $\pm$ 0.02 & 15.76 $\pm$ 0.02 & 15.02 $\pm$ 0.01 & 14.59 $\pm$ 0.01 & 14.20 $\pm$ 0.02 \\
        5  & 18 56 43.8 & +36 35 37.0 & 15.27 $\pm$ 0.01 & 15.39 $\pm$ 0.02 & 14.84 $\pm$ 0.01 & 14.46 $\pm$ 0.01 & 14.08 $\pm$ 0.02 \\
        6  & 18 56 38.8 & +36 37 01.2 & 16.32 $\pm$ 0.03 & 15.89 $\pm$ 0.01 & 15.12 $\pm$ 0.01 & 14.67 $\pm$ 0.02 & 14.26 $\pm$ 0.02 \\
        7  & 18 56 50.6 & +36 36 16.1 & 16.92 $\pm$ 0.02 & 16.44 $\pm$ 0.02 & 15.53 $\pm$ 0.02 & 15.02 $\pm$ 0.02 & 14.52 $\pm$ 0.02 \\
        8  & 18 56 56.2 & +36 40 03.1 & 16.01 $\pm$ 0.03 & 15.93 $\pm$ 0.01 & 15.30 $\pm$ 0.01 & 14.92 $\pm$ 0.01 & 14.54 $\pm$ 0.01 \\
        9  & 18 56 30.8 & +36 39 11.4 & 19.80 $\pm$ 0.10 & 18.57 $\pm$ 0.01 & 17.05 $\pm$ 0.02 & 16.01 $\pm$ 0.01 & 14.85 $\pm$ 0.01 \\
        10 & 18 56 52.4 & +36 38 25.6 & 16.75 $\pm$ 0.02 & 16.23 $\pm$ 0.03 & 15.42 $\pm$ 0.01 & 14.93 $\pm$ 0.01 & 14.48 $\pm$ 0.02 \\
\enddata
\end{deluxetable*}

\begin{deluxetable*}{lcrccccc}
\tabletypesize{\scriptsize}
\tablewidth{0pt}
\tablecaption{Optical $UBVRI$ photometric observations of SN 2020akf with HCT. \label{tab_mag_sn2020akf} }
\tablehead{
\colhead{Date} & \colhead{JD$^a$} & \colhead{Phase$^b$} & \colhead{$U$} & \colhead{$B$}  & \colhead{$V$} &  \colhead{$R$} & \colhead{$I$}
}
\startdata
        26/01/2020 & 8875.28 & $-$5.1 & 16.03 $\pm$ 0.01 & 16.49 $\pm$ 0.01 & 16.38 $\pm$ 0.02 & 16.25 $\pm$ 0.01 & 16.26 $\pm$ 0.04 \\
        06/02/2020 & 8886.32 & 5.9    & 16.39 $\pm$ 0.12 & 16.52 $\pm$ 0.02 & 15.87 $\pm$ 0.02 & 15.67 $\pm$ 0.02 & 15.54 $\pm$ 0.03 \\
        12/02/2020 & 8892.22 & 11.8   & 16.95 $\pm$ 0.04 & 16.96 $\pm$ 0.02 & 15.98 $\pm$ 0.02 & 15.64 $\pm$ 0.02 & 15.49 $\pm$ 0.02 \\
        13/02/2020 & 8893.18 & 12.8   & 17.03 $\pm$ 0.02 & 17.11 $\pm$ 0.02 & 16.08 $\pm$ 0.02 & 15.72 $\pm$ 0.02 & 15.56 $\pm$ 0.03 \\
        19/02/2020 & 8899.22 & 18.9   & 17.96 $\pm$ 0.03 & 17.72 $\pm$ 0.01 & 16.45 $\pm$ 0.01 & 15.94 $\pm$ 0.02 & 15.73 $\pm$ 0.02 \\
        21/02/2020 & 8901.20 & 20.9   &                 & 17.84 $\pm$ 0.02 & 16.54 $\pm$ 0.03 & 16.03 $\pm$ 0.03 & 15.78 $\pm$ 0.03 \\
        25/02/2020 & 8905.26 & 24.9   & 18.50 $\pm$ 0.04 &                 & 16.90 $\pm$ 0.02 & 16.31 $\pm$ 0.03 & 15.99 $\pm$ 0.03 \\
        02/03/2020 & 8911.14 & 30.8   & 18.86 $\pm$ 0.05 & 18.51 $\pm$ 0.02 & 17.24 $\pm$ 0.02 & 16.69 $\pm$ 0.03 & 16.28 $\pm$ 0.03 \\
        10/03/2020 & 8919.30 & 39.0   & 18.93 $\pm$ 0.08 & 18.75 $\pm$ 0.04 & 17.52 $\pm$ 0.02 & 16.94 $\pm$ 0.02 & 16.64 $\pm$ 0.04 \\
        14/03/2020 & 8923.17 & 42.8   & 19.19 $\pm$ 0.05 & 18.90 $\pm$ 0.02 & 17.68 $\pm$ 0.02 & 17.11 $\pm$ 0.02 & 16.61 $\pm$ 0.03 \\
        16/03/2020 & 8925.10 & 44.8   & 19.12 $\pm$ 0.05 & 18.90 $\pm$ 0.03 & 17.78 $\pm$ 0.03 & 17.18 $\pm$ 0.02 & 16.74 $\pm$ 0.02 \\
        20/03/2020 & 8929.27 & 48.9   & 19.13 $\pm$ 0.13 & 18.88 $\pm$ 0.02 & 17.76 $\pm$ 0.02 & 17.21 $\pm$ 0.03 & 16.71 $\pm$ 0.03 \\
        25/03/2020 & 8934.17 & 53.8   & 19.33 $\pm$ 0.06 & 18.94 $\pm$ 0.15 & 17.90 $\pm$ 0.02 & 17.38 $\pm$ 0.03 & 16.87 $\pm$ 0.02 \\
        28/03/2020 & 8937.17 & 56.8   & 19.37 $\pm$ 0.06 & 19.00 $\pm$ 0.02 & 17.90 $\pm$ 0.03 & 17.41 $\pm$ 0.03 & 16.90 $\pm$ 0.03 \\
        23/04/2020 & 8963.30 & 83.0   &                 & 19.29 $\pm$ 0.03 & 18.27 $\pm$ 0.02 & 17.77 $\pm$ 0.02 & 17.22 $\pm$ 0.03 \\
        19/05/2020 & 8989.19 &108.8   &                 &                 & 18.67 $\pm$ 0.02 & 18.07 $\pm$ 0.02 & 17.79 $\pm$ 0.02 \\
\enddata
\tablecomments{$^a$(JD $-$ 245\,0000); $^b$Observed phase with respect to the epoch of $B$ band maximum: JD = 245\,8880.4.
}
\end{deluxetable*}

\begin{deluxetable*}{lcrccccc}
\tabletypesize{\scriptsize}
\tablewidth{0pt}
\tablecaption{Optical $UBVRI$ photometric observations of SN 2021mxx with HCT. \label{tab_mag_sn2021mxx} }
\tablehead{
\colhead{Date} & \colhead{JD$^a$} & \colhead{Phase$^b$} & \colhead{$U$} & \colhead{$B$}  & \colhead{$V$} &  \colhead{$R$} & \colhead{$I$}
}
\startdata
        25/05/2021 & 9360.38 & $-$2.9 & 16.93 $\pm$ 0.12 & 17.05 $\pm$ 0.02 & 16.58 $\pm$ 0.01 & 16.44 $\pm$ 0.02 & 16.19 $\pm$ 0.02 \\
        26/05/2021 & 9361.41 & $-$1.9 & 16.68 $\pm$ 0.03 & 16.93 $\pm$ 0.03 & 16.45 $\pm$ 0.02 & 16.20 $\pm$ 0.02 & 16.03 $\pm$ 0.11 \\
        30/05/2021 & 9365.33 & 2.0    & 16.80 $\pm$ 0.06 & 16.95 $\pm$ 0.03 & 16.24 $\pm$ 0.02 & 15.95 $\pm$ 0.01 & 15.76 $\pm$ 0.03 \\
        10/06/2021 & 9376.31 & 13.0   &                  & 18.22 $\pm$ 0.01 & 16.89 $\pm$ 0.02 & 16.35 $\pm$ 0.02 & 16.11 $\pm$ 0.02 \\
        13/06/2021 & 9379.32 & 16.0   & 18.88 $\pm$ 0.13 & 18.52 $\pm$ 0.03 &                  & 16.65 $\pm$ 0.02 & 16.16 $\pm$ 0.02 \\
        14/06/2021 & 9380.26 & 16.9   & 19.08 $\pm$ 0.16 & 18.61 $\pm$ 0.02 & 16.97 $\pm$ 0.02 & 16.69 $\pm$ 0.01 & 16.25 $\pm$ 0.02 \\
        23/06/2021 & 9389.30 & 25.9   &                  & 19.13 $\pm$ 0.06 & 17.86 $\pm$ 0.03 & 17.28 $\pm$ 0.02 & 16.66 $\pm$ 0.08 \\
        25/06/2021 & 9391.34 & 28.0   & 19.20 $\pm$ 0.17 & 19.24 $\pm$ 0.09 & 18.04 $\pm$ 0.02 & 17.40 $\pm$ 0.02 & 16.62 $\pm$ 0.10 \\
        07/07/2021 & 9403.25 & 39.9   &                  & 19.56 $\pm$ 0.07 & 18.48 $\pm$ 0.02 & 17.85 $\pm$ 0.02 & 17.09 $\pm$ 0.02 \\
        14/07/2021 & 9410.29 & 46.7   & 19.88 $\pm$ 0.17 & 19.77 $\pm$ 0.04 & 18.60 $\pm$ 0.03 & 17.93 $\pm$ 0.03 & 17.13 $\pm$ 0.12 \\
        17/07/2021 & 9413.24 & 49.9   &                  & 19.98 $\pm$ 0.07 & 18.59 $\pm$ 0.03 & 17.96 $\pm$ 0.03 & 17.38 $\pm$ 0.05 \\
        21/07/2021 & 9417.19 & 53.9   & 19.90 $\pm$ 0.20 &                  & 18.78 $\pm$ 0.04 & 18.05 $\pm$ 0.03 & 17.29 $\pm$ 0.09 \\
        04/08/2021 & 9431.29 & 67.9   &                  & 20.05 $\pm$ 0.03 & 18.93 $\pm$ 0.02 & 18.28 $\pm$ 0.03 & 17.36 $\pm$ 0.10 \\
        11/08/2021 & 9438.28 & 74.9   &                  &                  & 18.92 $\pm$ 0.03 & 18.41 $\pm$ 0.03 & 17.50 $\pm$ 0.12 \\
        13/08/2021 & 9440.19 & 76.9   &                  &                  & 19.10 $\pm$ 0.02 & 18.46 $\pm$ 0.03 & 17.56 $\pm$ 0.14 \\
        16/08/2021 & 9443.27 & 79.9   &                  &                  & 19.01 $\pm$ 0.02 & 18.45 $\pm$ 0.03 & 17.78 $\pm$ 0.03 \\
        07/09/2021 & 9465.30 & 101.9  &                  & 20.52 $\pm$ 0.06 & 19.23 $\pm$ 0.05 & 18.75 $\pm$ 0.14 & 18.03 $\pm$ 0.20 \\
        13/09/2021 & 9471.21 & 107.9  &                  &                  & 19.44 $\pm$ 0.04 & 18.99 $\pm$ 0.04 & 18.12 $\pm$ 0.09 \\
        18/09/2021 & 9476.17 & 112.9  &                  & 20.89 $\pm$ 0.09 & 19.48 $\pm$ 0.04 & 18.94 $\pm$ 0.14 & 18.16 $\pm$ 0.11 \\
        19/09/2021 & 9477.25 & 113.9  &                  & 20.88 $\pm$ 0.13 & 19.56 $\pm$ 0.05 & 19.06 $\pm$ 0.07 & 18.20 $\pm$ 0.12 \\
        20/09/2021 & 9478.12 & 114.8  &                  &                  & 19.51 $\pm$ 0.05 & 19.05 $\pm$ 0.05 & 18.61 $\pm$ 0.07 \\
\enddata
\tablecomments{$^a$(JD $-$ 245\,0000); $^b$Observed phase with respect to the epoch of $B$ band maximum: JD = 245\,9363.3.
}
\end{deluxetable*}

\begin{deluxetable}{ccccc}
\tabletypesize{\scriptsize}
\tablewidth{0pt}
\tablecaption{Photometric parameters of SN~2020akf and SN~2021mxx. \label{tab_lc_parameter}}
\tablehead{ 
\colhead{Band} & \colhead{JD (max)$^a$} & \colhead{$m_{\lambda}^{\text{max}}$} &
\colhead{$M_{\lambda}^{\text{max}}$} &  \colhead{$\Delta m_{15}(\lambda)$}
}
\startdata
\multicolumn{5}{c}{\textbf{SN~2020akf}} \\
$U$ & 8878.6 $\pm$ 0.5 & 15.95 $\pm$ 0.09 & $-17.72 \pm 0.16$ & 1.20 $\pm$ 0.08 \\
$B$ & 8880.4 $\pm$ 0.5 & 16.34 $\pm$ 0.02 & $-17.38 \pm 0.15$ & 0.98 $\pm$ 0.02 \\
$V$ & 8886.2 $\pm$ 0.5 & 15.86 $\pm$ 0.02 & $-17.85 \pm 0.15$ & 0.69 $\pm$ 0.03 \\
$R$ & 8889.0 $\pm$ 0.5 & 15.64 $\pm$ 0.02 & $-18.06 \pm 0.15$ & 0.58 $\pm$ 0.03 \\
$I$ & 8889.7 $\pm$ 0.5 & 15.48 $\pm$ 0.03 & $-18.21 \pm 0.15$ & 0.48 $\pm$ 0.04 \\
\\[-2.5ex]
\multicolumn{5}{c}{\textbf{SN~2021mxx}} \\
$U$ & 9362.8 $\pm$ 0.5 & 16.64 $\pm$ 0.06 & $-17.58 \pm 0.16$ & 1.96 $\pm$ 0.10 \\
$B$ & 9363.3 $\pm$ 0.5 & 16.90 $\pm$ 0.03 & $-17.14 \pm 0.17$ & 1.52 $\pm$ 0.03 \\
$V$ & 9365.1 $\pm$ 0.5 & 16.25 $\pm$ 0.02 & $-17.60 \pm 0.15$ & 0.74 $\pm$ 0.03 \\
$R$ & 9366.8 $\pm$ 0.5 & 15.93 $\pm$ 0.02 & $-17.84 \pm 0.15$ & 0.89 $\pm$ 0.02 \\
$I$ & 9366.8 $\pm$ 0.5 & 15.74 $\pm$ 0.04 & $-17.87 \pm 0.16$ & 0.54 $\pm$ 0.05 \\
\enddata
\tablecomments{$^a$(JD $-$ 245\,0000)
}
\end{deluxetable}

\begin{deluxetable}{lccc}
\tabletypesize{\scriptsize}
\tablewidth{0pt}
\tablecaption{Log of Spectroscopic observations of SN 2020akf and SN 2021mxx with HCT. \label{tab_spec_2020akf_2021mxx} }
\tablehead{
\colhead{Date} & \colhead{JD$^a$} & \colhead{Phase$^b$} & \colhead{Range (\AA)}
}
\startdata
\multicolumn{4}{c}{\textbf{SN 2020akf}} \\
    26/01/2020 & 8875.27 & $-$5.1  & 3500--7800; 5200--9250 \\
    12/02/2020 & 8892.21 & 11.8   & 3500--7800; 5200--9250 \\
    13/02/2020 & 8893.17 & 12.8   & 3500--7800; 5200--9250 \\
    19/02/2020 & 8899.21 & 18.8   & 3500--7800 \\
    25/02/2020 & 8905.26 & 24.9   & 3500--7800; 5200--9250 \\
    09/03/2020 & 8918.22 & 37.8   & 3500--7800; 5200--9250 \\
    14/03/2020 & 8923.16 & 42.8   & 3500--7800; 5200--9250 \\
    25/03/2020 & 8934.16 & 53.8   & 3500--7800; 5200--9250 \\
    28/03/2020 & 8937.16 & 56.8   & 3500--7800; 5200--9250 \\
\multicolumn{4}{c}{\textbf{SN 2021mxx}} \\
    25/05/2021 & 9360.38 & $-$2.9  & 3500--7800; 5200--9250 \\
    26/05/2021 & 9361.40 & $-$1.9  & 3500--7800; 5200--9250 \\
    10/06/2021 & 9376.31 & 13.0   & 3500--7800; 5200--9250 \\
    13/06/2021 & 9379.31 & 17.0   & 3500--7800; 5200--9250 \\
    23/06/2021 & 9389.29 & 26.0   & 3500--7800; 5200--9250 \\
    25/06/2021 & 9391.34 & 28.0   & 3500--7800 \\
    07/07/2021 & 9403.25 & 39.9   & 3500--7800; 5200--9250 \\
    14/07/2021 & 9410.29 & 47.0   & 3500--7800 \\
    16/07/2021 & 9412.29 & 49.9   & 3500--7800; 5200--9250 \\
    28/08/2021 & 9455.12 & 91.8   & 5200--9250 \\
    07/09/2021 & 9465.29 & 102.0  & 5200--9250 \\
    18/09/2021 & 9476.17 & 112.9  & 5200--9250 \\
\enddata
\tablecomments{$^a$(JD $-$ 245\,0000); $^b$Phase in days relative to $B$-maximum.
}
\end{deluxetable}

\begin{deluxetable}{lll}
\tabletypesize{\scriptsize}
\tablewidth{0pt}
\tablecaption{Various parameters and priors used in the MOSFiT \textit{default} model for SN~2020akf and SN~2021mxx. ‘G’ denotes a Gaussian distribution and most of the priors are in log space; see corner plots in Figures \ref{fig_cakf_Mosfit} and \ref{fig_cmxx_Mosfit} for reference. \label{tab:mosfit_akf_parameters} }
\tablehead{
\colhead{Parameter} & \colhead{Priors type and Priors} & \colhead{Results} 
}
\startdata
\multicolumn{3}{c}{\textbf{SN~2020akf}}\\
$f_{\text{Ni}}$            & log-flat [10$^{-3}$, 1]                              & 0.06$^{+0.01}_{-0.01}$      \\
M$_{\text{ej}}$            & log-flat [10$^{-2}$, 100]                            & 3.16$^{+0.55}_{-0.47}$     \\
$\kappa_\gamma$            & G [$\mu$=0.06, $\sigma$=0.01, (10$^{-3}$, 10$^4$)]   & 0.06$^{+0.001}_{-0.001}$     \\
$v_{\text{ej}}$            & G [$\mu$=10$^4$, $\sigma$=2000, (2$\times$10$^3$, 15$\times$10$^5$)]  & 7519.88$^{+687}_{-644}$ \\
$T_{\text{min}}$           & flat [3.5$\times$10$^2$, 10$^5$]                     & 4074.03$^{+92.54}_{-86.25}$     \\
$\sigma$                   & log-flat [10$^{-2}$, 1]                              & 0.36$^{+0.03}_{-0.02}$     \\
$t_{\text{exp}}$ (days)    & flat [$-$15, 10]                                     & $-$7.59$^{+1.23}_{-1.44}$        \\
Score $\log(Z)$            &                                                      & 36.24           \\
\multicolumn{3}{c}{\textbf{SN~2021mxx}}\\
$f_{\text{Ni}}$            & log-flat [10$^{-2}$, 0.15]                           & 0.06$^{+0.01}_{-0.01}$                  \\
M$_{\text{ej}}$            & flat [10$^{-2}$, 10]                                & 1.53$^{+0.18}_{-0.17}$                  \\
$\kappa_\gamma$            & G [$\mu$=0.06, $\sigma$=0.01, (10$^{-2}$, 10$^4$]   & 0.06$^{+0.001}_{-0.001}$                  \\
$V_{\text{ej}}$            & G [$\mu$=10$^4$, $\sigma$=2000, (2$\times$10$^3$, 15$\times$10$^5$)] & 7958.62$^{+584}_{-531}$ \\
$T_{\text{min}}$           & log-flat [3.5$\times$10$^3$, 10$^6$]                    & 3981$^{+93}_{-91}$                      \\
$\sigma$                   & log-flat [10$^{-2}$, 1]                                 & 0.25$^{+0.02}_{-0.02}$                  \\
$t_{\text{exp}}$ (days)    & flat [$-$15, 10]                                    & $-$10.14$^{+0.71}_{-0.78}$              \\
Score $\log(Z)$            &                                                         & 113                                     \\
\enddata
\end{deluxetable}

\begin{figure*}[ht!]
\centering
\includegraphics[width=0.48\textwidth]{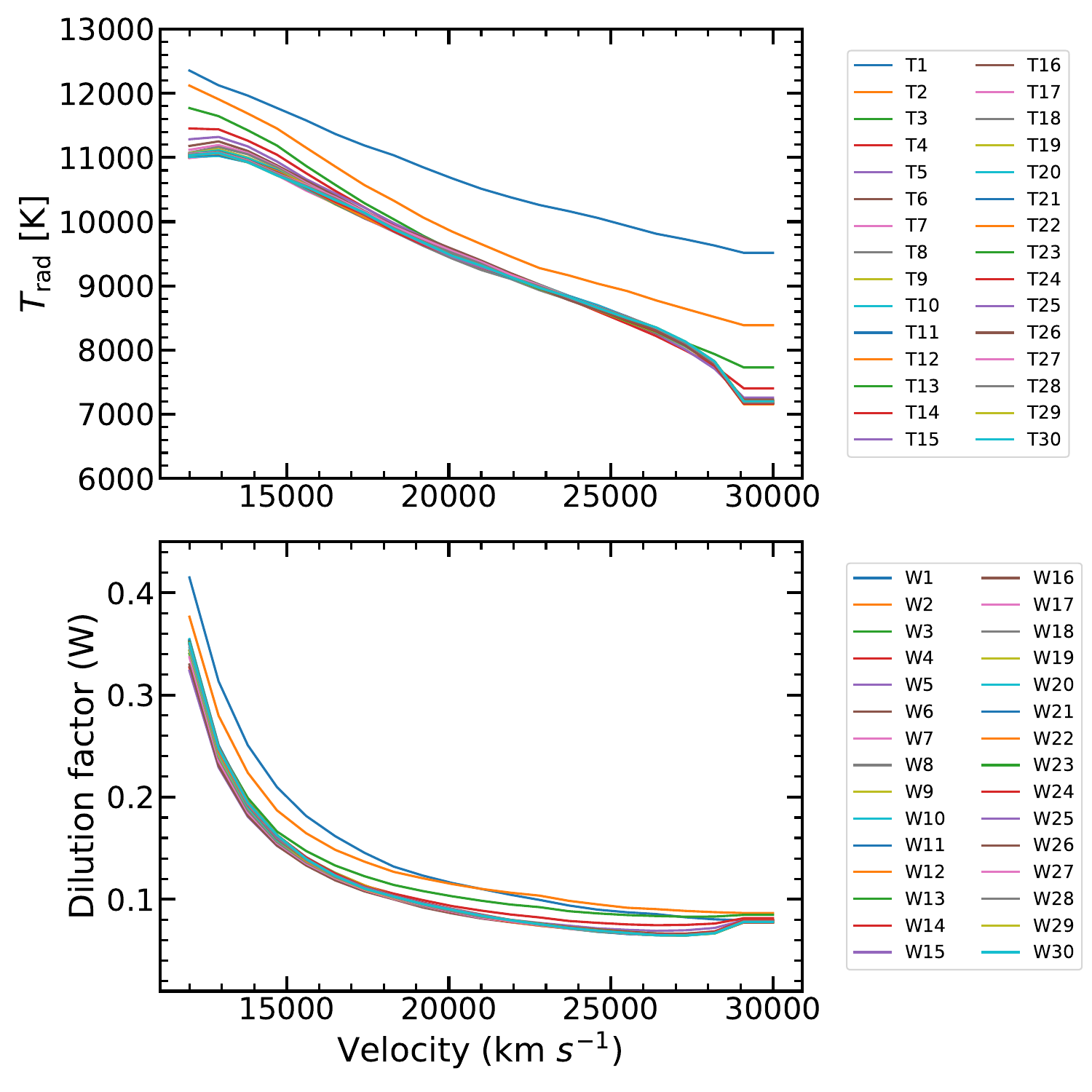}\hfill
\includegraphics[width=0.48\textwidth]{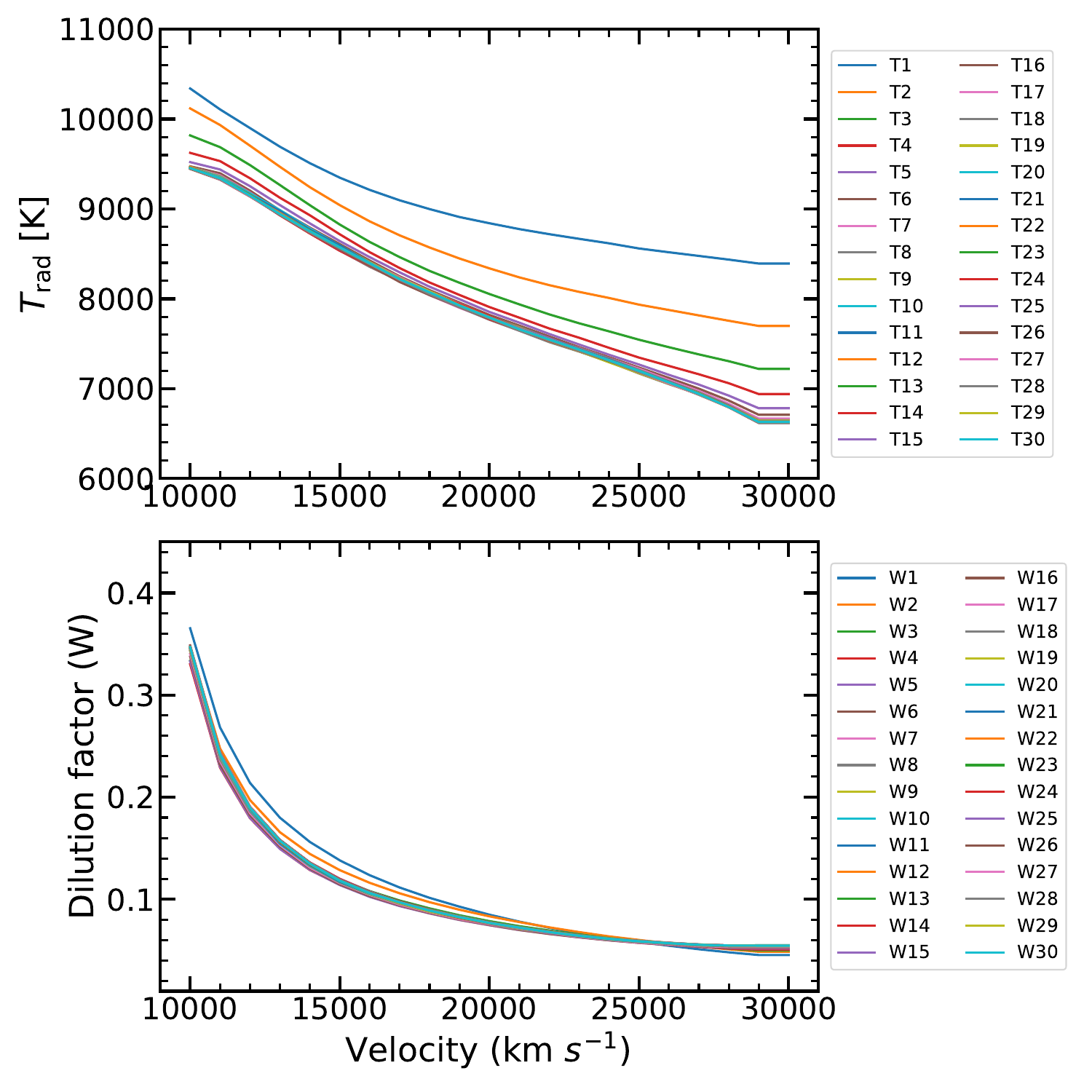}
\caption{Convergence profiles of the \texttt{TARDIS} model for SN 2020akf at $-$5 days (left column) and SN 2021mxx at $-$3 days (right column). The top panels show the distribution of the radiation temperature ($T_{\mathrm{rad}}$), and the bottom panels present the corresponding dilution factor ($W$) as a function of velocity.}
    \label{fig_converge_TW}
\end{figure*}

\begin{figure*}[ht!]
\centering
\includegraphics[width=0.48\textwidth, trim={0 0 0 0},clip]{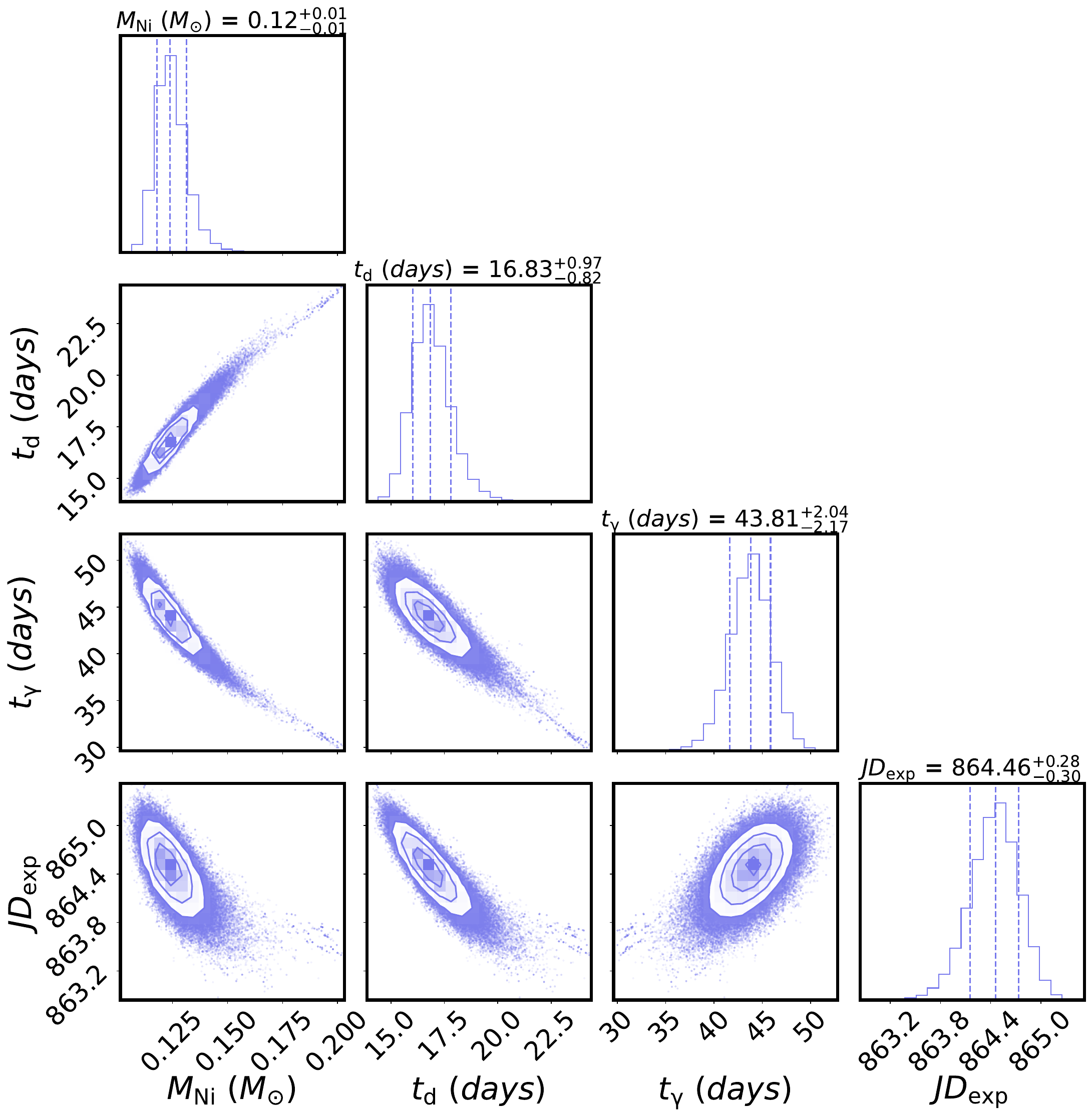}\hfill
\includegraphics[width=0.48\textwidth]{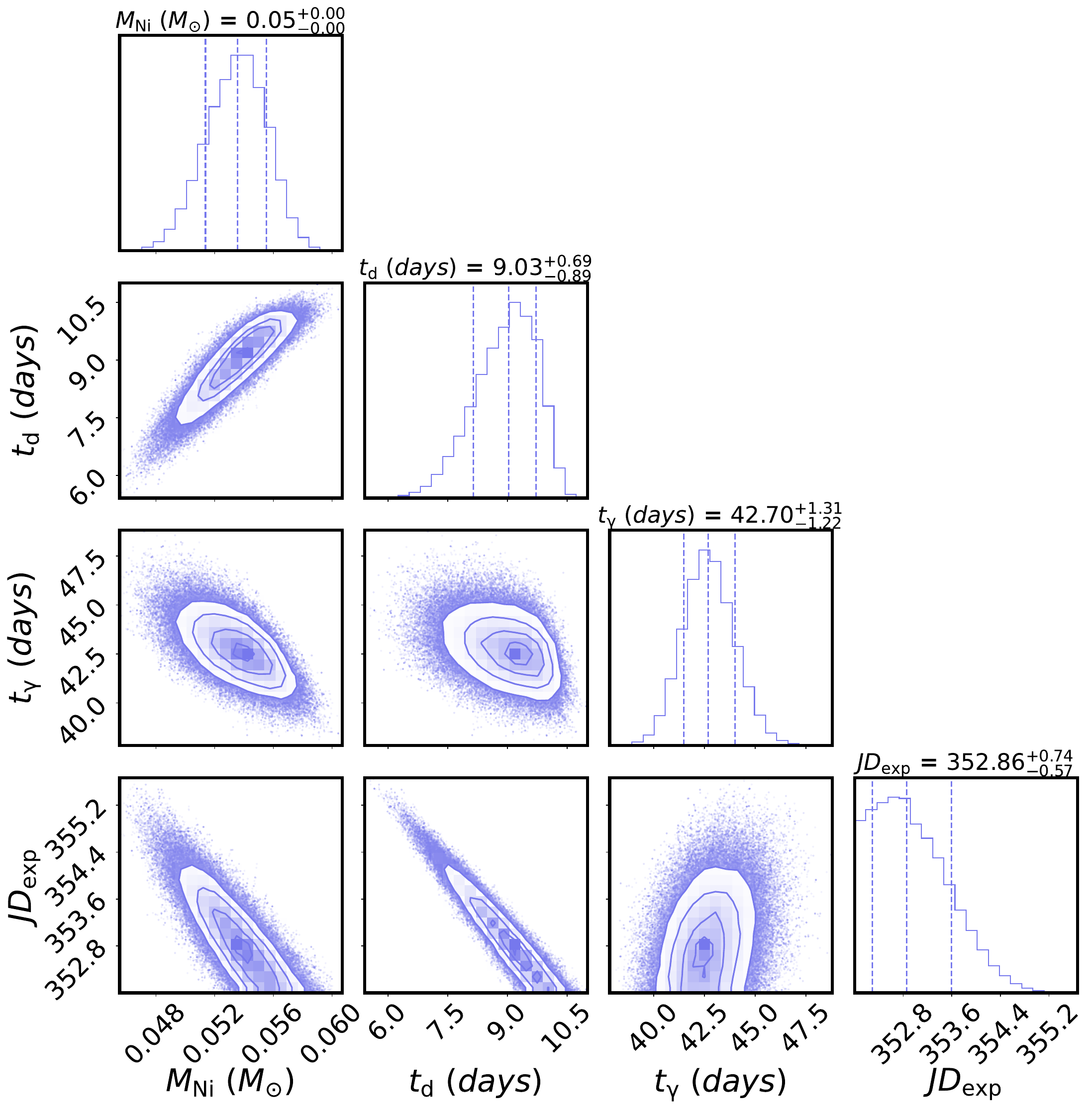}
\caption{The 1D and 2D posterior distributions of the parameters in the Arnett–Valenti model fit. The median values and 1$\sigma$ confidence intervals are marked, highlighting the best-fit values of key physical parameters, including the nickel mass (M$_{\text{Ni}}$), diffusion time ($t_{d}$), gamma-ray leakage timescale ($t_{\gamma}$), and explosion epoch (JD$_{\text{exp}}$) for SN 2020akf (left column) and SN 2021mxx (right column).}
\label{fig_20akf_21mxx_corner_arnett}
\end{figure*}

\begin{figure*}[ht!]
\centering
\includegraphics[width=0.8\textwidth]{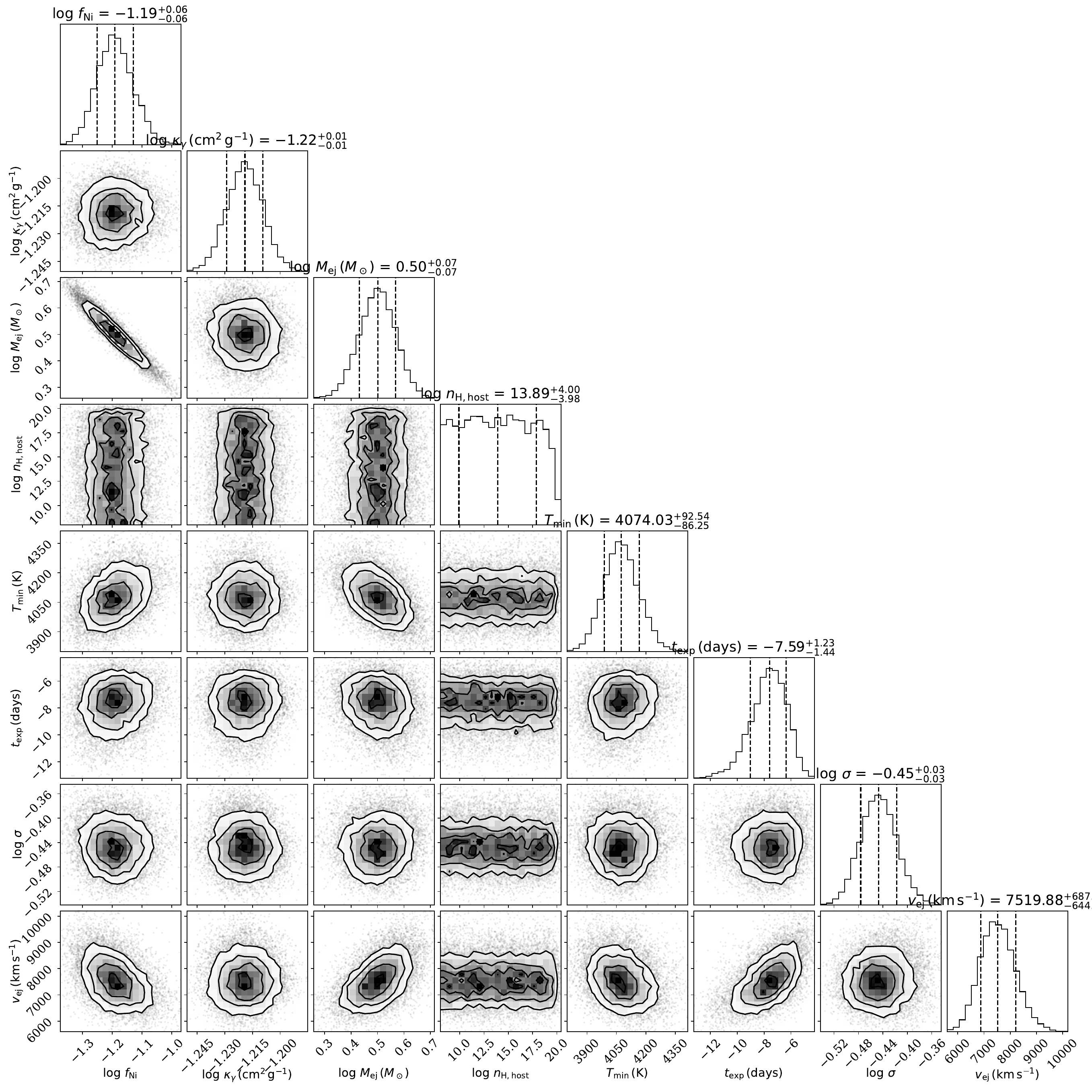}
\caption{The 1D and 2D posterior distributions of the {\textit{default}} $^{56}$Ni model parameters in MOSFiT used for SN 2020akf. The median and $1\sigma$ confidence intervals are marked to indicate the best-fit values of important physical parameters such as nickel fraction $f_{\text{Ni}}$, ejecta mass M$_{\text{ej}}$, ${k}$, $k_{\gamma}$, and $t_\text{exp}$, etc. }   
\label{fig_cakf_Mosfit}
\end{figure*}

\begin{figure*}[htbp]
\centering
\includegraphics[width=0.9\textwidth]{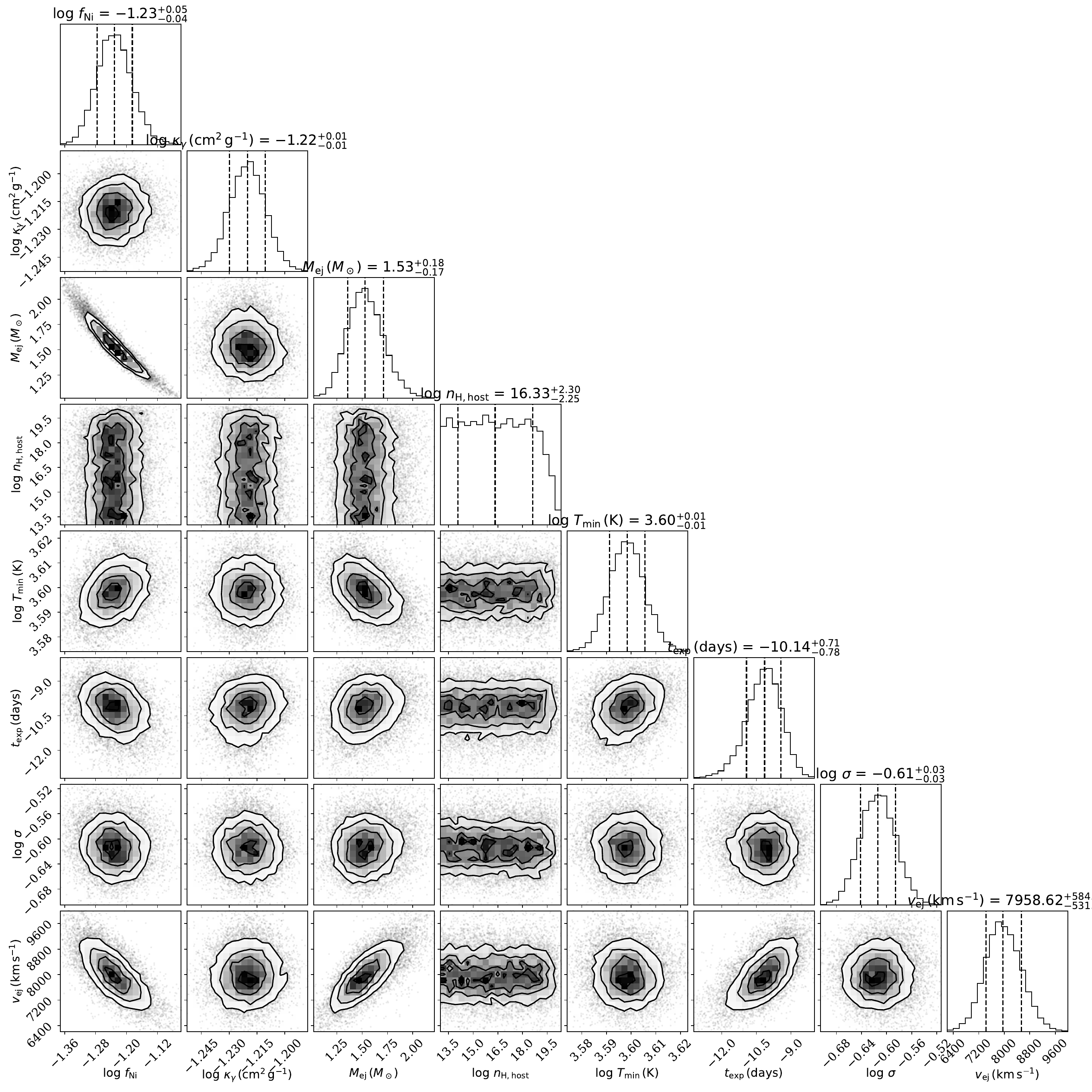}
\caption{The 1D and 2D posterior distributions of the \textit{default} $^{56}$Ni model parameters in MOSFiT used for SN 2021mxx. The median and $1\sigma$ confidence intervals are marked to indicate the best-fit values of important physical parameters such as nickel fraction $f_{\text{Ni}}$, ejecta mass M$_{\text{ej}}$, ${k}$, $k_{\gamma}$, and $t_\text{exp}$, etc.}   
\label{fig_cmxx_Mosfit}
\end{figure*}

\end{document}